\documentclass[bibyear]{aa}
\usepackage{graphicx}
\usepackage{gensymb}
\usepackage{amssymb}
\usepackage{tabulary}

\begin{document}
   \title{An HST/COS legacy survey of high-velocity UV absorption 
   in the Milky Way's circumgalactic medium and the Local Group
}

   \author{
          P. Richter \inst{1,2},
          S.E. Nuza \inst{2,3,4},        
          A.J. Fox \inst{5},
          B.P. Wakker \inst{6},
          N. Lehner \inst{7},
          N. Ben\,Bekhti \inst{8}
          C. Fechner \inst{1},
          M. Wendt \inst{1},
          J.C. Howk \inst{7},
          S. Muzahid \inst{9},
          R. Ganguly \inst{10},
       \and
          J.C. Charlton \inst{11}
          }
   \offprints{P. Richter\\
   \email{prichter@astro.physik.uni-potsdam.de}}

   \institute{Institut f\"ur Physik und Astronomie, Universit\"at Potsdam,
             Karl-Liebknecht-Str.\,24/25, 14476 Golm, Germany
   \and
   Leibniz-Institut f\"ur Astrophysik Potsdam (AIP), An der Sternwarte 16,
   14482 Potsdam, Germany
   \and
Instituto de Astronom\'{\i}a y F\'{\i}sica del Espacio (IAFE, CONICET-UBA), CC 67, Suc. 28, 1428 Buenos Aires, Argentina
\and
Facultad de Ciencias Exactas y Naturales (FCEyN), Universidad de Buenos Aires (UBA), Buenos Aires, Argentina
   \and
   Space Telescope Science Institute, 3700 San Martin Drive,
   Baltimore, MD 21218, USA
   \and
   Supported by NASA/NSF, affiliated with the Department of Astronomy, 
   University of Wisconsin-Madison,
   475 North Charter Street, Madison, WI 53706, USA
   \and
   Department of Physics, University of Notre Dame, 
   225 Nieuwland Science Hall, Notre Dame, IN 46556, USA
   \and
   FHR, Fraunhoferstr. 20, 53343 Wachtberg, Germany
   \and
   Leiden Observatory, University of Leiden, P.O. Box 9513, 2300 RA
   Leiden, Netherlands
   \and
   Department of Computer Science, Engineering, \& Physics,
   University of Michigan-Flint, Murchie Science Building,
   303 Kearsley Street, Flint, MI 48502, USA
   \and
   Department of Astronomy and Astrophysics, Pennsylvania State University,
   University Park, PA 16802, USA
   }



\abstract
{The Milky Way is surrounded by large amounts of diffuse gaseous matter
that connects the stellar body of our Galaxy with its large-scale
Local Group (LG) environment.}
{To characterize the absorption properties of this circumgalactic medium (CGM)
and its relation to the LG we present the so-far largest survey of metal
absorption in Galactic high-velocity clouds (HVCs) 
using archival ultraviolet (UV) spectra of extragalactic background sources.
The UV data are obtained with the Cosmic Origins Spectrograph (COS) onboard
the {\it Hubble Space Telescope} (HST) and are supplemented by
21 cm radio observations of neutral hydrogen.}
{Along 270 sightlines we measure metal absorption in the
lines of Si\,{\sc ii}, Si\,{\sc iii},  C\,{\sc ii}, and C\,{\sc iv} 
and associated H\,{\sc i} 21 cm emission in HVCs
in the velocity range $|v_{\rm LSR}|=100-500$ km\,s$^{-1}$. 
With this unprecedented large HVC sample we were able to improve the
statistics on HVC covering fractions, ionization conditions, small-scale 
structure, CGM mass, and inflow rate. For the first time, we determine 
robustly the angular two point correlation function of the high-velocity 
absorbers, systematically analyze antipodal sightlines on the celestial 
sphere, and compare the HVC absorption characteristics with that
of Damped Lyman $\alpha$ absorbers (DLAs) and constrained cosmological 
simulations of the LG (CLUES project).}
{The overall sky-covering fraction of high-velocity absorption
is $77\pm 6$ percent for the most sensitive ion in our survey, Si\,{\sc iii},
and for column densities log $N$(Si\,{\sc iii}$)\geq 12.1$.
This value is $\sim 4-5$ times higher than the covering fraction of
21 cm neutral hydrogen emission at log $N$(H\,{\sc i}$)\geq 18.7$ 
along the same lines of sight, demonstrating
that the Milky Way's CGM is multi-phase and predominantly ionized.
The measured equivalent-width ratios of Si\,{\sc ii}, Si\,{\sc iii},
C\,{\sc ii}, and C\,{\sc iv} are 
inhomogeneously distributed on large and small angular scales, suggesting
a complex spatial distribution of multi-phase gas that surrounds the
neutral 21 cm HVCs. We estimate that the total mass and accretion rate 
of the neutral and ionized CGM traced by HVCs is
$M_{\rm HVC}\geq 3.0 \times 10^9 M_{\sun}$ and 
d$M_{\rm HVC}/$d$t \geq 6.1 M_{\sun}$\,yr$^{-1}$, where the
Magellanic Stream (MS) contributes with more than $90$ percent
to this mass/mass-flow. If seen from an external vantage point, 
the Milky Way disk plus CGM would appear as a DLA
that would exhibit for most viewing angles an extraordinary 
large velocity spread of $\Delta v\approx 400-800$ km\,s$^{-1}$, 
a result of the complex kinematics of the Milky Way CGM that is 
dominated by the presence of the MS.
We detect a velocity dipole of high-velocity absorption at low/high
galactic latitudes that we associate 
with LG gas that streams to the LG barycenter. This scenario is supported 
by the gas kinematics predicted from the LG simulations.}
{Our study confirms previous results, indicating that the Milky Way CGM 
contains sufficient gaseous material to feed the Milky Way disk over the next
Gyr at a rate of a few solar masses per year, if the CGM gas can actually 
reach the MW disk.
We demonstrate that the CGM is composed of 
discrete gaseous structures that exhibit a large-scale kinematics together 
with small-scale variations in physical conditions. 
The MS clearly dominates both the cross section and mass flow of high-velocity
gas in the Milky Way's CGM.
The possible presence of high-velocity LG gas underlines
the important role of the local cosmological environment in the
large-scale gas-circulation processes in and around the Milky Way.
}


\titlerunning{The Milky Way CGM in absorption}
\authorrunning{Richter et al.}
\maketitle
%


\section{Introduction}

Observational and theoretical studies indicate that a substantial
(if not dominant) fraction of the diffuse gaseous material
in spiral galaxies is situated outside the disk in an
extended halo that reaches out to the virial 
radius ($R_{\rm vir}$). This extraplanar diffuse gas component is 
commonly referred to as the circumgalactic medium (CGM).
Beyond the virial radius there also exists a large amount 
of diffuse gas that is gravitationally bound to the large-scale
cosmological environment in which the galaxies reside (Wakker et al.\,2015).
Depending on the nature of this cosmological environment,
the diffuse gas beyond $R_{\rm vir}$ is called
the intergalactic medium, IGM, or the intragroup/intracluster medium 
in case of an IGM bound in galaxy groups and clusters.
The life cycle of such gas is determined by cosmological structure formation,
infall and outflow from galaxies, and galaxy merging. As a result,
gas far beyond the stellar bodies of galaxies spans large ranges
in physical conditions and chemical abundances, 
but also represents a major baryon reservoir (e.g., Fox et al.\,2014;
Tumlinson et al.\,2013; Werk et al.\,2013; Liang \& Chen 2014; 
Nuza et al.\,2014; Richter et al.\,2016).

The Milky Way is known to also be surrounded by diffuse gas that 
originates in the Galaxy's CGM and in the LG
(see Wakker \& van Woerden 1997;
Richter 2006; Putman et al.\,2012; Richter 2017 for reviews). Recent 
studies further indicate that M31 also has an extended gaseous halo
(Lehner et al.\,2015). 
In contrast to more distant galaxies, the gaseous outskirts of the 
Milky Way and M31 can be studied in great detail owing to the large amounts 
of emission and absorption-line data that are available from various 
instruments, as well as from simulations (e.g., Fox et al.\,2014; Nuza
et al.\,2014, hereafter refered to as N14).
Studies that focus on characterizing the nature of the Milky Way's
CGM and its connection to LG gas and M31 thus are
crucial for our general understanding of circumgalactic gas in the
local Universe.

As a result of our position within the disk of the
Milky Way, the local CGM/IGM can be identified most efficiently from
its kinematics, as the bulk of the gas does not participate in
the rotational motion of the Galaxy's thin and thick disk (e.g.,
Wakker \& van Woerden 1997). 
The characteristic Local Standard of Rest (LSR) velocity range of gas in the Milky Way's
CGM and in the LG thus lies between $|v_{\rm LSR}|=50$ and $450$ km\,s$^{-1}$ 
(Wakker \& van Woerden 1997), although CGM gas at lower 
velocities might exist (Peek et al.\,2009; Zheng et al.\,2015).
The Doppler-shifted CGM therefore can be observed either through emission or absorption lines 
of hydrogen and heavy elements with spectrographs that exhibit
sufficient spectral resolution. In addition to radio observations that
can be used to pinpoint the amount of predominantly neutral 
gas in the local CGM using the 21 cm hyperfine structure transition 
of H\,{\sc i}, ultraviolet (UV) and optical absorption spectra 
of extragalactic background sources are extremely useful to 
study metal-ion absorption in both neutral and ionized gas
down to very low gas column densities. UV satellites such as the 
{\it Far Ultraviolet Spectroscopic Explorer} (FUSE) and the
various UV spectrographs installed on HST
(e.g., the Space Telescope Imaging Spectrograph, STIS, and 
COS) have been very successful in providing
large amounts of UV absorption-line data to study the CGM of the Milky Way in great detail
(e.g., Wakker et al.\,2003; Sembach et al.\,2003; Lehner et al.\,2012; 
Herenz et al.\,2013; Fox et al.\,2014).

In the canonical classification scheme of the Milky Way’s 
circumgalactic gas components one defines the so-called 
``high-velocity clouds'' (HVCs) as gaseous structures that are 
observed in H\,{\sc i} 21 cm radio emission or in 
line-absorption against extragalactic background sources at 
LSR velocities $|v_{\rm LSR}|\geq 100$ km\,s$^{-1}$.
The ``intermediate-velocity clouds'' (IVCs) represent
extra-planar gaseous features at lower radial velocities 
($|v_{\rm LSR}|=50-100$ km\,s$^{-1}$).
IVCs are often related to gas in the disk-halo interface 
at low vertical distances from the disk 
($|z|\leq 2$ kpc; Wakker 2001). 


\begin{figure*}[t!]
\begin{center}
\resizebox{0.95\hsize}{!}{\includegraphics{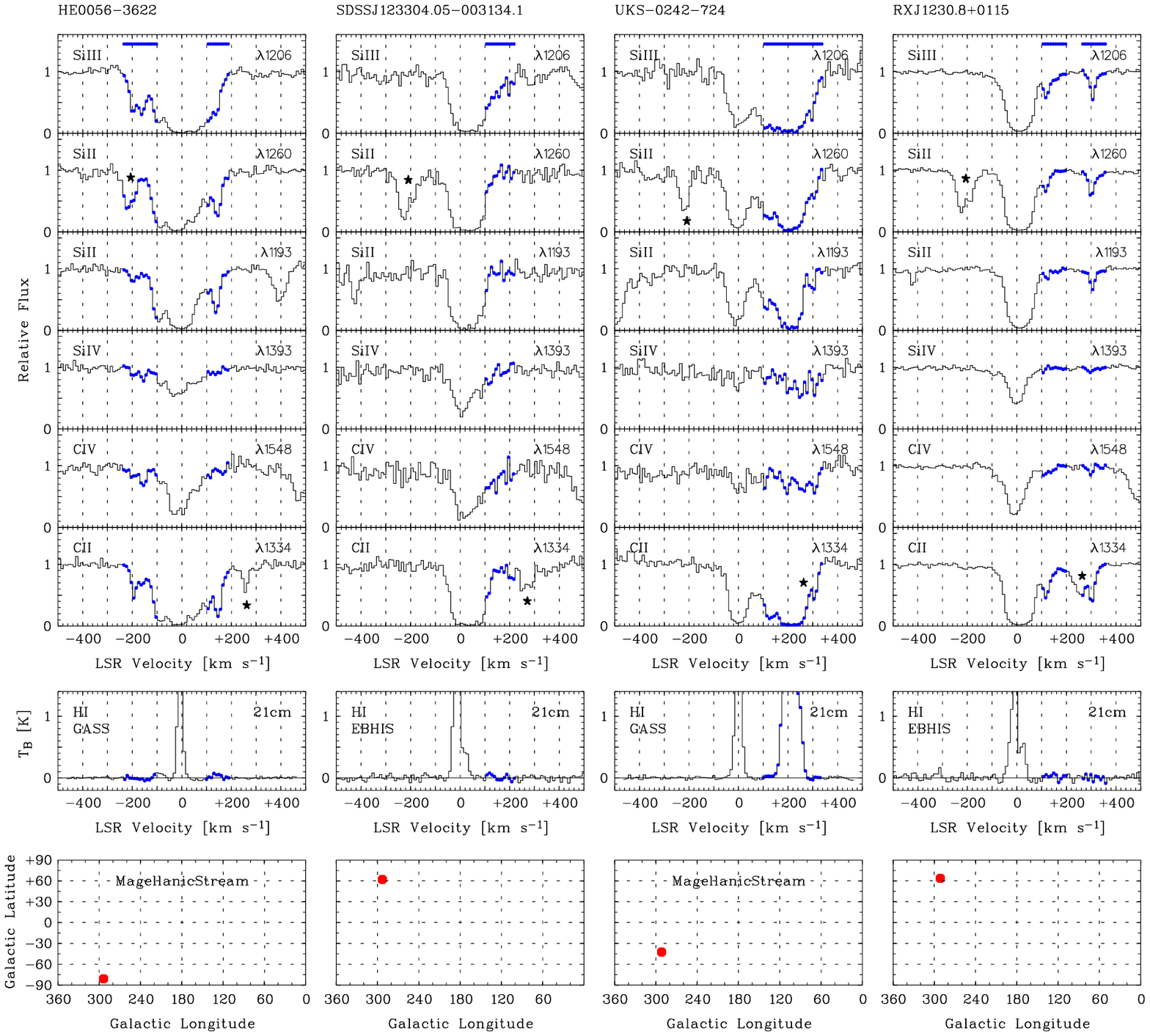}}
\caption[]{
Examples for HVC finding charts from our survey.
In the upper panel the absorption profiles
of Si\,{\sc iii}, Si\,{\sc ii}, C\,{\sc iv}, and C\,{\sc ii} are shown.
For comparison, Si\,{\sc iv} absorption is also displayed.
Regions with high-velocity absorption are indicated in blue,
with the velocity range indicated with the blue bar in the top
panel. We further show the 21 cm emission profile for the same
sightlines (middle panel) and the sightline position in the
$(l,b)$ plane (lower panel) together with the HVC identification
(see Sect.\,5.1). The star symbol indicates a blend with another ISM line
(S\,{\sc ii} $\lambda 1259$ in the Si\,{\sc ii} $\lambda 1260$ panel, 
C\,{\sc ii}$^{\star}\,\lambda 1335$ in the C\,{\sc ii} $\lambda 1334$ panel).
The Appendix contains this plot for all 270 sightlines.
}
\end{center}
\end{figure*}


Over the last decades there has been substantial progress in
characterizing the role of the Galactic CGM in the 
on-going evolution of the Milky Way. The combination of 
21 cm and UV spectral data allowed 
researchers to pinpoint the chemical composition of the 
gas and its physical conditions 
(Wakker et al.\,1999, 2003; Sembach et al.\,2003;
Richter et al.\,1999, 2001a, 2001b, 2009, 2012, 2013; Tripp et al.\,2012;
Fox et al.\,2005, 2010, 2013, 2014; Fox, Savage \& Wakker 2006; 
Ben\,Bekhti et al.\,2008, 2012;
Lehner et al.\,2010, 2011, 2012; Collins et al.\,2009; Shull et al.\,2009). 
In general, the 
Milky Way's CGM is metal-enriched with $\alpha$ abundances in
the range $0.1-1.0$ solar (e.g., Wakker et al.\,1999; Richter et
al.\,2001b; Fox et al.\,2013). 
The presence of low and high ions in the gas reflects its 
extreme multi-phase nature
with temperatures ranging from $T=10^2$ to $10^7$ K. Recent
studies imply that the bulk of the Milky Way CGM baryons reside in a diffuse 
(at gas densities $n_{\rm H}\leq 10^{-3}$ cm$^{-3}$), 
predominantly ionized gas phase that can be detected 
in UV and X-ray absorption lines of intermediate and high ions
against extragalactic background point sources (e.g., Sembach et al.\,2003; 
Wakker et al.\,2003; Shull et al.\,2009; Richter et al.\,2008, 2009;
Lehner et al.\,2012; Miller et al.\,2016).
Another important finding from absorption spectroscopy 
is that most of the spatially extended neutral gas features 
that are seen in H\,{\sc i} 21 cm emission are relatively nearby 
at distances $d<15$ kpc (Ryans et al.\,1997a, 1997b; Wakker et al.\,2007, 2008; 
Thom et al.\,2006, 2008; Lehner et al.\,2012; Richter et al.\,2015),
ruling that that these objects represent
pristine, extragalactic gas clouds (see discussion in Blitz et al.\,1999). 
Many of the {\it ionized} HVCs are located at similarly small distances 
deep within the potential well of the Milky Way (Lehner \& Howk 2011;
Lehner et al.\,2012), but some of observed high-ion absorbers 
possibly are related to LG gas (Sembach et al.\,2003; Nicastro et al.\,2003).

The only neutral high-velocity structure in the Galactic CGM 
that clearly reaches beyond 30 kpc distance is the MS,
a massive stream ($10^8-10^9$ $M_{\sun}$) of neutral and ionized gas 
that originates from the interaction of the two Magellanic Clouds
as they approach the Milky Way (Dieter 1971; Wannier \& Wrixon 1972;
Mathewson et al.\,1974; D'Onghia \& Fox 2016). Our recent studies of the MS based on
UV data from FUSE, HST/STIS and HST/COS 
(Fox et al.\,2010, 2013, 2014; Richter et al.\,2013) indicate that
the MS contains more gas than the Small and Large Magellanic
Clouds, feeding the Milky Way over the next $0.5-1.0$ Gyr
with large amounts of mostly metal-poor gas. 

In this paper, we present the so-far largest survey of UV absorption
features in the Milky Way's CGM and in the LG 
using archival HST/COS data of 270 extragalactic 
background sources. The aim of this paper is to pinpoint the
absorption characteristics of the Milky Way's CGM and 
gas in the LG based on proper statistics, to explore
the overall physical conditions and the distribution of different 
gas phases in the gaseous outskirts of our Galaxy, to 
estimate the total gas mass in HVCs and the Milky Way’s 
gas accretion rate, 
and to compare the absorption properties 
of the Milky Way disk+CGM with that of other galaxies.
The HVC absorption catalog presented here provides an excellent data
base for the comparison between the Milky Way CGM and circumgalactic gas 
around other low-$z$ galaxies, as traced by intervening absorption-line systems
(e.g., Prochaska 2017). Since massive gas streams from merger events like the
MS are rare in low redshift galaxies, we provide statistical results on 
the absorption properties of the Milky Way CGM including and excluding 
the MS contribution.
Note that in this paper we {\it do not} investigate the chemical 
composition of the Galactic CGM, dust depletion patterns
in the CGM, the role of outflows, or detailed ionization conditions in individual
HVCs; these aspects will be presented in forthcoming studies.

This paper is organized as follows:
in Sect.\,2 we describe the observations, the data reduction and
the analysis method.  
In Sect.\,3 we characterize the global absorption properties of the Milky
Way CGM and LG gas.
The distribution of equivalent widths and column densities of 
different metal ions is presented in Sect.\,4.
In Sect.\,5 we discuss structural properties of the CGM
(HVC complexes, relation to LG galaxies and LG gas, small-scale structure).
An estimate of the HVC ionized gas content, total mass,
and accretion rate is presented in Sect.\,6.
In Sect.\,7 we relate
the absorption characteristics of the Milky Way disk+halo to that of
DLAs and compare the observations with 
predictions from constrained cosmological simulations of the LG.
Finally, we conclude our study in Sect.\,8.


\begin{figure*}[t!]
\begin{center}
\resizebox{0.9\hsize}{!}{\includegraphics{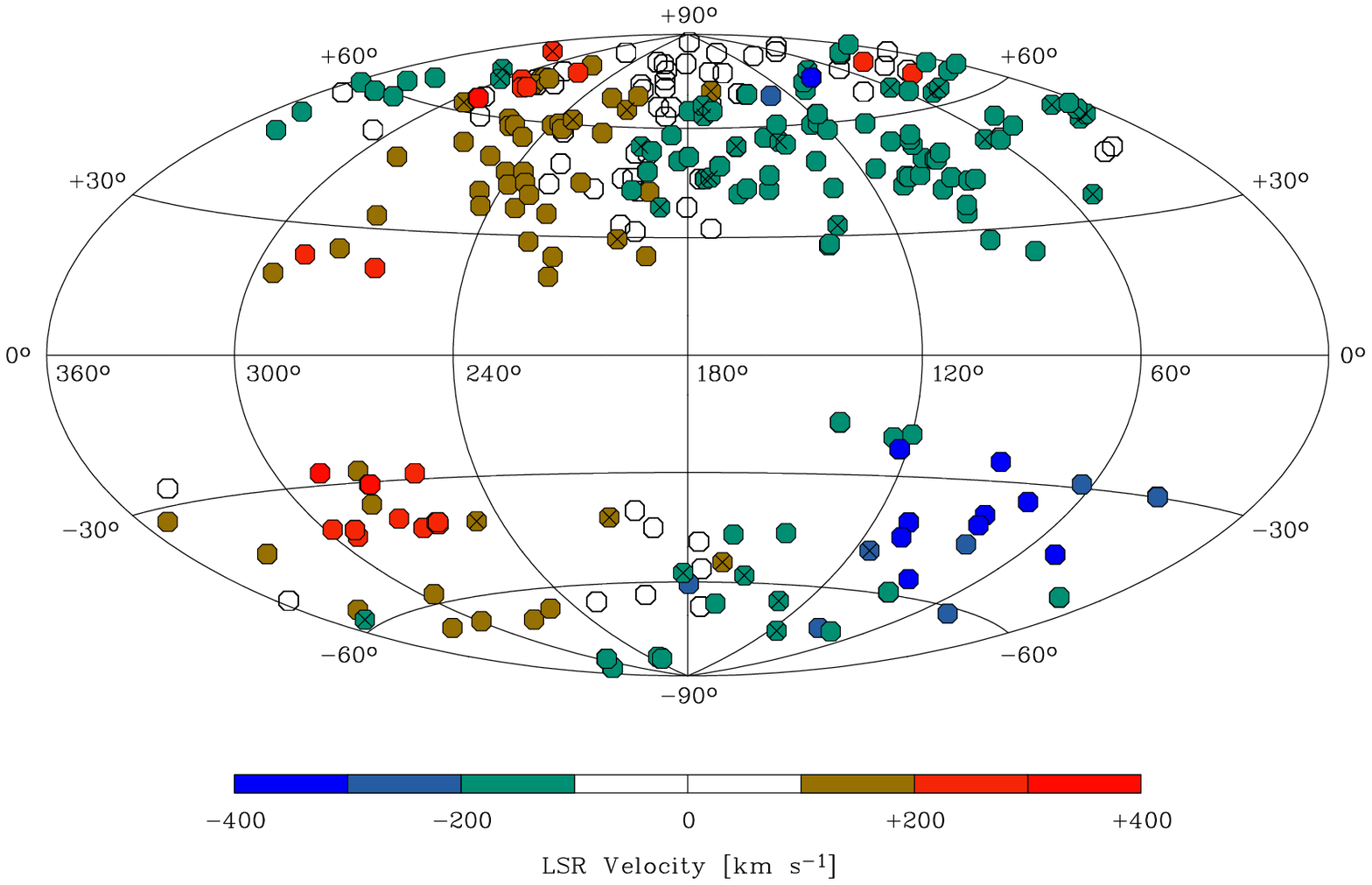}}
\caption[]{
Sky distribution of sightlines in our COS sample in a Hammer-Aitoff
projection of Galactic coordinates. Secure detections of high-velocity 
gas at $|v_{\rm LSR}|=100-500$ km\,s$^{-1}$ are indicated with filled 
circles, while tentative detections are labeled with crossed filled circles;
open circles mark the non-detections. The mean
absorption velocities of the dominant absorption components along each
sightline are color-coded according to the color scheme at
the bottom of the figure.
}
\end{center}
\end{figure*}


\section{Observations, data handling, and analysis method}

\subsection{COS spectra selection and data reduction}

In this study we use archival HST/COS data that were
retrieved from the HST Science Archive at the Canadian
Astronomy Data Centre (CADC). For our analysis we concentrate
on absorption lines of low, intermediate, and high ions from 
silicon and carbon (Si\,{\sc ii}, Si\,{\sc iii},
C\,{\sc ii}, and C\,{\sc iv}). 
Si\,{\sc ii}, Si\,{\sc iii}, and C\,{\sc ii} have strong transitions 
in the wavelength range between $1190$ and $1410$ \AA,
as listed in Table 1.
This wavelength range is covered by the COS G130M grating that operates
from $\lambda =1150-1450$ \AA\,at a spectral resolution
of $R\approx 15,000-20,000$ ($15-20$ km\,s$^{-1}$ FWHM
at a native pixel size of $3$ km\,s$^{-1}$; Green et al.\,2012; 
Debes et al.\,2016). For the study of the C\,{\sc iv} doublet 
at $1548.2,1550.8$ \AA\,we 
require data obtained with the COS G160M grating that 
covers the range $\lambda=1405-1775$ \AA\, (at a spectral resolution
similar to that of the G130M grating).
In Table 1 we summarize atomic data for the UV transitions of the metal
ions that we consider for our analysis (from Morton 2003).

In this paper, we do not analyze UV absorption of other available
ions, such as O\,{\sc i} and Si\,{\sc iv}. 
Neutral oxygen is an important ion
to determine the $\alpha$ abundance in Galactic HVCs (Wakker et al.\,1999; 
Richter et al.\,2001; Sembach et al.\,2004) and to study the distribution of
metal-enriched, neutral gas, as O\,{\sc i} and H\,{\sc i} have identical ionization
potentials. In our COS data, the only available O\,{\sc i} line
at $1302.17$ \AA\, is contaminated by airglow lines, however. They can be 
avoided only by using night-only data, which will usually reduce the S/N.
In addition, O\,{\sc i} $\lambda 1302.17$ is often saturated, so
that for most cases where this line is detected only lower limits on the 
O\,{\sc i} column density can be derived. The large range in S/N in our
data set together with the above listed restrictions limits the diagnostic
power of the O\,{\sc i} $\lambda 1302.17$ line to a relatively small sub-sample
of our sightlines and thus we refrain from considering O\,{\sc i} in
this study. A detailed analysis of O\,{\sc i} absorption in HVCs based
on HST/COS data instead will be presented in a forthcoming study.

Like C\,{\sc iv}, Si\,{\sc iv} is a useful tracer of highly ionized gas in and
around neutral HVCs, but because of the lower abundance of Si compared to
C (Asplund et al.\,2009) the detection rate 
of Si\,{\sc iv} at high velocities is relatively small ($<30$ percent in our 
data; see also Herenz et al.\,2013). Therefore, 
not much additional information is gained from this ion and therefore
we also exclude Si\,{\sc iv} from our analysis (but show the
Si\,{\sc iv} absorption in our finding charts; see Fig.\,1).

To study high-velocity UV absorption in Si\,{\sc ii}, Si\,{\sc iii},
C\,{\sc ii}, and C\,{\sc iv} we searched for all publicly available 
COS spectra from all types of extragalactic 
point sources using the CADC web interface. 
Suitable background sources include various types of AGN and galaxies.
G130M/G160M data sets for 552 targets were identified and downloaded 
by the end of February 2014.
The original (raw) data of the individual science exposures were processed 
by the CALCOS pipeline (v2.17.3) and transformed into standard {\tt x1d} fits 
files. In a second step, the individual
exposures then were coadded using a custom-written code that aligns
individual exposures based on a pixel/wavelength calibration. 
The code considers the relative position of line flanks (for spectra 
with a S/N per resolution element of $>5$) or line centers (for spectra
with lower S/N) of various
interstellar anchor lines that are distributed over the
wavelength range of the G130M and G160M gratings.
The heliocentric velocity positions of the anchor lines were
determined from supplementary H\,{\sc i} 21 cm data from 
the EBHIS and GASS surveys (see next section).
The individual spectra were rebinned and then coadded pixel-by-pixel
using the count rate in each pixel; pixels with
known artifacts were flagged accordingly and
the errors were calculated in the coadded spectra.
Finally, we performed 
a careful visual inspection of the final coadded
G130M and G160M spectra to check the quality of the 
data reduction process.

For the analysis of high-velocity absorption we selected only those spectra
that have a minimum S/N of $6$ per resolution element in
the wavelength range between $1208$ and $1338$ \AA. 
This selection criterion reduces the total sample
to $270$ lines of sight.
QSO names and Galactic coordinates
for these $270$ sightlines are listed in Table A.1 in the
Appendix.

\subsection{CGM identification and absorption-line analysis}


\begin{table}[t!]
\caption[]{Atomic data$^{\rm a}$ for considered UV transitions}
\begin{tabular}{lrr}
\hline
Ion & $\lambda_0$ [\AA] & $f$\\
\hline
Si\,{\sc iii}  & 1206.50 & 1.6690 \\
Si\,{\sc ii}   & 1190.42 & 0.2502 \\
               & 1193.29 & 0.4991 \\
               & 1260.42 & 1.0070 \\
               & 1304.37 & 0.1473 \\
               & 1526.71 & 0.2303 \\ 
C\,{\sc ii}    & 1334.53 & 0.1278 \\
C\,{\sc iv}    & 1548.19 & 0.1908 \\
               & 1550.77 & 0.0952 \\
\hline
\end{tabular}
\noindent
\\
{\small
$^{\rm a}$\,Taken from Morton (2003).\\
}
\end{table}


For the identification of high-velocity absorption along the 270 COS
sightlines we transformed the spectra into a LSR velocity frame. 
For all sightlines we determined (in an automated fashion) a global
continuum level by normalizing the velocity profiles to the highest 
flux level in the velocity range $|v_{\rm LSR}|\leq 500$ km\,s$^{-1}$.
In this way, we created HVC finding charts that were visually inspected 
to identify high-velocity absorption in the range 
$|v_{\rm LSR}|=100-500$ km\,s$^{-1}$. 
In Fig.\,1, we show four examples of such finding charts.
The full set of velocity profiles for all 270 sightlines is shown 
in the Appendix in Fig.\,B.2.
Note that absorption at lower (absolute) velocities
is not considered here (although such gas may belong to the halo), as
it would require a careful velocity modeling for each sightline to 
disentangle disk and halo components, which is clearly beyond the 
scope of this paper.  

We regard a high-velocity absorption feature as a 
definitive detection if
it is convincingly ($>4\sigma$ evidence) detected in at least two
of the above given metal transitions at 
$|v_{\rm LSR}|=100-500$ km\,s$^{-1}$, where we use the formalism 
to define a local detection limit described in Sect.\,3.1. This strategy is similar to our
previous surveys (Richter et al.\,2009, 2016; Lehner et al.\,2012; Herenz et al.\,2013), but 
is more restrictive than other studies
where also single-line detections are considered (e.g., Collins et al.\,2009;
Shull et al.\,2009).
If a high-velocity feature is seen only in one transition (e.g., as a result of low S/N,
lack of data, or due to blending with IGM lines)
we label it as HVC/CGM {\it candidate} absorber, but do not further
consider it in the statistical analysis unless stated otherwise.

The spectral features then were analyzed using the custom-written 
line analysis tool {\tt span}. 
For each high-velocity absorber the exact shape of the local continuum was determined 
by a low-order polynomial fit of the global continuum (see above; Fig.\,1; Fig.\,B.2)
within $|v_{\rm LSR}|\leq 1000$ km\,s$^{-1}$.
Equivalent widths (and their errors) were determined by a direct pixel
integration over the absorption profiles.
In a similar fashion, ion column densities (or lower limits) 
then were derived by 
integrating over the velocity profiles using the 
apparent optical depth (AOD) method, as originally described 
in Savage \& Sembach (1991). For possibly saturated absorption features
(i.e., features with absorption depths $>0.5$) the column density 
obtained from the AOD method is regarded as lower limit. 
To minimize the influence of saturation effects we adopted as final column density the
value obtained for the weakest available line for each ion that shows
well-defined high-velocity absorption features. Note that, because of the 
extended wings of the COS line-spread function, the equivalent widths and 
column densities derived in this way could be underestimated by a few percent
(see Wakker et al.\,2015; Richter et al.\,2013).

Since we here consider only absorption in the velocity range $|v_{\rm LSR}|=100-500$ km\,s$^{-1}$,
we cut away the low-velocity extensions of HVC features near $|v_{\rm LSR}|=100$ km\,s$^{-1}$.
While this partial velocity integration is unsatisfying, we decided to stick
to a strict velocity cut-off at $|v_{\rm LSR}|=100$ km\,s$^{-1}$
to avoid introducing a bias in our absorber statistics. Yet, the 
listed equivalent widths, column densities (and the 
ratios of these quantities) for absorbers near $|v_{\rm LSR}|=100$ km\,s$^{-1}$
need to be interpreted with some caution due to this velocity cut-off.

In view of the limited
spectral resolution of the COS instrument we do not further investigate 
in this survey the 
velocity-component structure of the detected HVC absorbers.
All measured equivalent widths and column densities for the high-velocity
absorbers in our sample are summarized in Table A.2 in the Appendix.
For those high-velocity features that have been studied previously with
HST/STIS and HST/COS data (e.g., Richter et al.\,2009; Herenz et al.\,2013;
Fox et al.\,2013, 2014) the values derived by us generally are in very good 
agreement with the previous results.

\subsection{Complementary H\,{\sc i} 21 cm data}

We complement our HST/COS absorption-line data with
21 cm data from different instruments and observing campaigns
to investigate the relation between UV absorption and 21 cm emission 
along each sightline and to 
infer the ionization state of the Milky Way CGM.

First, we use 21 cm data from the Galactic-All Sky Survey (GASS; McClure-Griffiths et al.\,2009; 
Kalberla et al.\,2010), which was carried out with the 64 m radio telescope at Parkes. 
The angular resolution of the GASS data is $\sim 15.6 \arcmin$ with an
rms of $57$ mK per spectral channel ($\Delta v=0.8$ km\,s$^{-1}$).
Secondly, we make use of 21 cm data obtained as part of the new {\it Effelsberg H\,{\sc i} Survey} 
(EBHIS), which was carried out on the Effelsberg 100 m radio telescope (Kerp et al.\,2011; 
Winkel et al.\,2010). Compared to GASS, EBHIS data has a slightly higher noise level
($\sim 90$ mK) and a somewhat lower velocity resolution (channel separation: 
$\Delta v=1.3$ km\,s$^{-1}$), but the angular resolution is higher
($\sim 10.5\arcmin$). 
The typical H\,{\sc i} column density limit in the 21 cm data used here is a few
times $10^{18}$ cm$^{-2}$. In general, our 21 cm data is complete for H\,{\sc i} column densities
$>5\times 10^{18}$ cm$^{-2}$ ($4 \sigma$ level).
Note that there are other, more sensitive 
21cm surveys for individual regions (e.g., Lockman et al.\,2002).
For each sightline we included the 21 cm velocity profile in the
HVC finding charts (Fig.\,1; Fig.\,B.2).

H\,{\sc i} column densities (and their limits) were determined by integrating the 21 cm emission profile
over the appropriate velocity range (defined by the UV absorption) using the relation


\begin{equation}
N({\rm H\,I})=1.823\times10^{18}\,{\rm cm}^{-2}\,\int^{v_{\rm max}}_{v_{\rm min}}
\,T_{\rm B}\,{\rm d}v,
\end{equation}


where $T_{\rm B}$ denotes the brightness temperature (in [K])
and the gas is assumed to be optically thin in 21 cm
(Dickey \& Lockman 1990).


\section{Characterization of HVC absorption}


\begin{table*}[t!]
\caption[]{Covering fractions for different ions (in percent)$^{\rm a}$}
\begin{small}
\begin{tabular}{lcccccccccc}
\hline
 & & & $f_{\rm c}$ & $f_{\rm c}$ & $f_{\rm c}$ & $f_{\rm c}$ & $f_{\rm c}$ & $f_{\rm c}$ & $f_{\rm c}$ & $f_{\rm c}$ \\
\hline
Ion & log $N_{\rm lim}$ & $\cal{C}^{\rm b}$ & all sky& $b>0\degree$ & $0\degree <b<75\degree$ &
$b\geq 75\degree$ & $b\leq 0\degree$ & $-75\degree <b \leq 0\degree$
& $b\leq -75\degree$ &  all sky \\
& & & & & & & & & & without MS$^{\rm c}$\\
\hline
Si\,{\sc iii}  & $12.1$ & $0.95$ & $77 \pm 6$ & $73 \pm 7$ & $79 \pm 7$ & $19 \pm 11$ & $89 \pm 13$ & $88 \pm 13$ & $55-100$ & $72 \pm 7$ \\
Si\,{\sc ii}   & $12.3$ & $0.95$ & $70 \pm 6$ & $66 \pm 6$ & $72 \pm 7$ & $19 \pm 11$ & $79 \pm 12$ & $76 \pm 12$ & $55-100$ & $65 \pm 6$ \\
C\,{\sc ii}    & $13.2$ & $0.96$ & $70 \pm 6$ & $66 \pm 6$ & $72 \pm 7$ & $13 \pm 9$  & $82 \pm 12$ & $80 \pm 13$ & $55-100$ & $66 \pm 6$ \\
C\,{\sc iv}    & $12.9$ & $0.94$ & $58 \pm 7$ & $44 \pm 7$ & $50 \pm 9$ & $0$         & $73 \pm 12$ & $70 \pm 13$ & $55-100$ & $44 \pm 7$ \\
\hline
\end{tabular}
\\
$^{\rm a}$\,Note that only those HVC absorbers are considered that are detected in at
least two transitions.\\
$^{\rm b}$\,Completeness level at $N_{\rm lim}$; see Sect.\,3.1.\\
$^{\rm c}$\,See Table 3 for adopted $(l,b)$ ranges.\\
\end{small}
\end{table*}


\subsection{Sky distribution}

In Fig.\,2 we show the sky distribution of the 270 sightlines in Galactic coordinates
($l,b$) using a Hammer-Aitoff sky projection centered on $l=180\degree$.
Filled circles indicate directions along which high-velocity absorption ($|v_{\rm LSR}|\geq 100$
km\,s$^{-1}$) is convincingly detected in at least two different UV lines, while
crossed filled circles label the tenative detections and open circles
the non-detections.
The mean radial velocity of the absorption is indicated with the color
scheme shown at the bottom of the plot.
There is a clear asymmetry in the sky distribution of the LOS: three quarters 
of the sightlines are located in the northern sky. 
About $60$ percent of the detected HVC absorption features have negative velocities. They are
found at $l<210\degree$ and $l>300\degree$ with only a few exceptions. Positive-velocity HVC
absorption concentrates in a strip in galactic longitude in the range $l=210-300\degree$. 
There is a striking lack of HVC absorption near the northern galactic pole
at $b>75\degree$. The low detection rate of low and intermediate ions in this region has been noted
previously (Lehner et al.\,2012), but this trend now becomes more significant
due to the improved statistics. 
In the southern part of the sky, gas from the MS dominates
the absorption characteristics of high-velocity gas (Fox et al.\,2013, 2014). However, 
some of the negative velocity gas at $l<130\degree$
possibly belongs to the CGM of M31 (Lehner et al.\,2015) and to intragroup gas in the 
Local Group filament, as will be discussed later.

Note that the observed velocity dipole of the high-velocity absorption at high galactic 
latitudes cannot be explained alone by galactic rotation, 
because along many sightlines cos\,($b$)$v_{\rm rot}$ is smaller than 
the observed absorption velocities, even if a high circular orbit speed 
of the Sun of $v_{\rm rot}=255$ km\,s$^{-1}$ is assumed 
(Reid et al.\,2014). For MS absorption in the south, in particular, 
the velocity dipole instead suggests a non-circular orbit of the 
Stream around the Milky Way (Putman et al.\,2003).

We securely detect $187$ HVC absorbers along the $270$ sightlines, translating
to a detection rate of high-velocity absorption of $f_{\rm det}=69\pm5$ percent. If
we add the $37$ HVC candidates (tentative detections, see above), the
detection rate increases to $83\pm 6$ percent. There are, however,
a large number of low-S/N spectra in our sample that are not particular
sensitive to detect weak HVC features.

To take into account the differen S/N ratios in our spectra and to 
compare our results with previous measurements we need to
consider the detection threshold of our COS data in more detail. 
For each of the four ions considered in this study 
(C\,{\sc ii}, C\,{\sc iv}, Si\,{\sc ii},
and Si\,{\sc iii}) we transformed the individual detection rates, $f_{\rm det}$,
into covering fractions, $f_{\rm c}(N_{\rm lim})$, where $N_{\rm lim}$
represents a specific (lower) column-density threshold.
As covering fraction we define the number of sightlines that exhibit 
securely detected high-velocity
absorption with a column density $N\geq N_{\rm lim}$, devided by the
total number of sightlines that are sensitive to detect HVC absorption
at $N\geq N_{\rm lim}$. For a spectrum with a given S/N ratio per resolution element, 
a spectral resolution $R$, the $4\sigma$ limiting column density threshold, $N_{\rm lim}$, for an
unresolved line with oscillator stength $f$ at laboratory wavelength $\lambda_0$ 
(Table 1) is given by the expression


\begin{equation}
N_{\rm lim}\geq1.13\times10^{20}\,{\rm cm}^{-2}\,
\frac{4}{R\,{\rm (S/N)}\,f\,(\lambda_0/{\rm A)}}.
\end{equation}


Because our COS spectra span a large range
in S/N, they are unequally sensitive to show high-velocity 
absorption features and thus $f_{\rm c}(N_{\rm lim})$ is 
expected to decrease with decreasing S/N (and increasing 
$N_{\rm lim}$). 

Having defined $N_{\rm lim}$, we can study the completeness, $\cal{C}$,
of our QSO sample. For this we relate for each ion 
the number of sightlines with a given column density threshold, ${\cal N}_{N>N_{\rm lim}}$,
to the total number of sightlines, $\cal{N}_{\rm tot}$, along which high-velocity
absorption in that ion can be detected:


\begin{equation}
{\cal C}(N_{\rm lim})=\frac{{\cal N}_{N>N_{\rm lim}}}
{{\cal N}_{\rm tot}} \leq 1.
\end{equation}


The most sensitive tracer for absorption in our survey is Si\,{\sc iii} $\lambda 1206.50$,
which has a very large oscillator strength (see Table 1; Morton 2003). Si\,{\sc iii}
resides in diffuse ionized gas and traces the CGM around galaxies for a
broad range of physical conditions (see Collins et al.\,2009; Shull et al.\,2009; 
Richter et al.\,2016).
In our sample, the Si\,{\sc iii} all-sky covering fraction in HVCs is $f_{\rm c}=77\pm 6$ percent
for $N_{\rm lim}=12.1$, where the completeness is ${\cal C}=0.95$ at that column-density level. 
This value is very similar to the covering fraction 
of highly-ionized gas traced by O\,{\sc vi} ($58-85$ percent; Sembach et al.\,2003).
The all-sky covering fractions for the other ions in our survey are
lower than for Si\,{\sc iii}; they are listed in the fourth row of Table 2 together
with log $N_{\rm lim}$ (second row) and $\cal{C}$ (third row).
Interestingly, the detection rate for C\,{\sc iv} is lower than for
the singly- and doubly-ionized species.
Note that for C\,{\sc ii} blending effects with C\,{\sc ii}$\star$ $\lambda 1335.71$ 
slightly reduce the sensitivity to detect high-velocity C\,{\sc ii} in the range
$v_{\rm LSR}=200-300$ km\,s$^{-1}$ (see Fig.\,1).

The values for $f_{\rm c}(N_{\rm lim})$ derived in our survey
are very similar to those presented in earlier studies using smaller data 
samples (Lehner et al.\,2012; Herenz et al.\,2013). 
The expected decline of $f_{\rm c}$ with increasing $N_{\rm lim}$ is shown
for all four ions in Fig.\,3 (filled boxes) together with the completeness 
function ${\cal C}(N_{\rm lim})$ (red solid line). 
For Si\,{\sc iii}, for instance, $f_{\rm c}$
declines from $77$ to $49$ percent if log $N_{\rm lim}$ is increased 
from $12.1$ to $13.0$. Fig.\,3 underlines the importance of high S/N data
for our understanding of the spatial distribution of low-column
density gas in the Milky Way's CGM. 


\begin{figure}[t!]
\begin{center}
\resizebox{1.0\hsize}{!}{\includegraphics{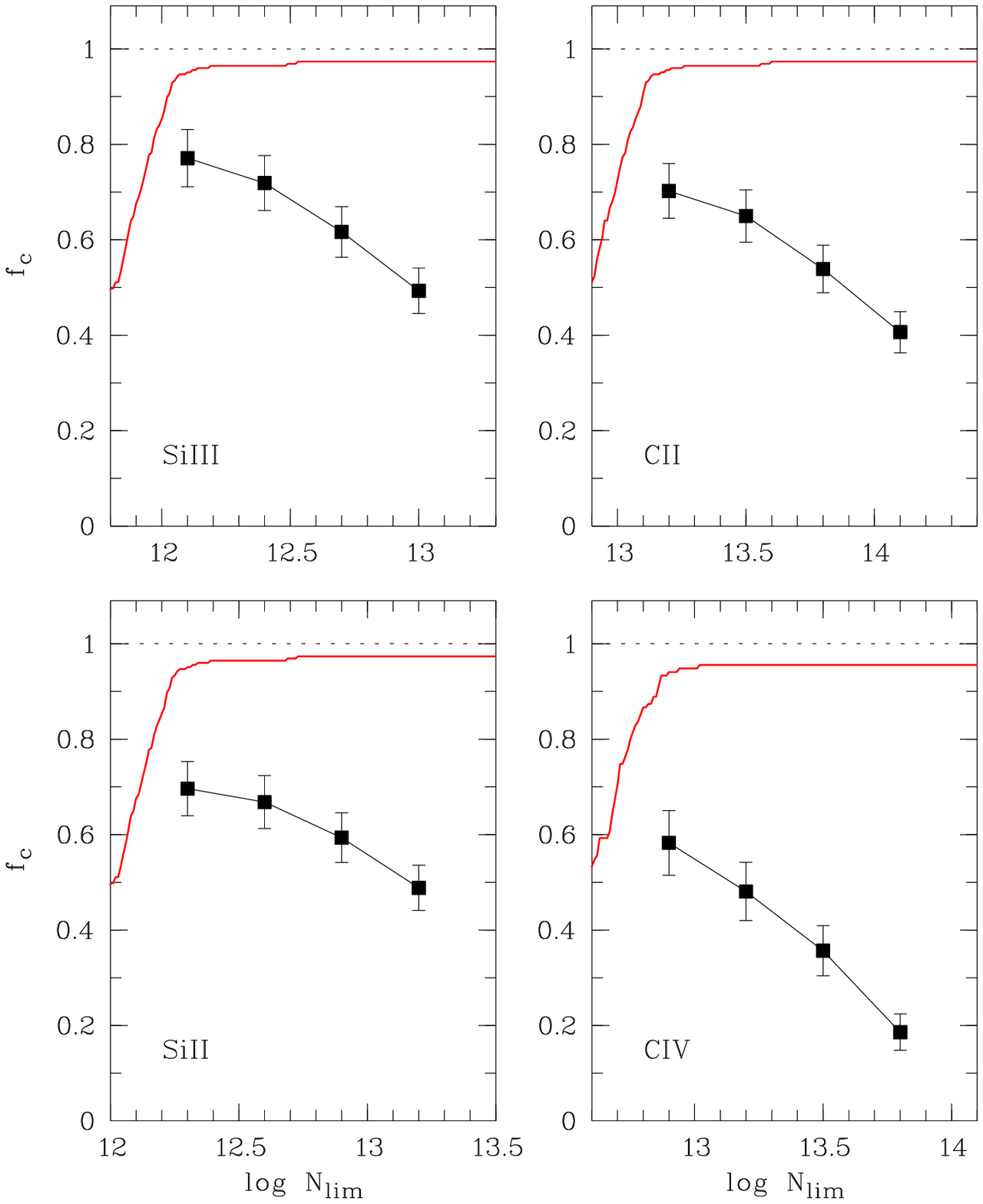}}
\caption[]{
Covering fractions of high-velocity gas for the four considered
ions as a function of limiting column density (filled squares; see
also equation 2). The red solid lines indicate the completeness
functions for each ion, as defined in equation (3).
}
\end{center}
\end{figure}


In Table 2 we provide additional information
on how $f_{\rm c}$ varies for different latitude
bins. In general, the covering fraction of high-velocity gas
is slightly smaller in north than in the south ($0.73\pm0.07$ vs.
$0.89\pm0.13$ for Si\,{\sc iii}), reflecting the fact that
the MS covers a significant portion of the southern sky (Fox et al.\,2014;
Lehner et al.\,2012).
The absorption fraction reaches 100 percent
near the southern galactic pole at $b\leq -75\degree$
(Table 2, last column).
Near the northern galactic pole for $b>75\degree$ (thus far away
from the MS) the covering fraction is instead only $< 20$ percent, marking
the most striking difference between the northern and southern high-velocity sky.
If we exclude the region covered by the MS (see Table 3 for adopted $(l,b)$ ranges),
then the covering fractions are slightly (but not substantially) smaller
than the all-sky values (Table 2, last column).

Our study demonstrates that more than three quarters of the sky is
covered by diffuse high-velocity gas.
Additional absorption-line data for $b<0\degree$ would be desirable to fill
the various gaps in the LOS distribution in the southern sky
(Fig.\,2). Yet, the observed large-scale trends for $f_{\rm c}$ are 
statistically robust, indicating an inhomogeneous distribution and
a north/south disparity of high-velocity gas on the Galactic sphere.


\begin{figure*}[th!]
\begin{center}
\resizebox{0.8\hsize}{!}{\includegraphics{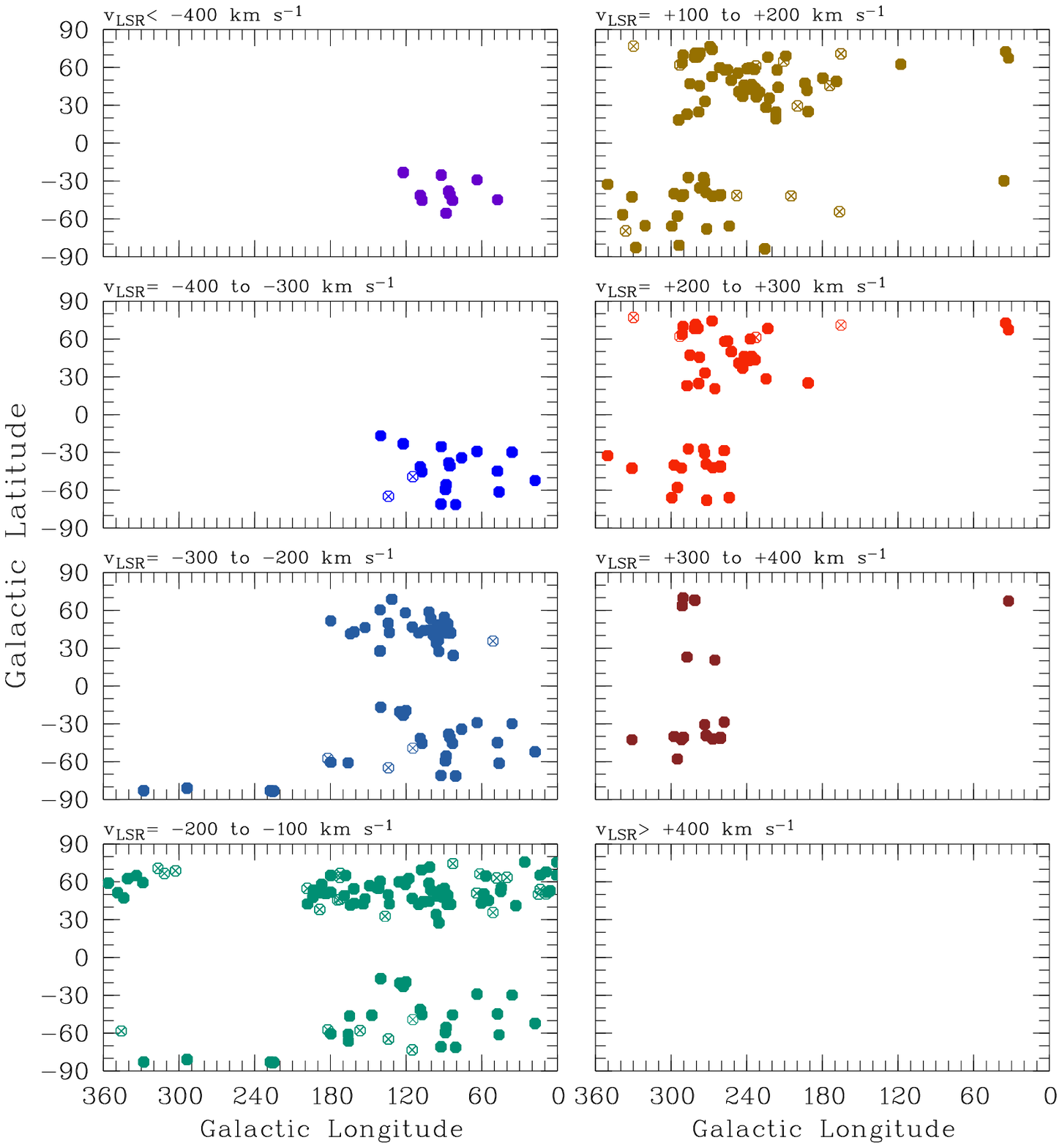}}
\caption[]{
Velocity channel maps of high-velocity CGM absorbers, indicating the
full velocity range in which absorption is observed along
each sightline. The velocity range for each panel is labeled
above each box. Filled circles indicate confirmed
high-velocity absorbers, crossed circles label tentative
detections (see Sect.\,2.2).
}
\end{center}
\end{figure*}


\subsection{Velocity distribution}

While Fig.\,2 gives a general overview of the distribution of high-velocity absorption 
on the sky at positive and negative radial velocities, it is useful to explore in more
detail the distribution of the absorption within different velocity bins.
In Fig.\,4 we therefore show velocity channel maps of high-velocity absorption in bins
of $\Delta v=100$ km\,s$^{-1}$. 

The main difference compared to Fig.\,2 is that here the entire 
velocity {\it range} of the detected high-velocity absorption 
is taken into account, while in Fig.\,2 only the mean absorption
velocity is considered.
High-velocity absorption at very high velocities $|v_{\rm LSR}|>400$ km\,s$^{-1}$ is seen
only at negative velocities, predominantly at $l<140\degree$ and $b<-10\degree$.
High-velocity absorption at negative velocities in the northern sky,
in contrast, is limited to $v_{\rm LSR}>-300$ km\,s$^{-1}$.
In Sect.\,5 we will further discuss the origin of these velocity
signatures with respect to the different HVC complexes and gas 
in the LG.

\subsection{Absorption fraction vs.\,radial velocity}

An important task is to investigate a possible radial change in the physical conditions
of the Milky Way CGM.
Ideally, one would study systematically diagnostic metal-ion ratios 
(such as Si\,{\sc ii}/Si\,{\sc iii}
and C\,{\sc ii}/C\,{\sc iv}) as a function of distance to  
the absorbers. 
Measuring HVC distances is very difficult, however, and reliable 
distance estimates have been determined only for a very limited
number of halo clouds (Ryans et al.\,1997a, 1997b; van\,Woerden et al.\,1998; 
Wakker et al.\,2007, 2008; Thom et al.\,2006, 2008; Lehner \& Howk 2011; Lehner 
et al.\,2012; Richter et al.\,2015).
An alternative approach in this context is to
investigate the detection rates (absorption fractions) of the various ions 
as a function of radial velocity, which provides at least some indirect 
information on the gas properties of nearby and more distant gas and its 
radial direction of motion. 

In Fig.\,5 we have plotted $f_{\rm abs}$ vs.
$v_{\rm LSR}$ for our HVC absorber sample. 
It is evident that there are substantial differences between the trends
at positive and negative radial velocities.
First, the absorption fractions are higher for gas at negative velocities
than for positive velocities (at similar sensitivity), a trend that is
valid for all of the four considered transitions. 
This implies that there is more absorbing CGM material that moves towards
the Sun than gas that is moving away from it.
Secondly, at negative velocities, the absorption fraction of 
Si\,{\sc iii} $\lambda 1206.50$ (red line) is always significantly 
higher than that of Si\,{\sc ii} $\lambda 1260.42$
(blue line), while at positive velocities the absorption fractions 
in both lines are very similar. 
Thirdly, there is a mild enhancement of $f_{\rm abs}$(C\,{\sc iv})
in the range $-300$ to $-200$ km\,s$^{-1}$
compared to positive velocities.

In section A.1 in the Appendix we discuss the relation between 
absorption fraction and LSR velocity excluding the contribution
from the MS.

\subsection{Interpretation of observed trends}

The covering fractions for the individual ions,
as discussed in the previous subsections, reflect a complex (projected)
spatial distribution of the different gas phases in the Galactic halo
that are traced by the various low, intermediate, and high ions in
our survey. These phases range from from cold/neutral gas at relatively
high densities ($n_{\rm H}\geq 10^{-3}$ cm$^{-3}$) 
to warm/hot ionized gas at low densities ($n_{\rm H}< 10^{-3}$ cm$^{-3}$) 
where the gas cannot recombine efficiently (see review by Richter 2017).

The sky covering fractions indicate that diffuse,
predominantly ionized gas, as traced by Si\,{\sc iii}, as well as
Si\,{\sc ii}, and C\,{\sc ii}, represents the
most widespread gas phase in the Milky Way's 
population of HVCs (see also Shull et al.\,2009), which
move as coherent circumgalactic structures through the ambient 
hot coronal gas. The high detection rate of 
Si\,{\sc iii} in circumgalactic absorption-line systems at low redshift
(Richter et al.\,2016) suggests that streams of predominantly
ionized gas represent typical features of low-redshift galaxy halos.
To reach the observed ion column densities at low gas densities, the 
absorption path lengths, $d$, in the diffuse ionized gas layers must be large 
(a few up to a few dozen kpc, typically, as $N_{\rm ion}=d\,n_{\rm ion}$.)
This, together with the large cross section, implies that 
the diffuse ionized gas phase occupies most of the volume in 
circumgalactic gas streams, 
whereas the bulk of the neutral gas (traced by H\,{\sc i}
21 cm emission) is confined to specific regions that
exhibit the largest gas densities (e.g., Lehner et al.\,2012; 
Joung et al.\,2012; Richter 2012).
The somewhat lower detection rate of C\,{\sc iv} compared to
Si\,{\sc iii} (Table 2) and O\,{\sc vi} (Sembach et al.\,2003; Wakker et al.\,2003) 
indicates that C\,{\sc iv} traces a gas phase in the Milky Way 
CGM that is not as widespread as the phase traced by the other two ions,
even if one takes into account the only moderate oscillator strengths of
the C\,{\sc iv} doublet lines (Table 1).

The apparent lack of HVC absorption near the northern galactic pole in our
survey and in the study by Lehner et al.\,(2012) suggests that this 
regions is devoid of neutral and diffuse ionized gas.
In contrast, Fox, Savage \& Wakker (2006) report the
detection of O\,{\sc vi} absorption at high positive velocities along several sightlines 
near the northern galactic pole (their Fig.\,1). Together, both results imply that the
halo gas near the northern galactic pole is predominantly {\it highly} ionized
and thus in a phase, that is not traced by the low and intermediate ions considered
in this survey.

In general, the physical conditions in the CGM around galaxies are
known to be diverse, with temperatures and densities spanning a 
large range (e.g., Joung et al.\,2012; Nuza et al.\,2014).
On the one hand, they are governed
by kinematically complex (and highly dynamic) gas circulation
processes (infall, outflow, tidal interactions) that create an inhomogeneous,
{\it irregular} distribution of
gaseous matter around Milky-Way type galaxies (see review by Richter 2017).
On the other hand, the physical conditions in the CGM are also expected to
change {\it gradually} from the inside-out owing to declining
depth of the gravitational potential at larger distances and
the resulting decreasing (equilibrium) gas pressure (see, e.g.,
Miller \& Bregman 2015). For isothermal gas one would expect to see
a declining (mean) gas density at larger distances, which - depending on
the radial decline of the ionizing radiation field - possibly results in
a higher degree of ionization in the outer halo.
Indeed, such a gradual increase in the degree of ionization with increasing
galactocentric distance is possibly visible in the CGM of
M31 (Lehner et al.\,2015) and other low-redshift galaxies (e.g., Werk et al.\,2013).

For the Milky Way halo, Lehner \& Howk (2011) demonstrated that the covering fraction of 
low and high ions in HVCs with $|v_{\rm LSR}|<180$ km\,s$^{-1}$ towards halo stars 
with $d<20$ kpc are similar to those derived against QSOs, implying that a large fraction
of the CGM gas at low velocities resides relatively nearby in the lower Milky Way halo,
while most of gas at very high velocities resides in the outer halo.
In view of this trend, Fig.\,5 suggests that the observed absorption fraction 
of ionized gas at high velocities in our survey is due to gas located at large distances
from the Galactic plane. Moreover, the observed excess of Si\,{\sc iii} and 
C\,{\sc iv} compared to Si\,{\sc ii} at $v_{\rm LSR}<-200$ km\,s$^{-1}$ suggests
that there is more diffuse ionized gas in the CGM at negative velocities compared 
to positive velocities.
As we show in the Appendix (Sect.\,A.1), much of this negative-velocity gas is related to
the MS, which has a very large cross section on the sky (see Sect.\,5.1; Fox et al.\,2014).
In addition, some of this ionized material
at high negative velocities possibly is related to UV-absorbing LG gas in
the general direction of the LG barycenter (see 
Sembach et al.\,2003). This scenario will be further discussed in Sect.\,5.4.


\begin{figure}[t!]
\begin{center}
\resizebox{1.0\hsize}{!}{\includegraphics{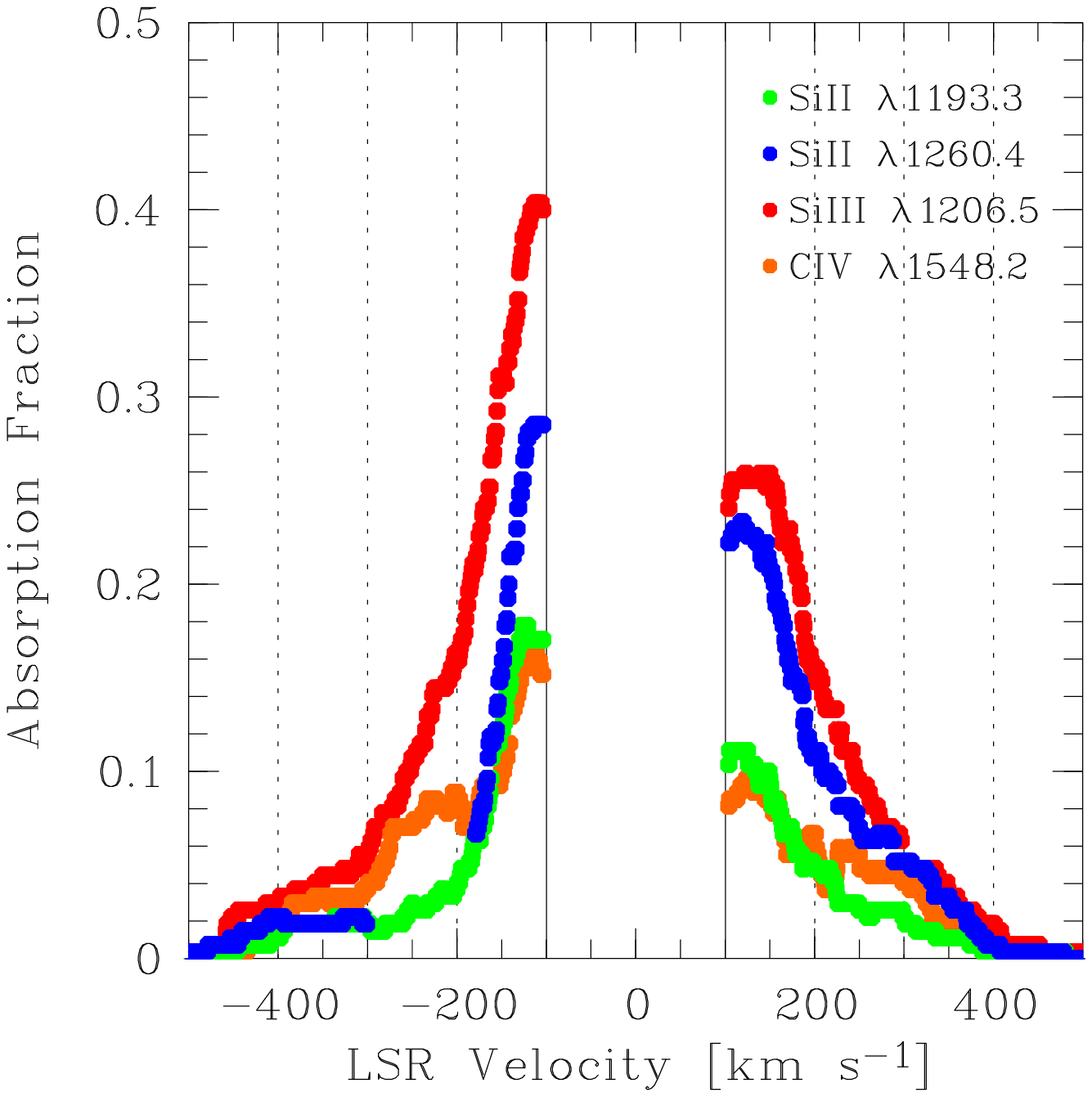}}
\caption[]{
Absorption fraction (detection rate) of the transitions of
Si\,{\sc iii} $\lambda 1206.5$ (red),
Si\,{\sc ii} $\lambda 1260.4$ (blue),
Si\,{\sc ii} $\lambda 1193.3$ (green), and
C\,{\sc iv} $\lambda 1548.2$ (orange)
as a function of LSR velocity. C\,{\sc ii}
$\lambda 1334.5$ is not included because of
severe blending with
C\,{\sc ii}$\star$ $\lambda 1335.7$
at positive velocities. The gap near
$-250$ km\,s$^{-1}$ for
Si\,{\sc ii} $\lambda 1260.4$
is due to blending with Galactic
S\,{\sc ii} $\lambda 1259.5$.
}
\end{center}
\end{figure}


\section{Distribution of equivalent widths and column densities}

\subsection{Equivalent widths and column densities of low and high ions}

In the upper panels of Fig.\,6 we show the equivalent width distribution of 
Si\,{\sc iii} $\lambda 1206.5$,
Si\,{\sc ii} $\lambda 1260.4$,
C\,{\sc ii} $\lambda 1334.5$, and
C\,{\sc iv} $\lambda 1548.2$ for the 187 securely
detected high-velocity absorption components in our survey.
Note that these equivalent widths are derived from integrating over
the entire velocity range without taking any component
structure into account. 
The equivalent-width distributions for these four ions are very 
similar to each other, with a peak at low equivalent widths between $0$ and $150$ m\AA\,
and a rapid decline towards larger equivalent widths.
The majority ($58$ percent) of the absorbers have equivalent widths 
$<200$ m\AA\,in the Si\,{\sc iii} $\lambda 1206.5$ line
($69$ percent for Si\,{\sc ii} $\lambda 1260.4$,
$56$ percent for C\,{\sc ii} $\lambda 1334.5$, and $78$ percent 
for C\,{\sc iv} $\lambda 1548.2$).
The equivalent-width distribution of Si\,{\sc iii} $\lambda 1206.5$ 
in Galactic high-velocity absorbers
mimics that of intervening Si\,{\sc iii} absorbers at $z=0-0.1$,
which are believed to trace the CGM of low-redshift galaxies
(Richter et al.\,2016; their Fig.\,2).

In the lower panels of Fig.\,6 we show histograms of the derived
ion column densities (green) and their lower limits (gray) based
on the AOD analysis (see Sect.\,2).
Only for Si\,{\sc ii} and C\,{\sc iv} is there more than one
transition available (and C\,{\sc iv} absorption is generally weak
in HVC absorbers), so that only for these two ions can saturation effects 
be minimized by using for each absorber the weakest detected
line for the determination of log $N$. As a result, the gray-shaded
area for these ions in Fig.\,6 is smaller than for Si\,{\sc iii}
and C\,{\sc ii}. 

\subsection{Equivalent-width ratios and column-density ratios}


\begin{figure*}[t!]
\begin{center}
\resizebox{1.0\hsize}{!}{\includegraphics{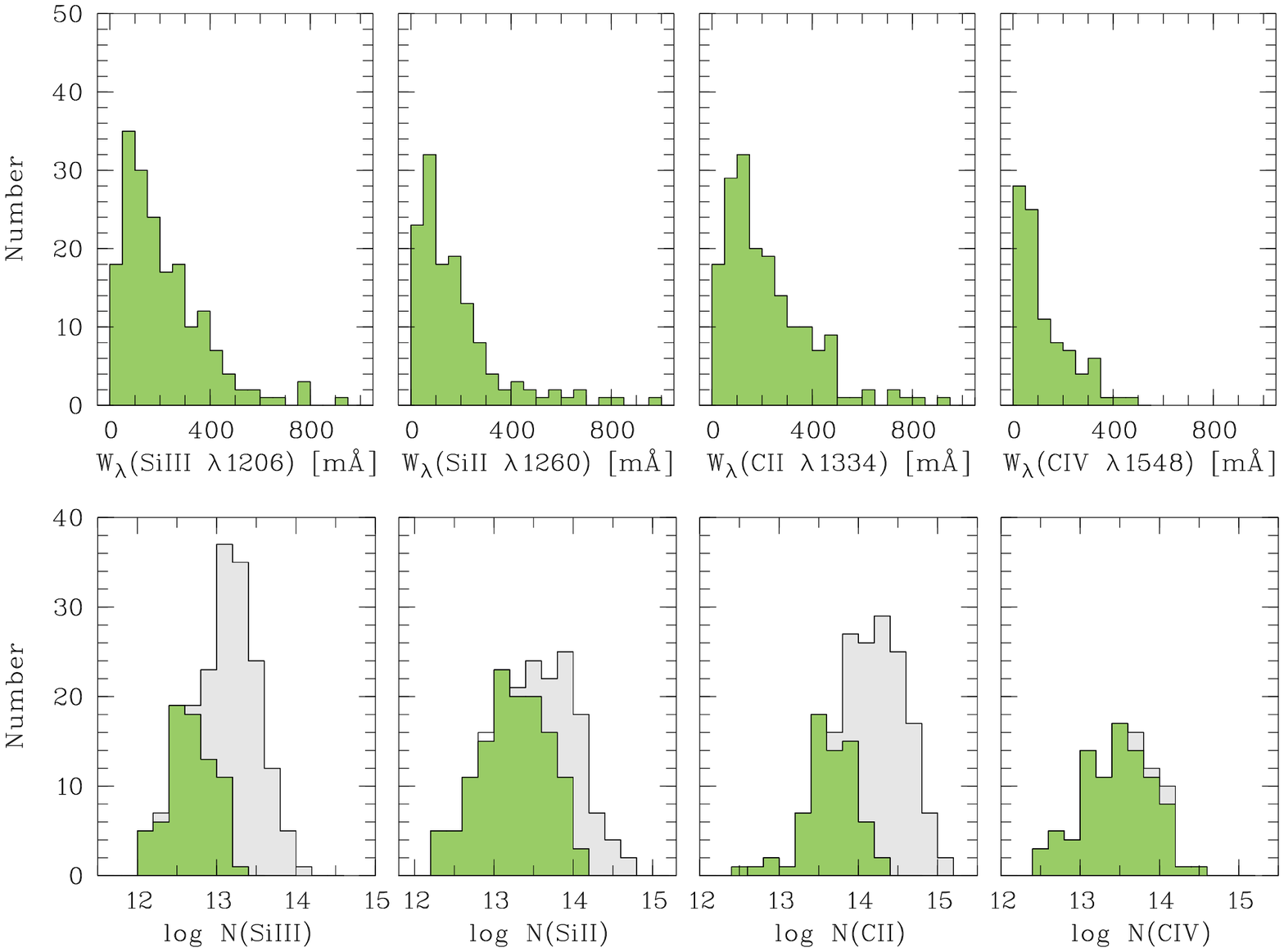}}
\caption[]{
{\it Upper panel:} equivalent-width distributions for the
high-velocity absorbers in our COS sample for four different ions.
{\it Lower panel:} histograms of logarithmic ion column densities,
as derived via the AOD method (Savage \& Sembach 1991). The green-shaded area denotes
absolute values, the gray-shaded area indicates lower
limits for log $N$ (saturated absorbers).
}
\end{center}
\end{figure*}


\begin{figure*}[ht!]
\resizebox{1.0\hsize}{!}{\includegraphics{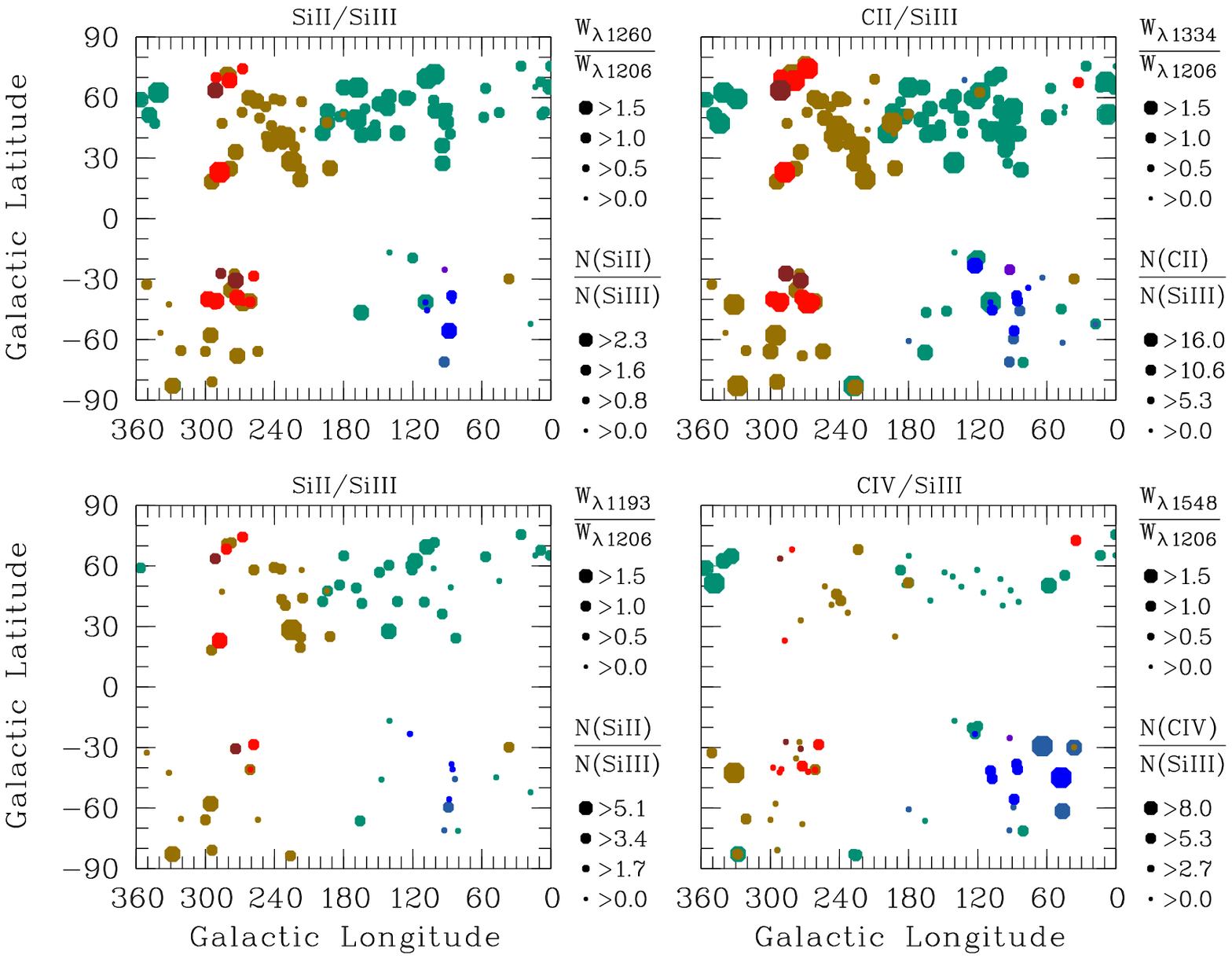}}
\caption[]{
All-sky distribution of equivalent-width ratios and column-density ratios
(assuming optically thin absorption) in high-velocity absorbers 
(see Sect.\,4.2). The color coding indicates
the radial velocity of the absorption (see colour scheme in Fig.\,8).
}
\end{figure*}


As discussed above, the absorption fractions for the different ions 
in high-velocity absorbers at different radial velocities (Fig.\,5) 
indirectly indicate a non-uniform radial distribution of 
different gas phases in Milky Way's circumgalactic environment.
Another strategy to explore the large-scale 
ionization structure of the absorbing gas in our sample is to 
investigate equivalent-width ratios of low/intermediate/high ions 
in different regions of the sky and/or in different velocity bins 
along sightlines, where 
these different ions are detected simultaneously.

In Fig.\,7 we show the spatial distribution of the equivalent-width
ratios (Si\,{\sc ii} $\lambda 1260.4$/Si\,{\sc iii} $\lambda 1206.5$),
(Si\,{\sc ii} $\lambda 1193.3$/Si\,{\sc iii} $\lambda 1206.5$),
(C\,{\sc ii} $\lambda 1334.5$/Si\,{\sc iii} $\lambda 1206.5$), and
(C\,{\sc iv} $\lambda 1548.2$/Si\,{\sc iii} $\lambda 1206.5$). 
We use Si\,{\sc iii} as reference ion as it arises in gas spanning
a wide range in physical conditions. Si\,{\sc iii} thus represents 
a robust tracer for the ionized gas column of metal-enriched circumgalactic 
gas at $T<10^5$ K (see Richter et al.\,2016).
For comparison, we also show the respective column-density ratios 
for these ions in Fig.\,7 assuming optically thin
absorption (i.e., saturation effects are ignored).

The sky distribution of the equivalent-width ratios shows some 
interesting trends.
In the northern hemisphere, there are many sightlines that exhibit
high Si\,{\sc ii}/Si\,{\sc iii} and C\,{\sc ii}/Si\,{\sc iii} ratios 
$>1$. 
As will be discussed in Sect.\,5.1 (Fig.\,8) many of these sightlines coincide
spatially with prominent northern 21 cm HVCs, such as Complex C and Complex A.
High Si\,{\sc ii}/Si\,{\sc iii} and C\,{\sc ii}/Si\,{\sc iii} ratios
are observed, however, also in the many positive-velocity absorbers 
at $l=220\degree-300\degree, b>35\degree$ that have 
little or no associated 21 cm emission
(see Fig.\,8). Also high-velocity absorbers with
substantially smaller Si\,{\sc ii}/Si\,{\sc iii} and C\,{\sc ii}/Si\,{\sc iii} 
ratios are present at $b>0\degree$. 
C\,{\sc iv} absorption is predominantly weak in the northern hemisphere when 
compared to Si\,{\sc iii}; only along a handful of sightlines (in directions
away from the 21 cm HVCs) are the C\,{\sc iv}/Si\,{\sc iii} equivalent ratios
high.

The situation is quite different in 
the southern hemisphere. Despite the fact that there are significantly fewer sightlines
to be analyzed at $b<0\degree$, it is possible to discern two distinct regions 
in Fig.\,7 with opposite trends.
For the sightlines at $l>240\degree, b<0\degree$, the observed 
Si\,{\sc ii} $\lambda 1260.4$/Si\,{\sc iii} $\lambda 1206.5$ equivalent ratios
are predominantly $>1$, tracing gas that is
associated with the 21cm emission from the MS
(see Sect.\,5).
For $l<180\degree, b<0\degree$, in contrast, the same ratio is predominantly {\it small},
while the C\,{\sc iv}/Si\,{\sc iii} ratio is coherently larger than in 
any other region in the northern or southern sky.
In Sect.\,4.4 we will further discuss the physical origin of these large
variations in the observed equivant-width ratios.


\begin{figure*}[th!]
\begin{center}
\resizebox{0.9\hsize}{!}{\includegraphics{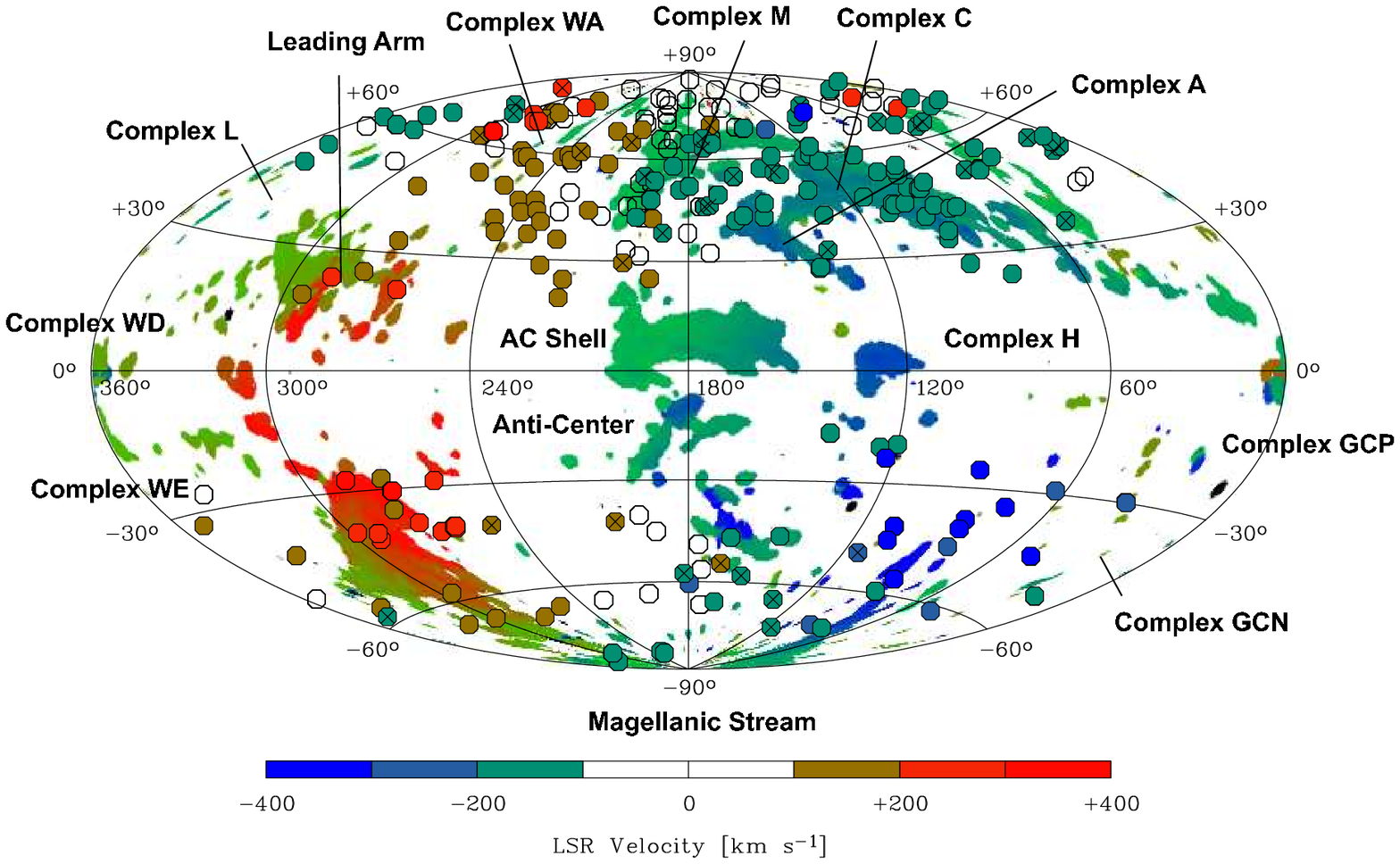}}
\caption[]{
Sky distribution of sightlines in our COS sample in an Hammer-Aitoff
projection of Galactic coordinates overlaid on a map of
21 cm emission for $|v_{\rm LSR}|\leq 500$ km\,s$^{-1}$ based
on the LAB survey (Kalberla et al.\,2005; courtesy Tobias Westmeier).
The color scheme is similar to that in Fig.\,2.
}
\end{center}
\end{figure*}


\begin{table*}[t!]
\caption[]{HVC complexes$^{\rm a}$}
\begin{scriptsize}
\begin{tabular}{lrrrrrrrrrrrr}
\hline
HVC complex    & $l_{\rm min}$ & $l_{\rm max}$ & $b_{\rm min}$ & $b_{\rm max}$ &
$v_{\rm min}$ & $v_{\rm max}$ & $d$ & $r_{\rm s}\,^{\rm b}$ & $f_{4\pi}$  & Det.\,rate &
$M_{\rm gas}$$^{\rm c}$ & d$M_{\rm gas}/$d$t$$^{\rm c}$ \\
               & [$\degree$] & [$\degree$] & [$\degree$] & [$\degree$] & [km\,s$^{-1}$] & [km\,s$^{-1}$] & [kpc] & [kpc] & &
               & [$M_{\sun}$] & [$M_{\sun}$\,yr$^{-1}$]\\
\hline
Complex GCN                &        0    &     60    &    -60   &     -30   &    -350    &   -100  &     ... &   5 & 0.028 &   2/2 &  $1.9\times 10^6$ & $0.039$ \\
Complex GCP (Smith Cloud)  &        0    &     60    &    -30   &       0   &      80    &    200  &  12     &  12 & 0.028 &   1/1 &  $2.7\times 10^6$ & $0.027$ \\
Magellanic Stream (MS)     &        0    &    180    &    -90   &     -60   &    -500    &    -80  &  20-100 &  55 & 0.333 & 41/46 &  $3.0\times 10^9$ & $5.593$ \\
                           &       60    &    120    &    -60   &       0   &    -500    &    -80  &         &     &       &       &                   &         \\
                           &      180    &    360    &    -90   &     -60   &      80    &    500  &         &     &       &       &                   &         \\
                           &      210    &    360    &    -60   &     -30   &     150    &    500  &         &     &       &       &                   &         \\
                           &      210    &    300    &    -30   &       0   &     150    &    500  &         &     &       &       &                   &         \\
Leading Arm (LA) of MS     &      300    &    360    &    -30   &       0   &     150    &    500  &  20-100 &  55 & 0.139 &  8/10 &  (see MS)         & (see MS)\\
                           &      240    &    360    &      0   &      60   &     150    &    500  &         &     &       &       &                   &         \\
Complex C                  &       40    &     90    &     15   &      45   &    -250    &    -80  &  10     &  10 & 0.062 & 31/32 &  $2.7\times 10^7$ & $0.277$ \\
                           &       60    &    110    &     25   &      65   &    -250    &    -80  &         &     &       &       &                   &         \\
                           &      110    &    150    &     35   &      65   &    -250    &    -80  &         &     &       &       &                   &         \\
Outer Arm (OA)             &       45    &     90    &      0   &      15   &    -150    &    -80  &   5     &   5 & 0.013 &   ... &               ... &         \\
                           &       90    &    160    &      0   &      20   &    -150    &    -80  &         &     &       &       &                   &         \\
Complex G                  &       60    &    105    &    -20   &       0   &    -150    &    -80  &     ... &   5 & 0.014 &   ... &               ... &         \\
Complex H                  &      105    &    150    &    -20   &      20   &    -250    &    -80  & $\geq 5$ &  5 & 0.028 &   2/2 &  $9.0\times 10^5$ & $0.018$ \\
Complex M                  &      120    &    200    &     45   &      75   &    -150    &    -80  &   3     &   3 & 0.037 & 12/24 &  $1.6\times 10^5$ & $0.006$ \\
Complex A                  &      145    &    175    &     30   &      45   &    -250    &   -140  &   8-10  &   9 & 0.010 &   5/6 &  $2.9\times 10^6$ & $0.033$ \\
                           &      130    &    145    &     20   &      35   &    -250    &   -140  &         &     &       &       &                   &         \\
Anti-Center (AC)           &      130    &    200    &      0   &     -60   &    -500    &    -80  &   3-8   &   5 & 0.065 &   2/6 &  $7.7\times 10^5$ & $0.016$ \\
AC Shell + ACO             &      160    &    210    &    -20   &      20   &    -130    &    -80  &     1   &   1 & 0.031 &   ... &               ... &         \\
Complex WE                 &      290    &    340    &    -30   &       0   &      80    &    150  & $\leq 12$ & 5 & 0.023 &   0/1 &               ... &         \\
Complex WA + WB            &      210    &    260    &      0   &      60   &      80    &    200  &   $8-20$ & 10 & 0.046 & 21/25 &  $7.4\times 10^6$ & $0.075$ \\
HVC toward LMC             &      275    &    285    &    -38   &     -28   &      80    &    150  & $\leq 13$ & 9 & 0.002 &   ... &               ... &         \\
Complex L                  &      300    &    360    &     20   &      60   &    -200    &    -80  &  0.5-1.5 &  1 & 0.037 &   5/7 &  $1.3\times 10^4$ & $0.001$ \\
\hline
\end{tabular}
\noindent
\\
$^{\rm a}$\,Note: HVC information compiled from various sources including Wakker (2001,2003); Wakker et al.\,(2007,2008); 
Thom et al.\,(2006,2008);  Richter (2006); Richter et al.\,(2015); Wakker et al.\,(2017, in prep.).\\
$^{\rm b}$\,Distance assumed in our model to calculate the total HVC mass and accretion rate (Sect.\,6).\\
$^{\rm c}$\, See Sect.\,6.2.\\
\end{scriptsize}
\end{table*}


\subsection{Relation between UV absorption and 21 cm emission}

In Fig.\,8 we again show the sky distribution of high-velocity absorbers
and their mean LSR velocities (as in Fig.\,2), but now overlaid on the
H\,{\sc i} 21 cm emission map from the LAB survey using the same
color coding for the LSR velocities (except for the range
$|v_{\rm LSR}|\le 100$ km\,s$^{-1}$, which is indicated in green
for the 21 cm data and ignored in our absorption-line analysis).
As expected, emission and absorption features coincide in 
velocity, demonstrating that they trace the same overall gas distribution.
In all regions of the sky, UV absorption is observed 
beyond the outer boundaries of 21 cm HVCs, indicating that the
neutral HVCs are surrounded by extended gaseous envelopes (traced
only in absorption) that have low neutral gas column densities.

There are regions that are almost devoid of high-velocity 
H\,{\sc i}, but show strong absorption in the UV. The most
prominent of such regions is, again, the region $l<180\degree$, 
$b<0\degree$,
which appears to contain mostly ionized gas at extreme 
negative velocities (see above).
Another prominent region in this context is at 
$l=200\degree -260\degree$, $b>0\degree$,
which shows UV absorption at high positive velocities,
but almost no high-velocity 21 cm emission. Most of the 
absorbers in this area have (mean) radial velocities 
$<200$ km\,s$^{-1}$ and belong to HVC Complex WA (see Sect.\,5.1).
However, there is also a group of absorbers in this region 
with $v_{\rm LSR}>200$ km\,s$^{-1}$ that possibly trace 
LG gas (Sect.\,7).

There are also a few absorbers above the Galactic center at 
$l<50\degree, b>30\degree$ that have no H\,{\sc i} 21 cm counterpart.
These absorbers might be related to a large-scale outflow
from the Galactic center region
(see Keeney et al.\,2006; Zech et al.\,2008; Lehner et al.\,2012;
Fox et al.\,2015).

To statistically compare the 21 cm properties of HVC-features with 
our UV absorption-line data we have determined the detection 
rate of 21 cm emission along the 270 COS sightlines. 
Out of the 187 high-velocity absorption components detected
in Si\,{\sc iii} absorption, only 46 show associated H\,{\sc i} emission
in the 21 cm data with $N$(H\,{\sc i}$)\geq 5\times 10^{18}$
cm$^{-2}$ (see Fig.\,1, middle column, for an example
where high-velocity H\,{\sc i} is detected and aligned with the UV 
absorption). 
This translates into an H\,{\sc i} covering fraction of $17$ percent
for log $N$(H\,{\sc i}$)\geq 18.7$, which
is a factor of $\sim 4-5$ lower than the covering fraction for Si\,{\sc iii}
at log $N$(Si\,{\sc iii}$)\geq 12.1$.
These numbers further demonstrate that the
neutral HVCs, as seen in 21 cm emission, just display the tips of the
icebergs of the Milky Way CGM, namely the regions with the highest
gas densities, while the bulk of the CGM structures has H\,{\sc i} 
column densities below $5\times 10^{18}$ cm$^{-2}$ and thus remains invisible
in our 21 cm data (see also Lehner et al.\,2012; 
Richter et al.\,2005, 2009; Fox et al.\,2006, 2014).
Note that the high-velocity H\,{\sc i} covering fraction increases to 
$\sim 30$ percent for log $N$(H\,{\sc i}$)\geq 17.9$, if more sensitive
21 cm data is considered (see Wakker 2004; Lockman et al.\,2002).

\subsection{Interpretation of observed trends}

It has been demonstrated in many previous studies that the
equivalent-width/column-density ratios of low, intermediate, and
high ions represent powerful diagnostic tools to study the ionization
conditions in the CGM (e.g., Fox et al.\,2014; Werk et al.\,2013; Richter et al.\,2016).
At higher gas densities, recombination is more efficient, and therefore
a high Si\,{\sc ii}/Si\,{\sc iii} ratio indicates halo gas at relatively 
high gas densities (same for C\,{\sc ii}/Si\,{\sc iii}). 
Smaller Si\,{\sc ii}/Si\,{\sc iii} and C\,{\sc ii}/Si\,{\sc iii} ratios, in contrast,
indicate regions with high ionization fractions at lower
gas densities.

In the northern sky, the observed equivalent widths ratios (Fig.\,7) indicate a
spatially complex, multi-phase structure of the absorbing gas with
many regions that appear to have relatively high densities (with
low ions dominating the absorption). Some (but not all) of the northern sightlines
with high Si\,{\sc ii}/Si\,{\sc iii} ratios are associated with prominent
21cm HVC complexes.
In the southern sky, the Si\,{\sc ii} $\lambda 1260.4$/Si\,{\sc iii} $\lambda 1206.5$ 
equivalent ratios $>1$ are predominantly related with relatively dense gas from 
the neutral gas body of the MS.
In the Appendix (Sect.\,A.2; Fig.\,A.2) we further discuss the equivalent ratios
shown in Fig.\,7 as a function of radial velocity.

The fact that UV absorption is observed beyond the outer boundaries of the 21cm HVCs
at similar radial velocities than the 21cm emission (Fig.\,8) underlines that the neutral
gas bodies are surrounded by extended ionized gas layers that build an interface
between the neutral HVCs and the ambient hot coronal gas 
(e.g., Sembach et al.\,2003; Miller \& Bregman 2015). 
Previous studies have demonstrated that the ionized envelopes of neutral HVCs contain 
substantially more mass than their neutral cores (Lehner et al.\,2012; Shull et al.\,2009;
Richter et al.\,2009; Sect.\,6).
A detailed analysis of UV absorption lines that trace the extended
ionized gaseous envelopes of 21 cm features from the MS recently has been
presented by Fox et al.\,(2014) using a subset of the COS data
sample considered here. 

As discussed in Sect.\,4.3, not all high-velocity UV absorbers are related
to known 21cm HVCs and the trends presented in Sect.\,4.2 indicate the 
presence of ionized gas streams in the halo with low neutral gas columns.
With its low Si\,{\sc ii}/Si\,{\sc iii} and high C\,{\sc iv}/Si\,{\sc iii} ratios,
the region ($b<0\degree$, $l<180\degree$) is distinct in its
ionization properties from any other large-scale structure in the
high-velocity sky (Fig.\,7). It apparently contains mostly moderately to highly 
ionized gas at low gas densities (resulting in low recombination rates). 
A high degree of ionization is also indicated by the presence of 
strong high-velocity O\,{\sc vi} absorption in this direction (Sembach et al.\,2003).
Sembach et al.\,(1999) modeled the ionization conditions of the high-velocity
gas towards PKS\,2155$-$304  and Mrk\,509 (see Tables A.1 and A.2)
and derived very low thermal gas pressures of $P/k<5$ K\,cm$^{-3}$. 
Gas at such low pressures would not survive long in the inner halo regions 
of Milky-Way size galaxies at $r\leq 50$ kpc, where the gas pressures are 
expected to be at least one order of magnitude higher (Wolfire et al.\,1995). 
From the density and temperature constraints of the Milky Way's hot coronal gas 
follows that the thermal gas pressure is expected to decline with radius as
$P/k\propto r^{-1.58}$ (Miller \& Bregman 2015), indicating that $P/k\approx 20$ K\,cm$^{-3}$
at $r=100$ kpc and $P/k\approx 10$ K\,cm$^{-3}$ at $r=200$ kpc, the latter being the 
Milky Way's assumed virial radius (Dehnen et al.\,2006; McMillan 2011). Similar pressure
gradients are also found in numerical simulations of hot gas around 
Milky Way-type galaxies (N14). The pressure limit of $P/k<5$ K\,cm$^{-3}$
for the high-velocity gas towards PKS\,2155$-$304  and Mrk\,509 therefore
implies that this absorber is located at very large distances 
to the disk ($r>250$ kpc) and possibly traces diffuse gaseous material 
outside the MW virial radius in the LG in the general direction of the LG 
barycenter at $l=147\degree, b=-25\degree$ (see Sembach et al.\,1999; 2003).

Further support for this scenario comes from the fact that the HVC absorbers in 
this region exhibit the highest radial velocities in our survey (extreme negative 
velocities; see Fig.\,4, upper left panel and Sect.\,3.4). 
Hydrodynamical simulations and observations indicate that the 
infall velocities of gas at large radii are higher than those close
to the disk (at $<5$ kpc), where the kinematics is predominantly determined 
by the hydrodynamical interaction and the reprocessing of infalling gas by the
surrounding hot corona (see Richter 2017 and references 
therein). We will further explore a LG origin of these HVC absorbers in Sect.\,7, 
where we study the gas flow in the LG using constrained cosmological simulations.


\section{Structural properties of the Milky Way CGM}

\subsection{Identification of HVC complexes}

The HVC 21 cm sky historically has been divided 
into a large number of apparently coherent structures, the so-called HVC "complexes", 
primarily to distinguish between gas clouds in the Milky Way halo that 
are spatially unrelated and may have different origins 
(Wakker \& van\,Woerden 1997; Richter 2005, 2017; Putman et al.\,2012).
The definition of these HVC complexes is based on the position 
in $l$ and $b$ of the detected 21 cm features and the radial velocity 
range in which they are observed (see, e.g., Wakker \& van\,Woerden 1997; Wakker 2001).

In Table 3 we summarize the definition of the most prominent Galactic HVCs,
based on the HVC compilation paper of Wakker (2001). 
For each HVC we list the relevant ranges in Galactic longitude, Galactic 
latitude, and LSR velocity range in the first seven columns.
We also list available distance information for each HVC, the sky covering
fraction, $f_{4\pi}$ (based on the angular definition in $l$ and $b$),
and the detection rate of absorption within the limits in $l$, $b$, and $v$
from our survey.
In the last two rows of Table 2 we further give for each HVC complex 
the estimated total gas mass and contribution to the CGM gas-accretion rate, as 
calculated in Sect.\,6.

Assuming the ($l,b$) ranges listed in Table 3, the MS together
with the Leading Arm, LA, and its extended gaseous environment, spans over
a total solid angle of $\sim 10,000$ deg$^2$. This gigantic stream of 
gas therefore represents the by far largest HVC complex, covering almost
a quarter of the entire sky. Note that our definition of the angular extent of the 
MS+LA is different from that used in our previous surveys.
This is because we use an angular grid together with an absorption-selected 
(in $l,b,v_{\rm LSR}$) sightline selection to define the HVC boundaries
in our sample, while other studies define the area of the MS+LA based
on the 21cm contours (e.g., Fox et al.\,2014). This aspect needs 
to be taken into account when comparing the HVC mass- and accretion-rate
estimates from different surveys (see review by Richter 2017).

For the MS+LA there are 56 sightlines available in our sample. For 
a detailed analysis of MS absorption see Fox et al.\,(2013, 2014). 
Other HVC complexes with more than 15 spectra are Complex C (32 sightlines),
Complex M (24 sightlines) and Complex WA+WB (25 sightlines).

In this study, we do not further investigate the chemical
composition of the individual HVC complexes or their internal 
kinematics. These aspects will be presented in a forthcoming paper.
Some interesting information on the multi-phase nature of the 
various HVCs can be obtained, however, by systematically 
studying the equivalent-widths/column-density ratios for those
HVC complexes, for which good spatial coverage in our
COS data set is available. 


\begin{figure}[th!]
\begin{center}
\resizebox{0.7\hsize}{!}{\includegraphics{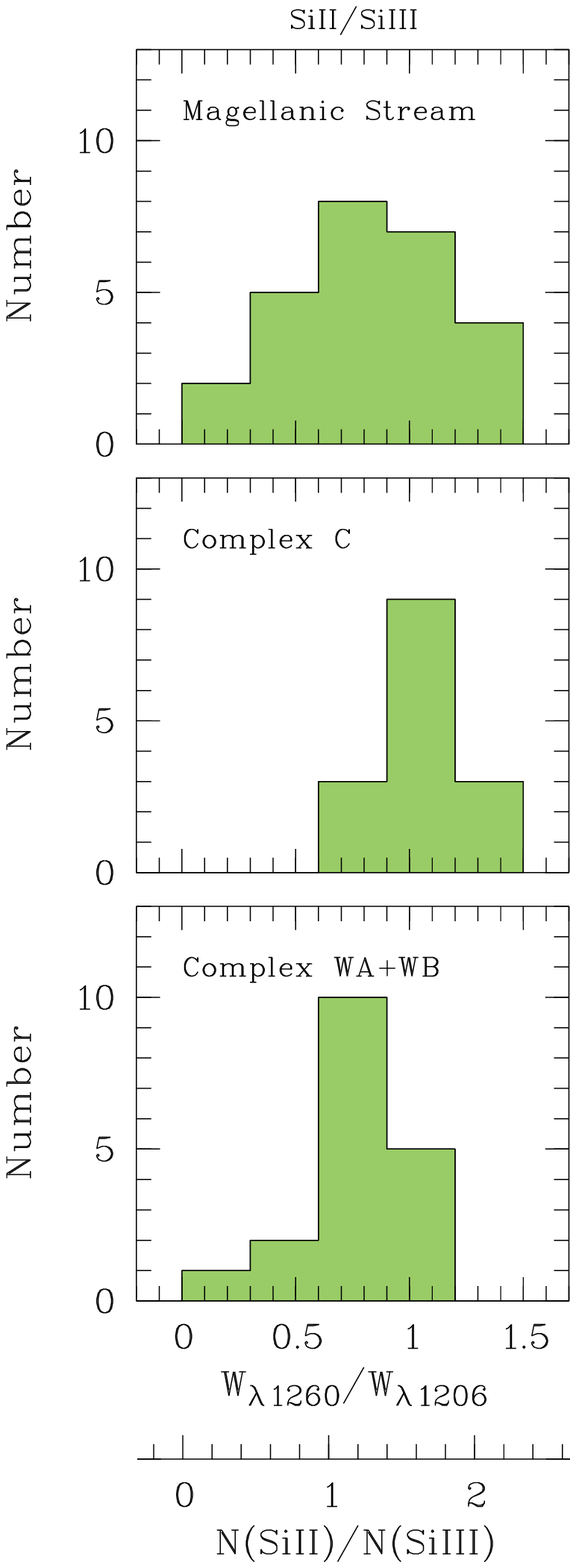}}
\caption[]{
Distribution of equivalent width ratios of
Si\,{\sc ii} $\lambda 1260.4$/Si\,{\sc iii} $\lambda 1206.5$
in three different HVC complexes.
}
\end{center}
\end{figure}


\subsection{Ion ratios in HVC complexes}

In Sect.\,4.4 we have discussed the equivalent-width/column-density ratios 
in our high-velocity absorbers in the context of the ionization structure in
the clouds. Fig.\,7 indicates that our large data sample allows us to
further investigate this aspect for a limited number of individual 
HVC complexes, for which sufficient spatial coverage is available.

In Fig.\,9 we show the distribution of measured equivalent-width/column-density ratios of
Si\,{\sc ii} $\lambda 1260.4$/Si\,{\sc iii} $\lambda 1206.5$
in the MS, Complex C, and Complex WA+WB. 
Lower values for Si\,{\sc ii}/Si\,{\sc iii} indicate lower-density gas
with a high degree of ionization (see Sect.\,4.4). 
For Complex C (100 percent) and Complex WA+WB (85 percent) the majority of the 
column-density ratios are $>1.0$. In contrast, the MS exhibits quite a number of absorbers 
(30 percent) with Si\,{\sc ii}/Si\,{\sc iii} column-density ratios $<1.0$. 
This trend indicates that the fraction of diffuse ionized gas is larger in the
MS than in the Complexes C and WA+WB. One the one hand, this may be partly 
related to the ab-initio definition of the HVC Complexes and their outer boundaries 
(Table 3). 
On the other hand, the observed trend in Fig.\,9 may be a result of the
different distances of these HVCs to the disk, which range 
from 10 kpc (Complex C), $8-20$ kpc (Complex WA+WB), to $20-100$ kpc
(MS; see Table 3, 8th row). Since the CGM gas pressure is expected to 
decline with increasing distance, HVCs in the outer halo can carry along
ionized envelopes with lower densities compared to clouds in the
inner halo. This would imply lower recombination rates in the 
HVC envelopes at large $d$ and thus lower values for 
Si\,{\sc ii}/Si\,{\sc iii} therein.
In addition, diffuse cloud layers around HVCs may be stripped away 
more efficiently in the inner halo than at larger distances 
(Heitsch \& Putman 2009). 

While these explanations are speculative, they are supported by the 
fact that low Si\,{\sc ii}/Si\,{\sc iii} column-density ratios are typical 
for CGM absorber at projected distances $\sim 25-150$ kpc around other 
low-redshift galaxies (Richter et al.\,2016; their Fig.\,15).


\subsection{Connection to Local Group galaxies}

While the majority of the high-velocity absorption features in
our survey are produced by gas that is gravitationally bound to the
Milky Way, several sightlines intersect the circumgalactic environment
of other LG member galaxies. Thus, some of the observed
high-velocity absorption features might be related to these galaxies. 
Previous absorption-line studies have indeed demonstrated that UV absorption 
is observable in the extended halos of the most nearby 
LG members, such as the Magellanic Clouds (e.g., de\,Boer \& Savage 1980;
Lehner \& Howk 2007; Richter et al.\,2014) and Andromeda
(Rao et al.\,2013; Lehner et al.\,2015; Barger, Lehner \& Howk 2016).


\begin{figure*}[th!]
\begin{center}
\resizebox{0.93\hsize}{!}{\includegraphics{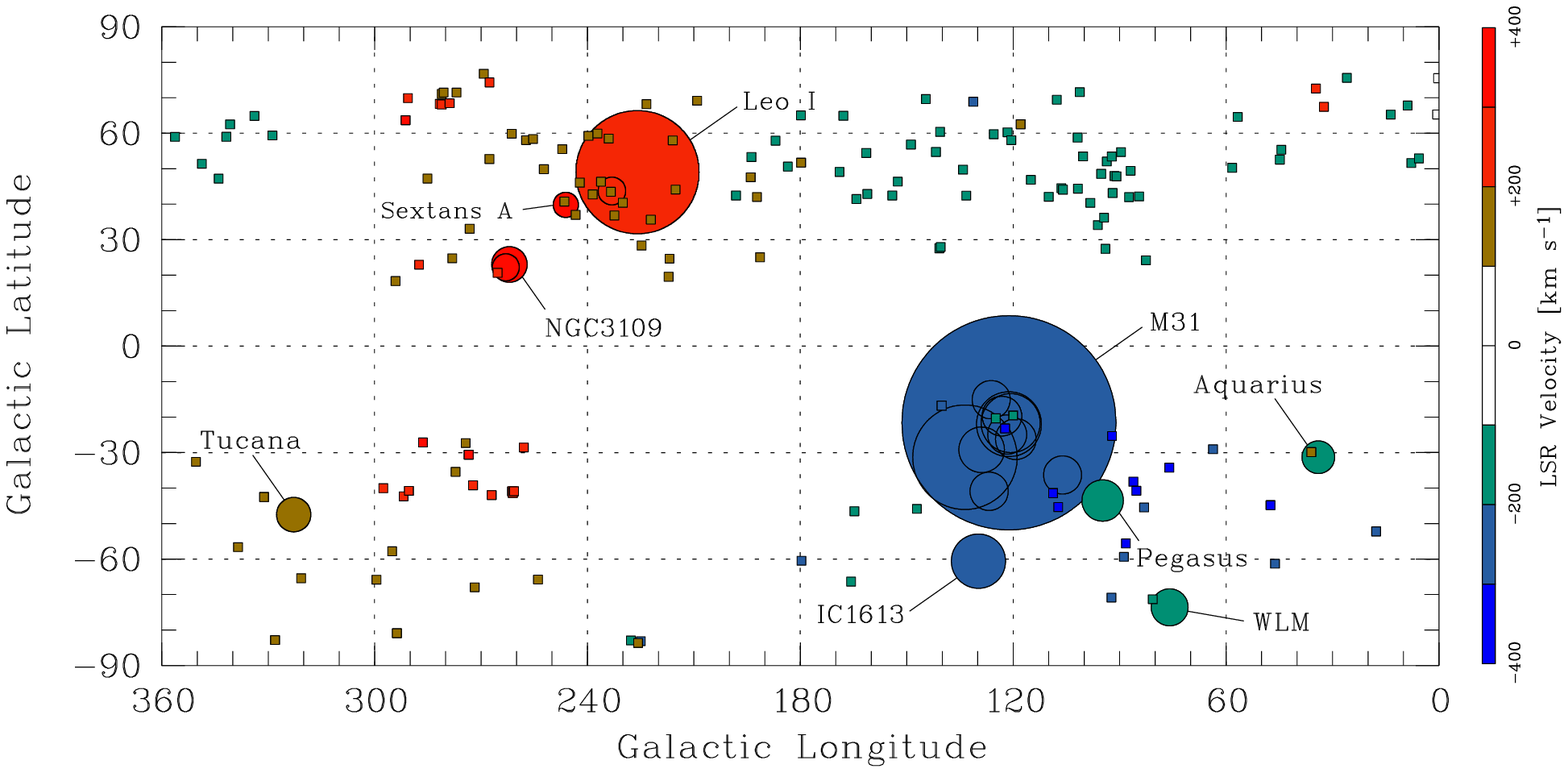}}
\caption[]{
Sky distribution of 19 Local Group galaxies that are located outside
the virial radius of the MW and that have systemic velocities
$|v_{\rm sys,LSR}|=100-500$ km\,s$^{-1}$, $|b|\geq 15\degree$, and
$M_V>20$. The size of the circles indicate
the size of their halos at $2\,R_{\rm vir}$ and the color indicates
their radal velocity (see color scale).
The filled small boxes indicate the position of the COS sightlines with
HVC detections using same color coding for the mean absorption velocities.
}
\end{center}
\end{figure*}


A list of the LG member galaxies is presented in the Appendix 
in Table A.3. 28 of the listed LG galaxies have radial velocities in the 
range $|v_{\rm LSR}|=100-500$ km\,s$^{-1}$ and are located at $|b|\geq 15\degree$.
To investigate the relation between high-velocity absorption and gaseous
halos of LG galaxies we have calculated the projected (angular) size of 
these galaxies.
We here concentrate on LG galaxies {\it beyond} the virial radius of 
our own galaxy ($R_{\rm vir,MW}\approx 200$ kpc; Dehnen et al.\,2006; 
McMillan 2011), thus excluding 
very nearby galaxies such as the Magellanic Clouds and other MW satellite 
galaxies. We also only consider the brightest galaxies with $M_V<20$ mag 
that have the largest angular cross section.
For the calculation of $R_{\rm vir}$ and the angular size we use the
information on the galaxies' absolute magnitudes and distances,
which we transform into an estimate of their luminosities. We then calculate
$R_{\rm vir}$ from the $R_{\rm vir}/L$ relation from Stocke et al.\,(2013),
which is based on recent halo matching models (Trenti et al.\,2010), and
then determine the angular size at $2\,R_{\rm vir}$. 
In Fig.\,10 we show the sky distribution and angular size at
$2\,R_{\rm vir}$ of the 19 LG galaxies 
selected in this manner together with the COS sightlines along
which HVC absorption is detected.
In the finding charts in the Appendix (Fig.\,B.2) we have indicated 
the galaxy velocities for those sightlines that pass within 
$2\,R_{\rm vir}$ of the selected 19 LG galaxies.

There is an interesting match between the general velocity 
distribution of the galaxies and the absorption velocities. For 
$l>180 \degree$ all LG galaxies have positive radial velocities 
from the perspective of the Sun, while for $l\leq 180 \degree$ the
radial velocities are negative. This trend reflects the general flow
of galaxies in the LG towards the LG barycenter (with the MW and M31 being
part of this flow), as concluded from studies of the overall dynamics
of the LG members (Peebles et al.\,2001,2011; Whiting 2014). 
As shown in Whiting (2014; his Fig.\,1), the absolute space
velocities of the LG galaxies are expected to be larger if they are
closer to the barycenter. From the perspective of the Milky Way,
LG galaxies that lie in the antipodal direction with respect to the LG 
barycenter ($l=147\degree, b=-25\degree$)
thus are expected to lag behind the MW flow speed, thus having 
positive relative velocities, while galaxies located on the opposite side
of the LG barycenter will move towards the Milky Way, thus having
negative relative velocities.
We would like to emphasize that the effect of galactic rotation only plays 
a minor role for the observed radial velocities at $|b|>30\degree$.
As we will discuss below, our observations as well as constrained LG 
simulations suggest that the LG gas (i.e., gas outside the halos of LG member 
galaxies) is expected to show an identical velocity-flow pattern. 

Fig.\,10 implies that there are only two general directions in the sky for which
the distribution of COS sightlines allows us to systematically study the 
connection between high-velocity absorption and nearby LG galaxies based
on {\it multiple} lines of sight.

In the northern sky there are 12 sightlines that pass through the Leo\,I dwarf
galaxy ($l=226.0\degree, b=+49.1\degree$) at impact parameters $\rho\leq 2R_{\rm vir}$. 
High-velocity absorption is detected in this direction at positive velocities 
in the range $+100$ to $+250$ km\,s$^{-1}$ (Complex WA), but not near the systemic velocity
of the Leo\,I at $v_{\rm LSR}\approx 285$ km\,s$^{-1}$. Thus,
there is no evidence for an extended gaseous halo around Leo\,I that could
be detected in the ionization states discussed here, supporting earlier
results presented by Bowen et al.\,(1997) based on low-resolution spectra
of three sightlines passing the inner halo of Leo\,I. The lack of absorption 
does not necessarily mean a lack of CGM gas around Leo\,I.
Since the stellar metallicity of Leo\,I is low ($0.05$ solar; Bosler et al.\,2007),
the metal content of the CGM might be too small to be detectable in
C\,{\sc iv} and Si\,{\sc iii} within such a small dark matter halo. Alternatively, the 
gas might be highly ionized and thus invisible in the UV lines traced
by COS. In any case, there is no evidence for the presence of cold streams that
could feed Leo\,I with fresh material to form stars. 
This is not surprising, however, because it is known from HST observations 
that the star-formation activity in Leo\,I dropped dramatically 
$\sim 1$ Gyr ago (Gallart et al.\,1999). Possibly, the lack of cold gas
around Leo\,I and the quenching of star-formation in this dwarf galaxy are 
related.

In the southern sky, M31 and its companion galaxies are distributed at $l<150\degree$,
all of them having negative radial velocities. Fig.\,10 shows that there are seven sightlines 
that pass M31 within $2 R_{\rm vir}$ and that show absorption at high negative 
velocities.
The interpretation of blue-shifted absorption along these sightlines is tricky, however, 
as there are large amounts of gas from the MS at high negative velocities 
in the general direction of M31.
Disentangling these two components (MS and M31) can only be done based on specific assumptions 
about the kinematics of the M31 CGM and the velocity distribution of MS gas.
Lehner et al.\,(2015; hereafter L15) present a detailed analysis of the COS spectra in the direction
of M31. They came to the conclusion that the M31 CGM is visible in the spectra of 
HS\,0058+4213, HS\,0033+4300, RX\,J0048.3+3941, UGC\,12163, 3C\,66A, MRK\,335, PG\,0003+158,
and NGC\,7469 in the LSR velocity range between $-300$ and $-150$ km\,s$^{-1}$, while
material from the MS is seen predominantly at $v_{\rm LSR}\leq -400$ km\,s$^{-1}$. 
Since we are using the same data in our study, the corresponding absorption features 
associated with the M31 CGM are visible in the velocity plots in Fig.\,B.2.
Assuming that the L15 velocity model for the MS and M31 is correct, the COS data 
for the above mentioned sightlines indeed suggest that M31 is
surrounded by a massive circumgalactic envelope of gas. 
Additional COS data for QSO sightlines that are located within $2 R_{\rm vir}$ will
help to distentangle gaseous material stemming from the M31 CGM and the MS
(AMIGA project; Lehner et al.\,2017, in prep.).

There are other, {\it individual} sightlines in our COS sample that pass the halos of other
LG dwarf/satellite galaxies at $\rho\leq 2 R_{\rm vir}$ (see Fig.\,10).
The sightline towards Mrk\,509
passes Aquarius at $\rho \sim 1 R_{\rm vir}$, but no absorption
is found within $140$ km\,s$^{-1}$ of the systemic velocity of Aquarius.
Similarly, no absorption is found at $v_{\rm LSR}>300$ km\,s$^{-1}$ 
towards the background source NGC\,3125, where circumgalactic gas from the nearby 
($\rho < 1.7 R_{\rm vir}$) LG galaxies NGC\,3109 and Antila would
be expected to be seen. 
The sightline towards PG1011$-$040  
passes Sextans\,A at $\rho \sim 0.5 R_{\rm vir}$, but, again, no absorption
is found within $160$ km\,s$^{-1}$ of the systemic velocity of Sextans\,A.
In contrast to the previous cases, absorption is found 
towards LBQS$-$0107$-$0235 at $v_{\rm LSR}\approx -200$ km\,s$^{-1}$,
a sightline that passes IC\,1613 ($v_{\rm sys}=-236$ km\,s$^{-1}$) 
at $\rho\sim 1.5 R_{\rm vir}$.
The high-velocity absorption towards LBQS$-$0107$-$0235 ($l=134\degree,
b=-65\degree$) is, however, 
more likely associated with gas from the MS, which
shows absorption (and emission) at similar velocities along other
sightlines in this general direction of the sky. No firm conclusion 
about the origin of this (weak) absorption feature and its possible relation
to IC\,1613 can be made based on this single line of sight.

Summarizing, only for the case of M31 do the COS data provide compelling evidence 
for the presence of an extended gaseous galaxy halo. For the LG dwarfs/satellites 
outside the virial radius of the MW
there are no hints of a circumgalactic gas component.

\subsection{Local Group gas}

One important aspect of our study is the search for a possible connection of
high-velocity UV absorption with gas bound to the Local Group as a whole,
gas far away from individual galaxies. A number of previous studies
have suggested that part of the observed UV and X-ray absorption of high ions
such as O\,{\sc vi}, O\,{\sc vii}, and O\,{\sc viii} at
$|v_{\rm LSR}|\leq 500$ km\,s$^{-1}$ is caused by LG gas outside the virial radius of the Milky Way
(e.g., Sembach et al.\,2003; Collins, Shull \& Giroux 2005; Gupta et al.\,2012), gas that may harbor the
dominating baryon fraction in the Local Group. 
A direct proof of this hypothesis would require a reliable distance measurement
of the absorbing (ionized) gas in the LG via the bracketing method 
(e.g., Wakker et al.\,2007, 2008), which is extremely challenging due
to the lack of suitable background sources.

Absorption-line studies of gas in other galaxy groups, however, do provide compelling
evidence for the presence of discrete gas structures in groups outside the virial radii
of group member galaxies, structures that give rise to absorption in intermediate and
high ions (e.g., Stocke et al.\,2014). Yet, the interpretation of
individual absorbers in group environments often remains inconclusive 
(see, e.g., the Dorado group; Richter et al.\,2016).

From our absorption-line analysis and the comparison between 
UV absorption and 21 cm emission we have identified two
distinct regions, at $l>240\degree, b>60\degree$ 
in the northern sky and at $l<120\degree, b=-60\degree-0\degree$ in the
southern sky, for which high-velocity gas with $|v_{\rm LSR}|>200$ km\,s$^{-1}$
has distinct properties:

\begin{itemize}

\item
both regions contain very little H\,{\sc i}, as indicated by the
lack of large-scale 21 cm emission (Fig.\,8);

\item
towards these directions the high-velocity O\,{\sc vi} absorption 
is particularly strong (Sembach et al.\,2003;
Wakker et al.\,2003)

\item
the detected H\,{\sc i} 21 cm clumps at $l<120\degree, 
b=-60\degree -0\degree$ have properties that are unusual compared to other HVCs,
as they are lacking cold cloud cores and do not show extended diffuse H\,{\sc i} 
emission (see Winkel et al.\,2011)

\item
in the same region, the C\,{\sc iv}/Si\,{\sc iii} ratio is enhanced compared to
other regions, indicating a high degree of ionization (Fig.\,7, lower right panel)

\item
high-velocity gas near $l=36\degree$ and $b=-36\degree$ has a very low thermal
gas pressure of $P/k<5$ K\,cm$^{-3}$, incompatible with pressures expected in the
inner halo of the Milky Way (Wolfire et al.\,1995).

\end{itemize}

As mentioned in Sect.\,4.4 these trends provide evidence that the 
gas at $l<120\degree, b=-60\degree-0\degree$, $v_{\rm LSR}<-300$ km\,s$^{-1}$
and possibly also the gas $l>240\degree$, $b>60\degree$, $v_{\rm LSR}>200$ km\,s$^{-1}$
is relatively diffuse and predominantly ionized, possibly residing
at large distances from the disk in outer halo or even in the LG 
(see dicussions in Sembach et al.\,1999; 2003; Wakker et al.\,2003;
Nicastro et al.\,2003; Richter et al.\,2009; Winkel et al.\,2011).
Throughout the following, we refer to the absorber regions $l>240\degree,
b>60\degree$, $v_{\rm LSR}>200$ km\,s$^{-1}$ and
$l<120\degree, b=-60\degree-0\degree$, $v_{\rm LSR}<-300$ km\,s$^{-1}$
as northern ionized region (NIR) and southern ionized region (SIR),
respectively.


\begin{figure}[t!]
\begin{center}
\resizebox{0.85\hsize}{!}{\includegraphics{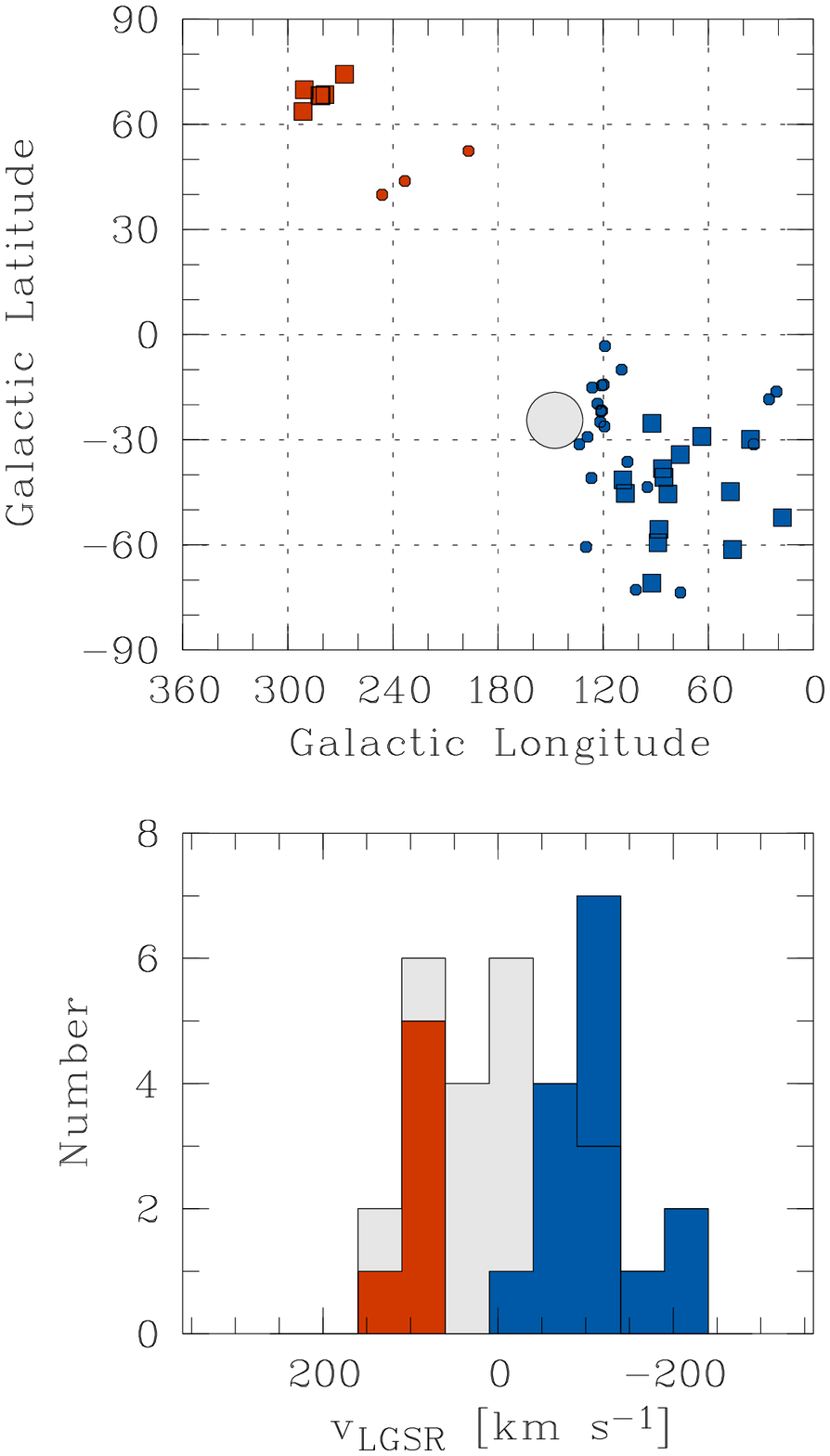}}
\caption[]{
{\it Upper panel:} sky distribution of high-velocity absorbers 
with $l>180\degree$, $b>30\degree$, $v_{\rm LSR}>200$ km\,s$^{-1}$ 
and $l<180\degree$, $b<0\degree$, $v_{\rm LSR}<-200$ km\,s$^{-1}$
(filled boxes) together with LG galaxies 
(filled circles) in the same $(l,b)$ range.
{\it Lower panel:} velocity distribution of these absorbers (colored
histograms) and galaxies (gray-shaded histrograms) after
transforming their velocities into the LGSR velocity frame.
}
\end{center}
\end{figure}


In the northern sky, Complex WA is located at $d=8-20$ kpc in a  
($l,b$) range similar as the one given above, but at lower radial velocities 
($v_{\rm LSR}<200$ km\,s$^{-1}$; see Table 3). Thus, the NIR is not associated with Complex WA.
The southern region at $l<120\degree, b=-60\degree -0\degree$ contains negative-velocity 
gas from the MS (formally defined for $l>60\degree$) and from Complex 
GCN ($l< 60\degree$; Table 3). In contrast to the MS, origin and distance of Complex GCN are 
unknown. Therefore, at least some of the detected high-velocity absorption features in the SIR
could be located deep in the LG, as there are no observational
data that {\it favor} a location within the Milky Way halo.

It is striking that the NIR and SIR (together with other nearby 
high-velocity absorbers) form 
a velocity dipole on the sky that mimics 
the one seen for the LG galaxies (Fig.\,10). 
To further emphasize this, we 
plot in the upper panel of Fig.\,11 the sky distribution of all 
high-velocity absorbers with $l>180\degree$, $b>30\degree$, 
$v_{\rm LSR}>200$ km\,s$^{-1}$ and $l<180\degree$, $b<0\degree$, 
$v_{\rm LSR}<-200$ km\,s$^{-1}$ (filled boxes) together with 
LG galaxies (filled circles) in the same $(l,b)$ range. 

LG gas that follows the same flow towards the LG barycenter as the galaxies (Sect.\,5.3)
would show exactly the kind of velocity dipole that is seen in Fig.\,11, upper panel. 
The Milky Way would move faster towards the barycenter than the LG gas that would 
lag behind in the general anti-barycenter direction, so 
that positive {\it relative} velocities are expected. In the barycenter direction, 
the Milky Way halo would ram into LG gas that that is at rest at the barycenter or that 
flows to the barycenter 
from the opposite side of barycenter (i.e., from the direction of M31 and its 
large-scale environment), so that it would have high negative velocities. 
Similar arguments already have been used by Blitz et al.\,(1999) to model the expected 
kinematics of an extragalactic HVC population (see also Collins, Shull \& Giroux 2005).

In the lower panel of Fig.\,11 we show the velocity distribution of the
NIR and SIR absorbers after 
transforming their velocities into the Local Group Standard of Rest (LGSR)
velocity frame. In general, the absolute velocity spread in the sample of the
absorbers is substantially reduced from $\Delta v=720$ km\,s$^{-1}$ in the LSR frame 
to $\Delta v=360$ km\,s$^{-1}$ in the LGSR frame. Similarly, the standard deviation
of the velocity distribution reduces from 260 km\,s$^{-1}$  in the LSR velocity frame
to 110 km\,s$^{-1}$ in the LGSR frame (see also Sembach et al.\,2003 and Nicastro et 
al.\,2003 for identical trends in FUSE O\,{\sc vi} absorption-line data).
In view of the
latitude range, this trend can only partly be related 
to the removal of the Galactic rotation effect in the LGSR frame.

A more detailed interpretation of this plot is difficult, however, without 
knowing the contribution of the MS to the velocity distribution at high
negative velocities. In addition, the overall kinematics of the LG gas is 
expected to be more complicated than the kinematics of the galaxies because 
some of the gas is being accreted by the LG galaxies, thus having 
different directions of motion. 
The 3D velocity distribution of the gas might also be influenced 
by ram-pressure forces and large-scale turbulent flows. To
critically evaluate the importance of these processes and 
to investigate whether a simple flow scenario towards the group barycenter is 
realistic for LG gas around MW and M31, hydrodynamical simulations of the LG and its
gaseous environment are required (N14).
In Sect.\,7 we analyze such simulations to further
study the expected large-scale kinematics of gas in the LG. 


\begin{figure}[t!]
\begin{center}
\resizebox{1.0\hsize}{!}{\includegraphics{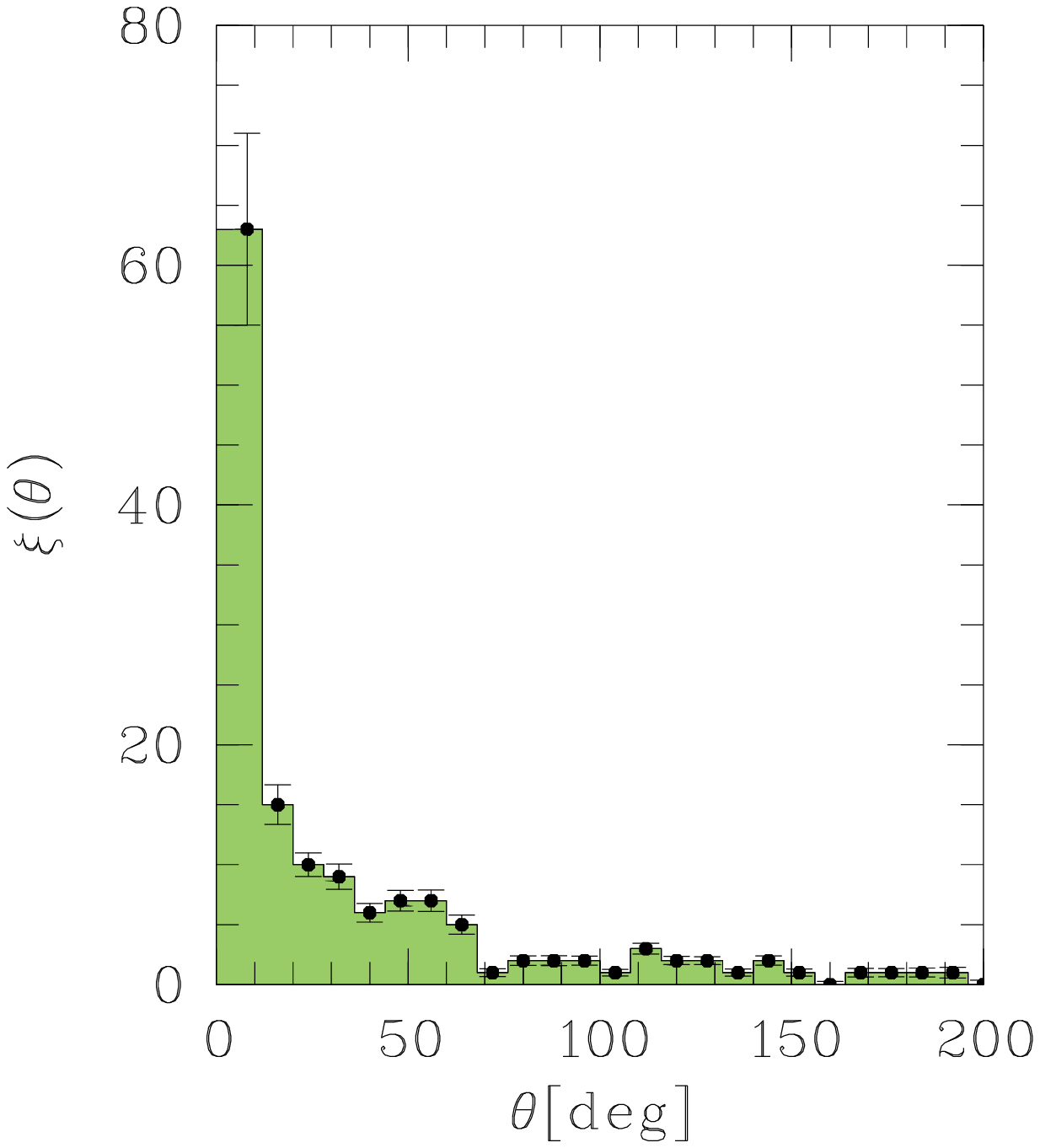}}
\caption[]{
Two-point correlation function $\xi(\theta,\Delta v)$ of
HVC absorbers along sightlines with angular separations of
$d\leq 200$ deg.
}
\end{center}
\end{figure}


\begin{figure*}[t!]
\begin{center}
\resizebox{0.93\hsize}{!}{\includegraphics{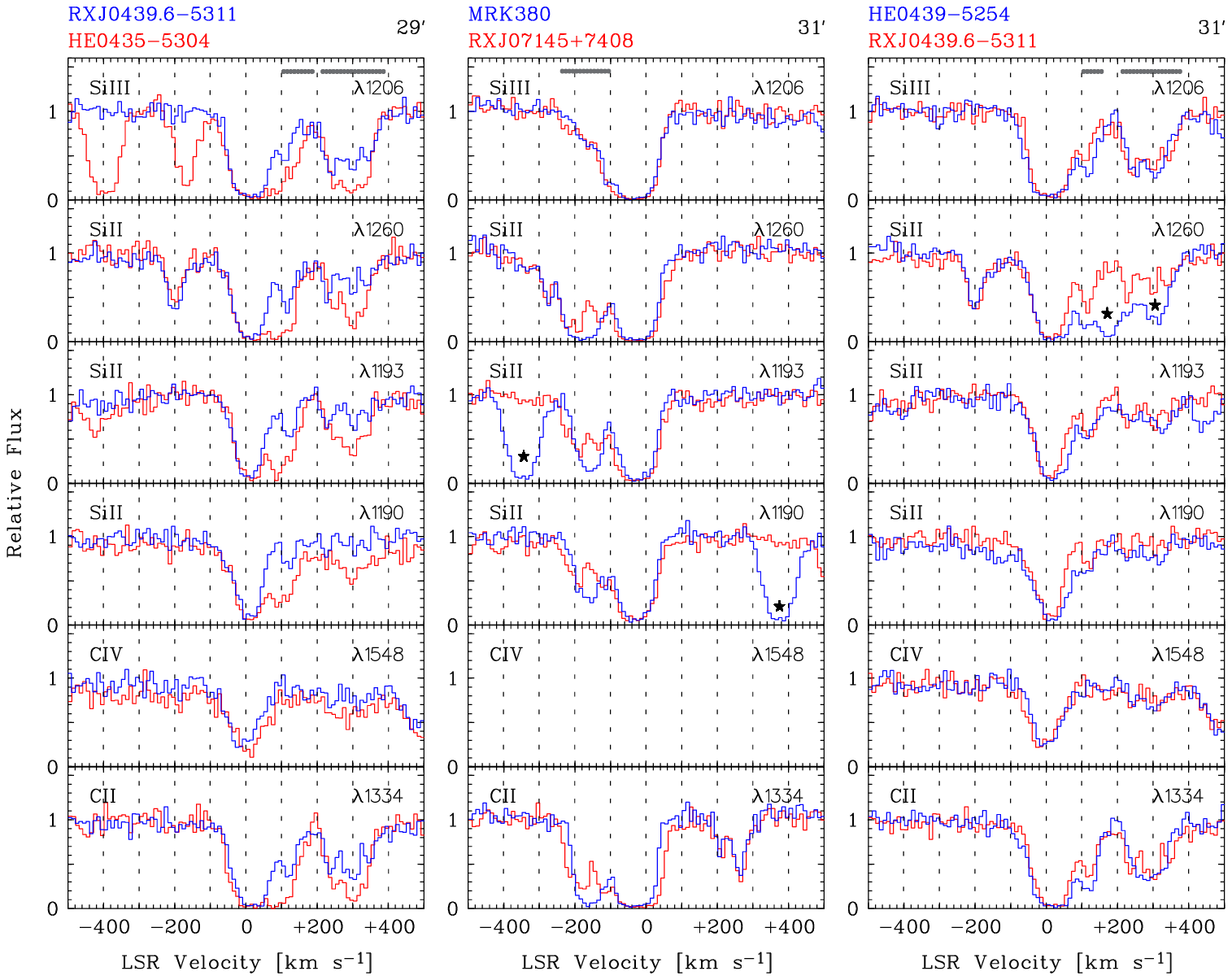}}
\caption[]{
Three examples for sightline pairs with angular separations
$\leq 1 \degree$. Differences in the absorption profiles indicate small-scale
structure in the absorbing gas. High-velocity absorption components are
indicated with the gray bar. Blends with intergalactic
absorption features are indicated with a star symbol.
QSO names and angular separations between the sightlines are given
on top of each panel.
The Appendix contains this plot for all 12 sightline-pairs.
}
\end{center}
\end{figure*}


\subsection{Spatial clustering and sub-structure}

To characterize the amplitude of spatial clustering of HVC absorption
components on the sky we analyze the two-point correlation function, 
$\xi(\theta,\Delta v)$, of HVC absorbers, where $\theta$ 
represents the angular distance between
any two sightlines and $\Delta v$ is the selected velocity bin.
We determined $\xi(\theta,\Delta v)$ in our data set by 
counting the number of HVC absorption components,
${\cal N_{\rm HVC}}$, in a given bin $\Delta v$ along all sightline 
pairs in our sample and comparing the result 
to the number of absorbers derived for a random distribution of HVC components,
${\cal N_{\rm random}}$. Using the Davis \& Peebles (1983) standard 
estimator, we can write

\begin{equation}
\xi(\theta,\Delta v)=\frac{{\cal N_{\rm HVC}}(\theta,\Delta v)}
{{\cal N_{\rm random}}(\theta,\Delta v)}-1.
\end{equation}

The random HVC absorber distribution was calculated from a Monte-Carlo simulation,
in which we simulated 100 realizations of random distributions in $(l,b)$ and 
$v=[-500,-100][+100,+500]$ along 270 sightlines.

The result of our analysis is shown in Fig.\,12, where we plot $\xi(\theta,\Delta v)$ against
$\theta$ for all sightline pairs with $\theta\leq 200\degree$. 
A strong clustering signal 
is evident for small angular separations $\theta\leq 30\degree$, while $\xi$ 
decreases rapidly for larger angles. Obviously, the high-velocity absorption
is caused by coherent gas structures that typically span $<30\degree$ on the 
sky (see also Lehner et al.\,2012).
This result is in line with the visual appearance of typical 
angular sizes of the 21 cm HVC Complexes displayed in Fig.\,8.
At a distance of $d=10$ kpc (e.g., Complex C) an angular separation 
of $\theta=1\degree$ on the 
sky corresponds to a linear separation of $l\approx 0.0175\,d\,(\theta /{\rm 1\degree})=175$ pc.

The sky distribution of the sightlines shown in Fig.\,2 indicates that there
are sightline pairs with small angular separations that can be used
to study the internal structure of the Milky Way CGM on pc scales by
comparing along two adjacent sightlines the absorption depths and velocity centroids 
of high-velocity absorbers.
We have selected 12 sightline pairs in our COS sample with angular separations
$\leq 1\degree$ to identify such differences in the absorption patterns.
In Fig.\,13 we show three examples of these pairs, where the angular separation
between the sightlines is indicated on top of each panel. 
The first sightline pair in Fig.\,13 (RX\,J0439.6$-$5311/HE\,0435$-$5304) has an angular 
separation of only $29 \arcmin$. These sightlines trace gas 
from the MS at high positive velocities. There are significant differences
in the absorption depth of all plotted ion transitions, indicating 
changes in the gas properties on (relatively large) scales of $l\approx450$ pc
if a MS distance of $d=55$ kpc is assumed. The sightlines towards Mrk\,380 and RXJ\,01745+7408 
(Fig.\,13, middle panel) are separated by $\theta =31 \arcmin$ and 
pass through negative-velocity gas in HVC Complex A at 
$d\approx10$ kpc distance (van\,Woerden et al.\,1999; Wakker 2001). 
Significant differences are seen in the singly-ionized species Si\,{\sc ii} and C\,{\sc ii},
indicating differences in the gas distribution on scales of $l\approx 90$ pc.
The third sightline pair in Fig.\,13 (HE\,0439$-$5254/RX\,J0439.6$-$5311) 
again shows differences in the absorption properties in the MS on scales
of several hundred pc.
The full set of these plots is presented in the Appendix in Fig.\,B.1.
In addition, we discuss in the Appendix (Sect.\,A.3; Fig.\,A.3) small-scale
structure in the equivalent-widths maps of Complex C and Complex WA and
previous results on small-scale structure in HVCs.

The observed variations in the UV absorption patterns of 
HVCs along adjacent sightlines further support the scenario, in which 
HVCs represent coherent gaseous structures with large-scale kinematics 
(on kpc scales) and small-scale variations (on pc and sub-pc scales; 
see Sect.\,A.3) in physical conditions.


\section{Mass estimate of the Milky Way's CGM}

\subsection{Total gas columns and ionization fractions}

A commonly used approach to study the ionization conditions in 
individual high-velocity clouds in the Milky Way halo is the 
use of column density ratios of low, intermediate, and high
ions together with Cloudy (Ferland et al.\,2013) ionization models 
(e.g., Richter et al.\,2009, 2013; Fox et al.\,2013, 2014, 2015; Herenz et al.\,2013).
Such ionizations models provide useful constraints on the gas densities,
ionization parameters, and metal abundances in the gas, but
they require accurate information on the column densities of 
different ions, the velocity-component structure, and the 
distance of the absorbing gas from the Milky Way disk. The latter
aspect is important, because the ionizing radiation field in the 
Milky Way halo is expected to be anisotropic with contributions
from the overall UV background and from stars in the Milky Way 
disk (see Fox et al.\,2005).

Because our COS absorber sample is limited in S/N, spectral resolution, 
and because model-constraining distance information is available for only 
a sub-set of the HVC complexes, we refrain from using Cloudy
models in this all-sky survey, as these would be afflicted with large 
systematic uncertainties.
Instead, we estimate for each high-velocity absorber a lower limit for the total column 
density of hydrogen, $N$(H$)=N($H\,{\sc i}$)+N($H\,{\sc ii}), from the 
column densities of Si (Si\,{\sc ii}+Si\,{\sc iii}) and C (C\,{\sc ii}+C\,{\sc iv})
and combine the values for $N$(H) with the 
neutral gas column densities derived from the 21 cm data
(see also Fox et al.\,2014). To obtain a limit for $N$(H) we define


\begin{equation}
{\rm log}\,N'({\rm H})=\,
{\rm log}\left[\frac{N({\rm Si}\,{\rm II})+N({\rm Si}\,{\rm III})}
{X\,{\rm (Si/H)}_{\sun}}\right]
\end{equation}


and


\begin{equation}
{\rm log}\,N''({\rm H})=\,
{\rm log}\left[\frac{N({\rm C}\,{\rm II})+N({\rm C}\,{\rm IV})}
{X\,{\rm (C/H)}_{\sun}}\right].
\end{equation}


and consider the larger of the two values for each 
absorber; in 90 percent, $N'({\rm H})>N''({\rm H})$.

In the above equations, (Si/H$)_{\sun}=3.24\times 10^{-5}$ and 
(Si/H$)_{\sun}=2.69\times 10^{-4}$ are the solar abundances
of Si and C (Asplund et al.\,2009). The parameter $X$ denotes the mean metallicity 
of the gas in solar units. We assume $X=0.1$,
motivated by the fact that most of the prominent HVCs in the Milky Way halo,
such as the MS, Complex C, Complex A, have $\alpha$ abundances
of $\sim 10$ percent solar (Wakker et al.\,1999; Richter et al.\,2001;
Sembach et al.\,2004; Fox et al.\,2013, 2015). 


\begin{figure}[t!]
\begin{center}
\resizebox{0.9\hsize}{!}{\includegraphics{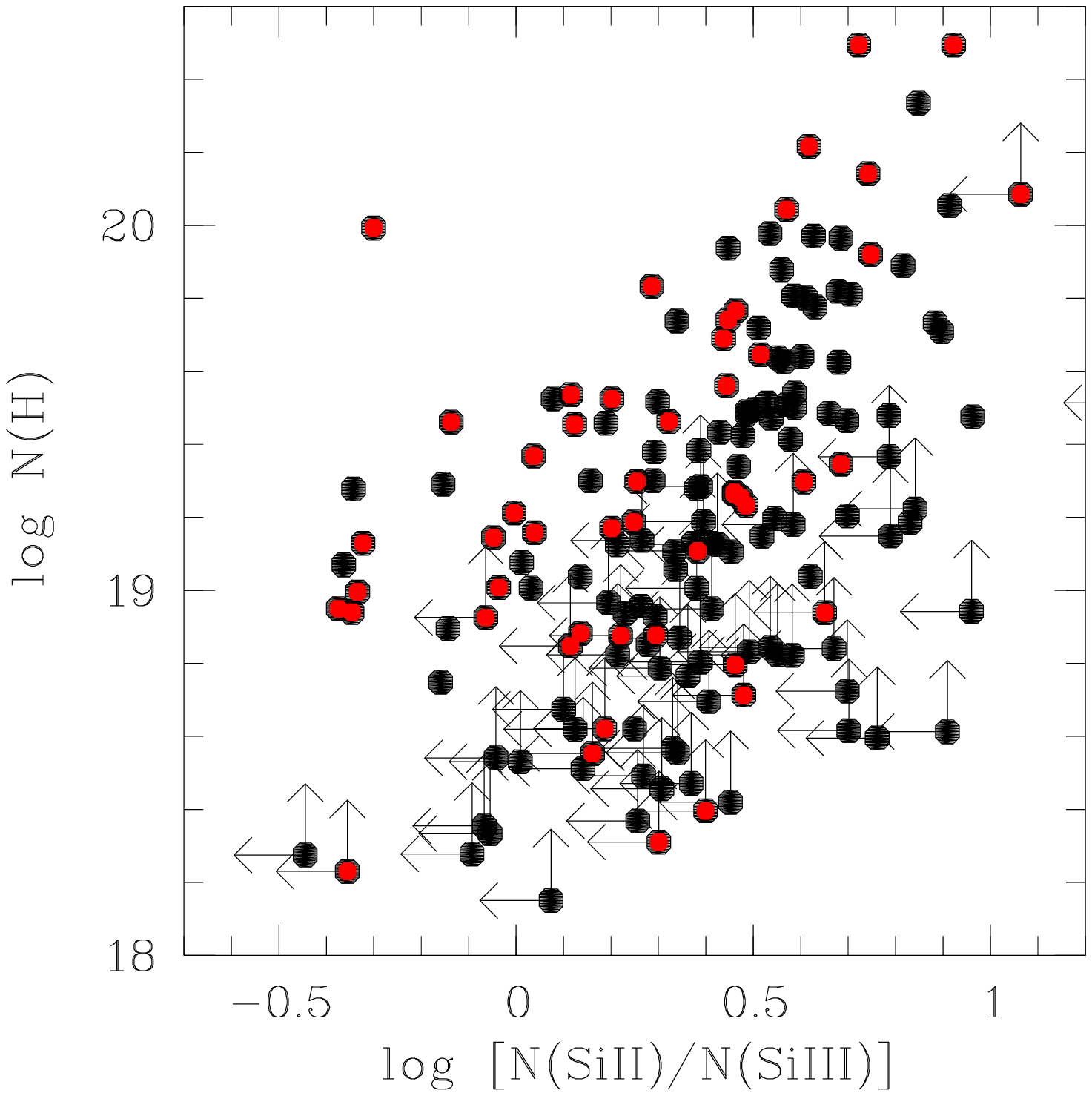}}
\caption[]{
Scatter plot showing the logarithmic total hydrogen column density
for the 187 HVC absorbers plotted against the logarithmic
Si\,{\sc ii}/Si\,{\sc iii} column-density ratio. Values for 
MS sightlines are indicated in red.
}
\end{center}
\end{figure}


We need to emphasize at this point that the values for $N'$(H) and $N''$(H) derived in this way
reflect only the amount of hydrogen that is traced by Si\,{\sc ii}, Si\,{\sc iii},
C\,{\sc ii}, and C\,{\sc iv}.
Photoionized and shock-heated H\,{\sc ii} at lower gas densities and/or higher gas temperatures
that is traced by higher metal ions such as O\,{\sc vi}, O\,{\sc vii}, and O\,{\sc viii}
is known to exist in the Milky Way's CGM as well (Sembach et al.\,2003; Wakker et al.\,2003;
Gupta et al.\,2012; Miller et al.\,2016), but is not sampled in our data. In addition, 
because the Si\,{\sc ii}, C\,{\sc ii},
Si\,{\sc iii} absorption is saturated in the strongest absorbers, 
$N'$(H) and $N''$(H) are systematically 
underestimated in these systems. However, $N'$(H) and $N''$(H) cannot be smaller than 
$N$(H\,{\sc i}) and thus the H\,{\sc i} column density measured from the 21 cm data
sets the lower limit for the total hydrogen column.
Consequently, we adopt as final value $N$(H$)=$\,MAX[$N$(H\,{\sc i}), $N'$(H),$N''$(H)],
which represent conservative lower limit for the total hydrogen column density
in each absorber.

In Fig.\,14, we show the total hydrogen column density plotted against
the Si\,{\sc ii}/Si\,{\sc iii} column-density ratio for all
187 high-velocity absorbers, where limits for these parameters are 
indicated with arrows (red dots indicate values for MS sightlines). 
Despite the large scatter there is a clear trend of log $N$(H) increasing with 
increasing Si\,{\sc ii}/Si\,{\sc iii} ratio. 
{\it The intrinsic scatter of this distribution would be significantly 
smaller if the many data points with limits in Si\,{\sc ii}/Si\,{\sc iii} 
and H\,{\sc i} would be removed.}
The observed trend implies that
high-column density HVCs (which are also seen in H\,{\sc i} 21 cm
emission) have a smaller mass fraction of
diffuse gas (traced by Si\,{\sc iii}) than low-column density
HVCs, which are predominantly diffuse and ionized.
Although Si\,{\sc iii} is the most sensitive Si ion to trace
diffuse, multi-phase gas in HVCs, the dominating ionization
state of Si in our HVC absorber sample (in terms of column density)
is Si\,{\sc ii}. In our sample, the Si\,{\sc ii} column density exceeds
that of Si\,{\sc iii} for gas columns log $N$(H$)>10^{19}$ cm$^{-2}$, typically.

In Fig.\,15, upper panel, we show the distribution of the measured 46 neutral 
hydrogen column densities (log $N$(H\,{\sc i}$)>18.7$, upper panel) in our 
HVC sample. The majority (87 percent) of the measured H\,{\sc i} column
densities lie below $10^{20}$ cm$^{-2}$ and for log $N$(H\,{\sc i}$)>19.4$ 
the distribution shows a decline in the form of a power-law, similar
to what has been found in previous H\,{\sc i} HVC surveys (see, e.g., 
Lockman et al.\,2002; Wakker 2004).
In the lower panel of Fig.\,15 we display the distribution of the estimated
total hydrogen column densities for all 187 HVC absorbers. Also
the total hydrogen column densities predominantly are $<10^{20}$ cm$^{-2}$
with a median value of log\,$\langle N($H$)\rangle =19.13$
Because of the observational restrictions in constraining absolute values for 
$N$(H\,{\sc i}) and $N$(H) for our HVC sample, only upper limits for 
the neutral gas fraction log $f_{\rm HI}=\,$log\,$N$(H\,{\sc i})$-\,$log\,$N$(H) 
can be given for most sightlines. 
Values/limits for log $N$(H\,{\sc i}), log $N$(H), and log $f_{\rm HI}$ for
each high-velocity absorber in our sample are listed in Table A.4 in the Appendix.


\begin{figure}[t!]
\begin{center}
\resizebox{0.8\hsize}{!}{\includegraphics{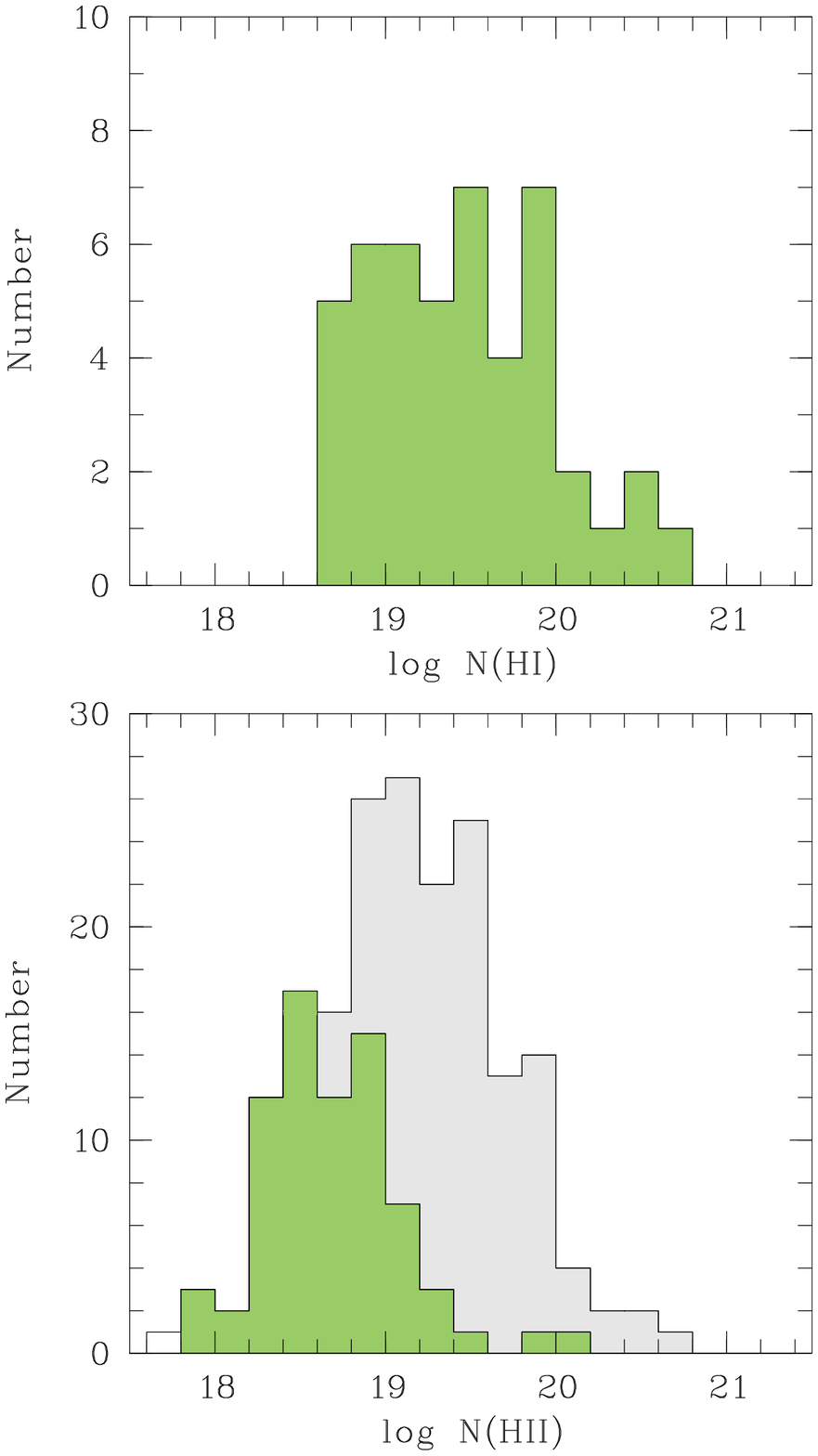}}
\caption[]{
{\it Upper panel:}  
distribution of the measured neutral
hydrogen column densities with log $N$(H\,{\sc i}$)>18.7$ 
for 43 HVC absorbers.
{\it Lower panel:} distribution of 
total hydrogen column densities.
The gray-shaded area shows the distribution of log $N$(H) derived for 
all 179 HVC absorbers. The green-shaded area indicates 
the distribution of log $N$(H) for systems with unsaturated 
Si\,{\sc ii}/C\,{\sc ii}/Si\,{\sc iii} absorption (see Sect.\,6.1).
}
\end{center}
\end{figure}


\subsection{The total gas masses of HVCs}

To estimate the total gas mass of the entire population of 
Galactic HVCs we assume
that HVCs are gas patches that are ellipsoidally/spherically distributed 
around the Milky Way disk. Let $r_{\rm disk}=20$ kpc be the fixed radius of the
Milky Way H\,{\sc i} disk and $r_{\rm s}$ be the distance of 
the absorbing structures from the Galactic center. 
We adopt a spherical 
distribution of HVCs for $r_{\rm s}\geq r_{\rm disk}$ and a
ellipsoidal distribution for $r_{\rm s}< r_{\rm disk}$ (i.e.,
in the latter case, the height above the disk is smaller than
the disk radius).

For each sphere, the area covered by HVCs at distance $r_{\rm s}$ 
then is simply
$A(r_{\rm s})=4\pi r_{\rm s}^2\,f_{4\pi}(r_{\rm s})\,B$,
where $f_{4\pi}(r_{\rm s})$ is the covering fraction of
HVC gas on that sphere. The parameter $B$ denotes the
correction factor for ellipsoidal symmetry and is defined
as $B=1+(r_{\rm s}/r_{\rm disk})^2\,({\rm arctanh}\,{\epsilon})/{\epsilon}$ 
with the ellipticity $\epsilon=(1-(r_{\rm s}/r_{\rm disk})^2)^{1/2}$.
Based on our observations, we estimate the mean total hydrogen column density,
$N$(H), for each HVC Complex listed in Table 3 using the method described in 
the previous subsection.

In general, $N$(H) is the product of the hydrogen volume density
and the absorption path length, $N$(H$)=n_{\rm H}d$.
In our HVC model, $d$ can be regarded as the thickness of the absorbing
gas layer. The mass per particle is $\mu m_{\rm H}$ with
$\mu=1.4$ as factor that corrects for the presence of helium
and heavy elements in the gas. Combining these equations,
we obtain an expression for the gas mass of an HVC at 
radius $r_{\rm s}$ then in the form


\begin{equation}
M_{\rm HVC}(r_{\rm s})=
\mu m_{\rm H}\,4\pi r_{\rm s}^2\,f_{4\pi}(r_{\rm s})\,B\,
\langle N({\rm H)}\rangle_{\rm HVC}.
\end{equation}


To calculate the total gas mass in each HVC,
we adopt the values for $r_{\rm s}$ and 
$f_{4\pi}(r_{\rm s})$ listed in Table 3 and take
the median total gas column density, $\langle N$(H$)\rangle$ 
in each HVC complex.
The total gas mass in the CGM traced by the ions
in our survey then is obtained by summing over all
$M_{\rm HVC}(r_{\rm s})$ (Table 3, column 12).

For the MS (including the LA), for instance, we assume
$r_{\rm s}=55$ kpc and $f_{4\pi}=0.47$ and from
our measurements, together with equations (5) and (6), we obtain
$M_{\rm MS}= 3.0 \times 10^9\,M_{\sun}$. This value is higher than
the one estimated by Fox et al.\,(2014), which is due
to larger assumed sky covering fraction of the MS in our
study ($f_{4\pi}=0.47$ vs. $0.27$; see Sect.\,3.1 in 
Fox et al.\,2014).
The total mass of the MS would be even higher, if some of the gas 
was located at larger distances than the canonical 55 kpc assumed here.
In fact, several authors suggested that the bulk of the MS is located at 
$d=100$ kpc rather than at $d=55$ kpc
(Besla, et al.\,2012; Jin \& Lynden-Bell 2008; Bland-Hawthorn et al.\,2013),
implying that the MS mass could be as large as $>5 \times 10^9\,M_{\sun}$.

The contribution of all other HVCs to the total mass is small;
their gas mass sums up to a total value of no more than
$M_{\rm HVCs without MS}= 4.3 \times 10^7\,M_{\sun}$.
The total mass of high-velocity gas around the Milky Way thus is 
clearly dominated by material that stems from the interaction
of the Milky Way with the Magellanic Clouds, while the contribution
of gas from other origins is negligible ($\sim 1$ percent; see also Fox et al.\,2014).
The total mass of HVCs thus is essentially identical to the estimated mass of the MS,
$M_{\rm HVC}= 3.0 \times 10^9\,M_{\sun}$, which is $\sim 43$ percent
of the ISM gas mass in the Galactic disk ($\sim 7\times 10^9\,M_{\sun}$; 
Ferriere 2001). Because in our survey we do not account for
the highly-ionized gas phase in HVCs as traced, for instance, by 
O\,{\sc vi} (e.g., Sembach et al.\,2003; Fox et al.\,2004), our estimates 
for the HVC masses represent conservative lower limits.

For the calculation of the Milky Way's gas-accretion rate of 
high-velocity gas (Table 2, last column) we set 


\begin{equation}
\frac{{\rm d}M_{\rm HVC}}{{\rm d}t} = 
\frac{M_{\rm HVC}\,\eta\,v_{\rm infall}}{r_{\rm s}},
\end{equation}


where we assume $v_{\rm infall}=100$ km\,s$^{-1}$ as characteristic 
value for the infall velocity of the gas and define 
$v_{100}=v_{\rm infall}/(100$ km\,s$^{-1}$)
as according scaling parameter.
The choice of $v_{\rm infall}=100$ km\,s$^{-1}$ 
is motivated by the results from hydrodynamical simulations of 
neutral HVCs infalling into hot, coronal halo gas (Heitsch \& Putman 2009;
Joung et al.\,2012). These studies demonstrate that the bulk of infalling 
gas clouds have velocities far below the expected ballistic velocities
due to the hydrodynamical interaction between the infalling gas and the
ambient hot medium.
The parameter $\eta\leq 1$ denotes the fraction of the HVC gas that
actually arrives in the Milky Way disk (in whatever form and state).
Estimating realistic values for $\eta$ is challenging, as this requires 
full-fledged hydrodynamical models that describe the passage of 
neutral and ionized gas through the hot coronal gas of the Milky Way 
(see again Heitsch \& Putman 2009; Joung et al.\,2012).
We here consider the range $\eta=0.1-1.0$
as realistic, but carry along this scaling parameter for the 
following estimates.

The lower limit of the total gas accretion rate from all HVCs
listed in Table 3 
comes out to $6.1\,v_{100}\eta\,M_{\sun}$\,yr$^{-1}$.
Here, the MS contributes with $5.5\,v_{100}\eta\,M_{\sun}$\,yr$^{-1}$,
while the contribution of all other HVCs sums up to
$0.6\,v_{100}\eta\,M_{\sun}$\,yr$^{-1}$.
If much of the MS would be located at $d=100$ kpc (as discussed above) or
the infall velocity of the gas would be higher,
the gas accretion rate would be increased accordingly.
For comparison, the current star-formation rate of the Milky Way
is estimated to be SFR(MW)$\leq 1.9 M_{\sun}$\,yr$^{-1}$ (Chomiuk \& Povich 2011;
Robitaille \& Whitney 2010), thus very similar. The infall of gas towards the disk in the form
of HVCs thus provides a significant mass flow that is exptected to substantially 
influence the future star-formation rate of the Milky Way.
Because we consider the velocity range $|v_{\rm LSR}|=100-500$ km\,s$^{-1}$
in our analysis, our calculations do not cover the contribution of 
low-velocity halo clouds (LVHCs; Peek et al.\,2009; Zheng et al.\,2015) 
to the total gas mass in the Milky Way halo and to the global gas accretion rate.

If we re-scale $f_{4\pi}$(MS) to previously used, smaller sky-covering
fractions of the Stream, our values for $M_{\rm HVC}$ and d$M_{\rm HVC}/$d$t$ are 
in good agreement with our previous results (Fox et al.\,2014) and with
estimates from earlier studies (e.g., Putman, Peek \& Joung 2012;
Richter 2012; Lehner \& Howk 2011). We would like to note
that these previous estimates are based on other data sets and 
different model assumptions, so that a direct comparison remains difficult.
For a detailed discussion of observational and theoretical studies of the
Milky Way's gas accretion we refer to the review by Richter (2017).
 

\begin{figure}[ht!]
\begin{center}
\resizebox{0.7\hsize}{!}{\includegraphics{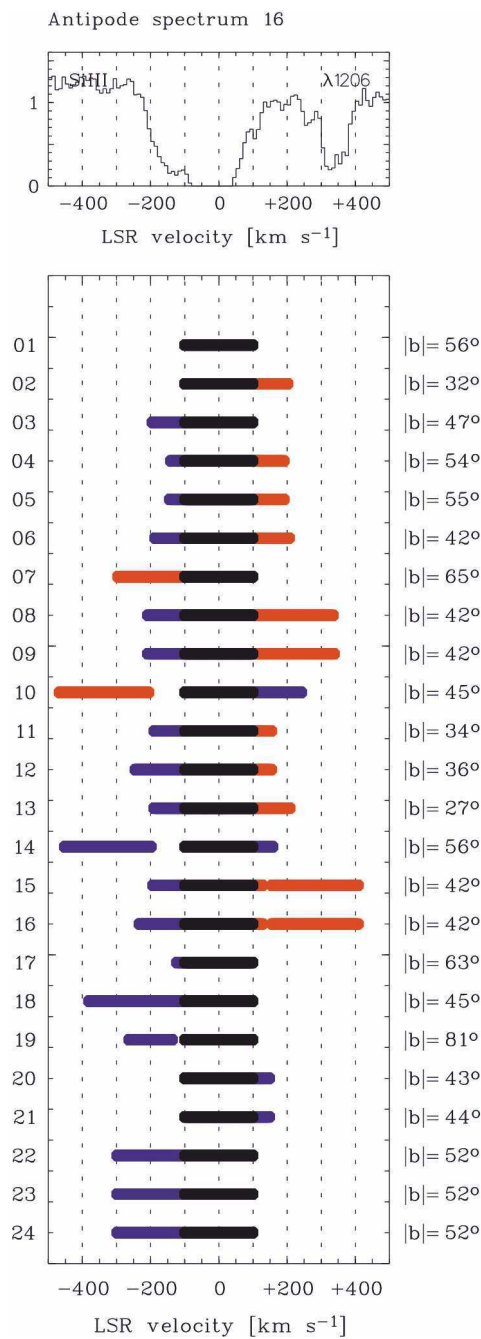}}
\caption[]{
{\it Upper panel:} example for a composite
absorption profile from two antipodal sightlines
(from antipode pair no.\,16). 
{\it Lower panel:} velocity distribution of the
Si\,{\sc iii} $\lambda 1206.50$ absorption in 24 antipodal
composite profiles with disk gas ($|v_{\rm LSR}|\leq 100$ km\,s$^{-1}$)
indicated in black and CGM absorption ($|v_{\rm LSR}|>100$ km\,s$^{-1}$)
indicated in red (MS absorption) and blue (other HVCs).
}
\end{center}
\end{figure}


\begin{figure}[ht!]
\begin{center}
\resizebox{0.8\hsize}{!}{\includegraphics{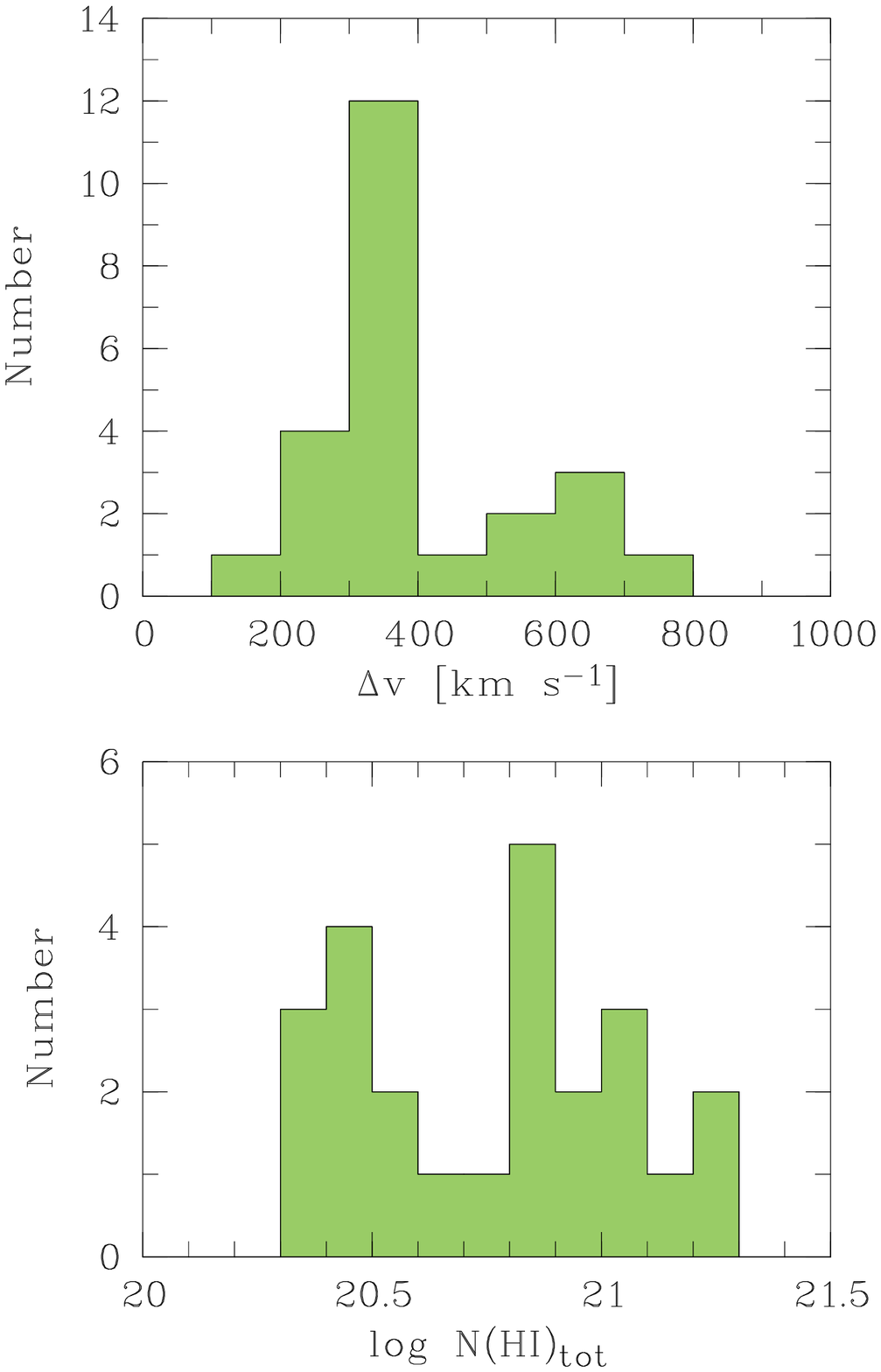}}
\caption[]{
{\it Upper panel:} distribution of velocity widths ($\Delta v$, in
bins of 100 km\,s$^{-1}$) in 24 antipodal
composite spectra sampling Milky Way disk and halo gas. {\it Lower panel:}
distribution of total H\,{\sc i} column densities along these 24 sightline-pairs
derived from 21 cm data.
}
\end{center}
\end{figure}


\begin{figure}[t!]
\begin{center}
\resizebox{0.8\hsize}{!}{\includegraphics{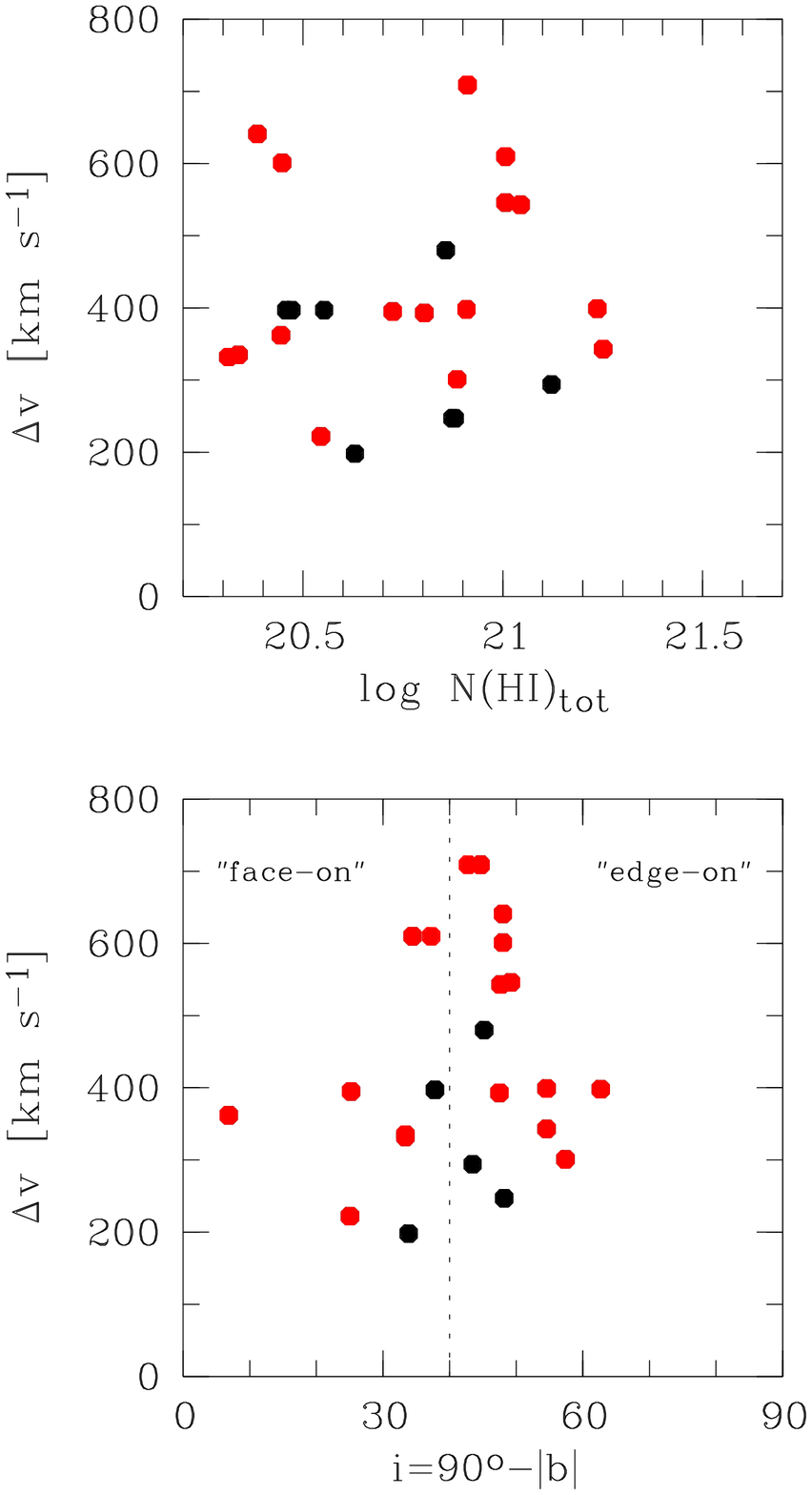}}
\caption[]{
{\it Upper panel:} total H\,{\sc i} column densities are plotted against the
velocity widths, $\Delta v$, for the 24 antipodal composite spectra.
No systematic trend is visible.
{\it Lower panel:} the velocity widths are plotted against the inclination $i$
of the Milky Way disk (assuming an external vantage point). The velocity widths
are systematically higher for $i>40\degree$ (``edge-on'' case) than for $i\leq 40\degree$ 
(``face-on'' case). In both panels, MS sightlines are indicated in red.
}
\end{center}
\end{figure}


\begin{figure}[t!]
\begin{center}
\resizebox{1.0\hsize}{!}{\includegraphics{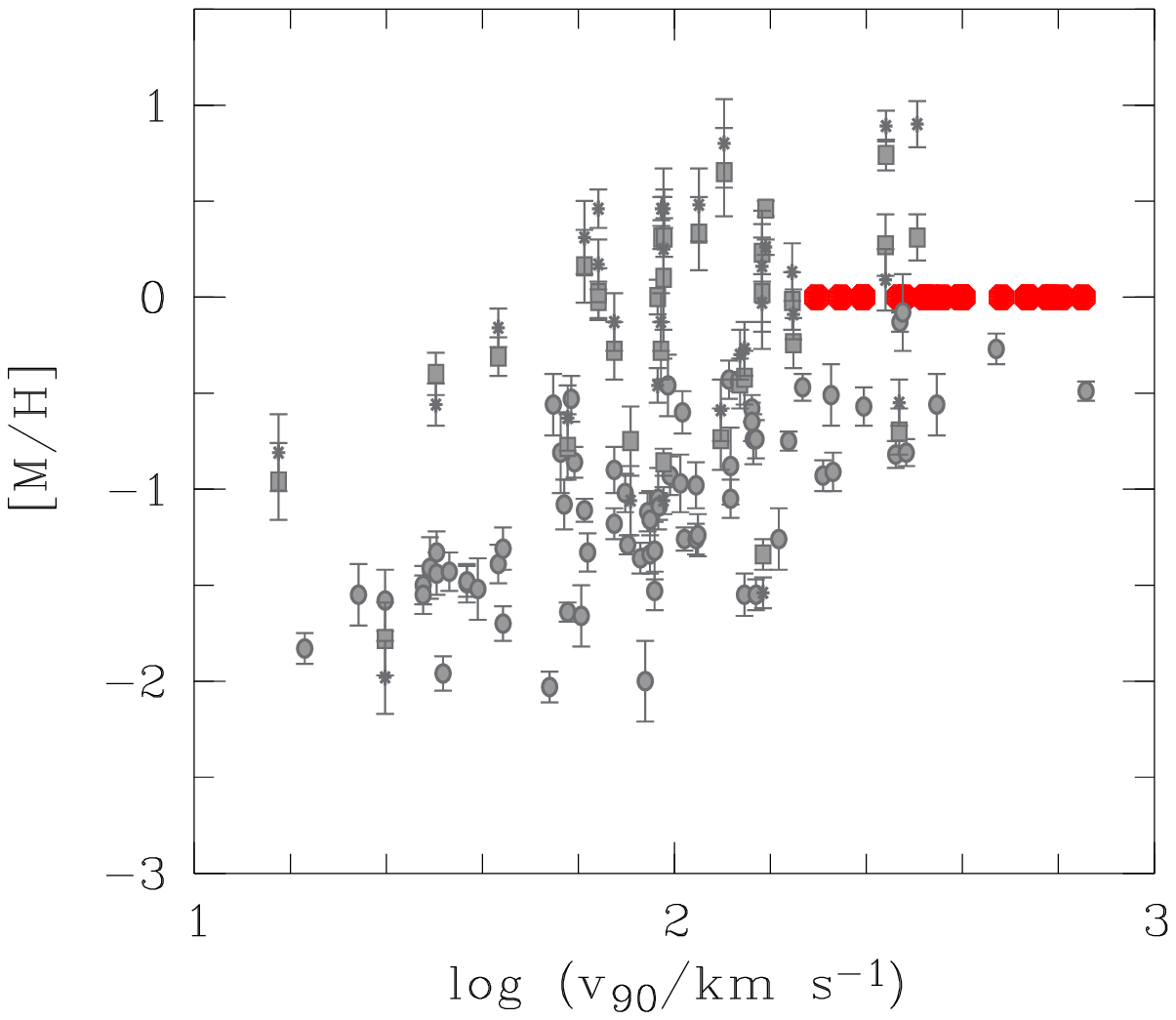}}
\caption[]{
The Milky Way's $\Delta v$ range (red filled circles) from the analysis (Fig.\,16)
of antipodal sightlines plotted on the velocity width/metallicity relation of
DLAs (Som et al.\,2015). The parameter $v_{90}$ denotes the velocity width in which 
90 percent of the integrated optical depth is contained
(Wolfe \& Prochaska 1998). We here set $v_{90}\equiv \Delta v$.
}
\end{center}
\end{figure}


\section{The Milky Way's CGM in a cosmological context}

\subsection{A search for antipodal sightlines}

With 270 more or less randomly distributed sightlines at hand, it is
promising to search for sightline pairs in our COS data sample
that represent antipodes on the celestial sphere.
A systematic study of antipodal sightlines and their spectra allows us to
produce composite spectra that reflect the absorption characteristics of
{\it complete} sightlines through the Milky Way disk and halo environment at
the position of the Sun from different external vantage points
(see Marggraf \& de\,Boer 2000 for an example towards LMC/M81).
Such antipodal composite absorption spectra then can be readily compared
with spectra of intervening absorption-line systems, such as 
Damped Lyman $\alpha$ absorbers (DLAs). They provide an idea of how the 
Milky Way and its gaseous environment would look like if it was seen as 
a QSO absorption-line system from far away (Herenz et al.\,2013).

We consider antipodal sightline pairs in cones with opening angles of
$\theta=6\degree$ in our sample and find 24 such pairs (48 QSO spectra);
they are listed in the Appendix in Table A.5. The basic concept of identifying 
antipodal QSO sightlines at $(l_{\rm A},b_{\rm A})$,$(l_{\rm B},b_{\rm B})$ 
is illustrated in Fig.\,A.4. To systematically study the velocity
width of disk and halo absorption along the antipodal sightline pairs we
combined the COS data for
individual antipodal pairs by multiplying the continuum-normalized spectra.
In this way, we created composite velocity profiles for 
for the strongest ion transition, Si\,{\sc iii} $\lambda 1206.50$, in the 
velocity range $v_{\rm LSR}=-500$ to $500$ km\,s$^{-1}$ (see Fig.\,16,
upper panel, for an example).
For each of the 24 sightline pairs we define the total velocity width
of the absorption in the composite spectrum as $\Delta v=v_{\rm max}-v_{\rm min}$.
In the lower panel of Fig.\,16 we show the velocity distribution of
Si\,{\sc iii} $\lambda 1206.50$ absorption along these 24 antipodal
composite profiles (with the contribution of the MS labeled in red). 
Twenty-three out of the 24 antipodal profiles show halo absorption at either high negative
or positive LSR velocities, and in 12 spectra absorption is seen at high positive {\it and}
negative velocities. The plot illustrates the diversity of the absorption characteristics
of the Milky Way disk and halo gas components, if seen from different vantage points.

This diversity is also seen in the upper panel of Fig.\,17, 
where we show the distribution of
$\Delta v$ in a histogram. $\Delta v$ ranges from 100 to 800 km\,s$^{-1}$, while
the distribution shows a strong peak at $\Delta v=300$ km\,s$^{-1}$. The 
antipodal sightline pairs with the largest values for $\Delta v$ are those
passing the MS (see Fig.\,16). This further highlights the importance
of the MS for the overall absorption characteristics of the Milky Way as
an absorption-line system.

\subsection{The Milky Way as a DLA}

We combined the kinematic information discussed in the previous section with information
on the total H\,{\sc i} column densities along the 24 antipodal sightlines 
to compare in detail the absorption characteristics of Milky Way disk and halo
gas with the absorption properties of low-redshift QSO absorbers.
In the lower panel of Fig.\,17 we show the distribution of total (logarithmic) H\,{\sc i} 
column densities along the 24 antipodal sightlines, as derived from integrating the combined 
21 cm profiles. The total column densities vary considerably in a range between 
log $N$(H\,{\sc i})$=20.31$ and $21.25$ with 71 percent of the sightlines having 
log $N$(H\,{\sc i}$)\geq 20.5$. This distribution suggests that, 
from an external vantage point, log $N$(H\,{\sc i}) (and thus 
the absorber type) would strongly depend on the viewing {\it angle}.
The Milky Way disk and halo gas would be seen either as low-column density DLA 
(log $N$(H\,{\sc i}$)<20.5$) or high-column density DLA (log $N$(H\,{\sc i}$)\geq 20.5$),
if the sightline would pass the disk at the position of the Sun
at an impact parameter of $b\approx 8$ kpc. 

The contribution of halo H\,{\sc i} to the total H\,{\sc i} column density is generally
small. For 71 percent the column-density fraction of halo H\,{\sc i} with 
$|v_{\rm LSR}|\geq 100$ km\,s$^{-1}$ with respect to the total neutral gas column along
the antipodal sightline, $f_{\rm HI,halo}$, is less than 10 per cent.
The median value of $f_{\rm HI,halo}$ is $0.055$.
This implies that the total H\,{\sc i} column density along the antipodal sightlines is
determined predominantly by the intrinsic variations in the neutral gas columns in
the Milky Way disk, if seen at different viewing angles. The total velocity spread
of the absorption, in contrast, is governed predominantly by the large-scale distribution 
of Milky Way halo gas components along the different sightlines, which should be 
independent of the H\,{\sc i} disk properties. In fact, we do not find any correlation 
between $\Delta v$ and log $N$(H\,{\sc i}$)_{\rm tot}$ (Fig.\,18, upper panel) in our sample.

Recent studies of strong metal absorbers and their host galaxies suggest that 
the velocity spread of the absorbing gas is related to the inclination $i$ of the
galaxy and its absorbing ISM and CGM components with respect to the observer
(Nielsen et al.\,2015). The general trend in the Nielsen et al.\, sample 
is that $\Delta v$ has a maximum
for blue (star-forming) "face-on" cases (with small values for $i$), possibly 
reflecting a kinematic imprint from outflows and winds along the projected minor axis.

To investigate the angular dependence of the absorption characteristics 
of Milky Way disk and halo gas along the different antipodal sightlines
we transformed the absolute latitude $|b|$ into an inclination $i$ by using the
relation $i=90\degree -|b|$. In this way, sightlines passing through the Galactic poles 
are regarded as "face-on" sightlines, while those with small values of $|b|$ are 
considered as "edge-on" cases (see Nielsen et al.\,2015, their Fig.\,1).
In Fig.\,18, lower panel, we plot the velocity spread $\Delta v$ 
(from Si\,{\sc iii} $\lambda 1206.50$) against $i$ for the 24
antipodal sightlines.

The velocity spread has its maximum in the range 
$i=40\degree-50\degree$, extending to very large values above $\Delta v\sim 700$ km\,s$^{-1}$,
while for the "edge-on" case ($i>50\degree$) and the "face-on" case ($i<40\degree$)
the values for $\Delta v$ are $<420$ km\,s$^{-1}$.
The observed velocity spreads in the Milky Way thus are substantially 
(by a factor of a few) larger than those presented in Nielsen et al.\,(2015), who
report values for $\Delta v$ of $<300$ km\,s$^{-1}$. 
It needs to be kept in mind that the DLAs in the Nielsen et 
al. sample spans a large range of impact parameters (with $b$ typically
larger than 8 kpc). They also trace different types of DLA host galaxies
that, in general, do not mimic the chemical and kinematic properties of 
nearby disk gas in the solar neighborhood.
The primary origin for the discrepancy between the Milky Way $\Delta v$ range
and that of DLAs is the MS, however. The MS alone spans a LSR velocity 
range of more than 800 km\,s$^{-1}$ along the 24 antipodal sightlines (see Fig.\,16).

In general, CGM features 
that arise from the interaction and merging of (satellite) galaxies presumably dominate
the velocity spread of gas around Milky-Way type galaxies, as the velocity dispersion 
of satellite galaxies is expected to be higher than the gas infall- and outflow velocities. 
As a result, the kinematic signatures of outflows, winds, and accretion most likely are 
obscured by those of galaxy mergers. The cross section of merger-related
gas streams around the local galaxy population is small, however 
(Gauthier et al.\,2010; Martin et al.\,2012), so that the observed large velocity dispersion 
of the Milky Way CGM is an exception, but not the rule. Possibly existing outflowing gas structures
above the Milky Way disk, such as those belonging to the Fermi Bubbles (Fox et al.\,2015),
thus do not show up as separate features in the $i$/$\Delta v$ diagram.

Another regularly used diagnostic tool to study strong absorption-line systems is the velocity 
width/metallicity relation for DLAs.
In Fig.\,19 we plot the $\Delta v$ range from the
antipodal sightline analysis on the velocity width/metallicity relation 
presented by Som et al.\,(2015). For the Milky Way gas, we assume solar
metallicities (i.e., [M/H$]=0$), which is justified because the mean metallicity
should be a column-density weighted mean value, so that for the Milky Way
(as DLA) it should be close to the value representative for the solar
neighborhood. For the Milky Way antipodal sightlines the
measured values for $\Delta v$ range from $100$ to $800$ km\,s$^{-1}$
(see Fig.\,17), thus being broadly consistent with
values of $\Delta v$ measured for high-metallicity DLAs.
The majority of the MW data points lie, however, at the high end of the 
$\Delta v$ distribution for [M/H$]=0$ absorbers, which we again
attribute to the fact that large values for $\Delta v$ in the MW are 
caused by gas from the MS, i.e., from merger processes, which are 
believed to be rare in DLA samples.
If such a large scatter in $\Delta v$ was typical for low-redshift 
$L\star$ galaxies, the {\it individual} data points 
on DLA scaling relations must be interpreted with great
caution, as the observed parameters (H\,{\sc i} column density, velocity spread)
may not reflect at all {\it representative} properties of the absorber host galaxy.


\begin{figure}[t!]
\begin{center}
\resizebox{0.95\hsize}{!}{\includegraphics{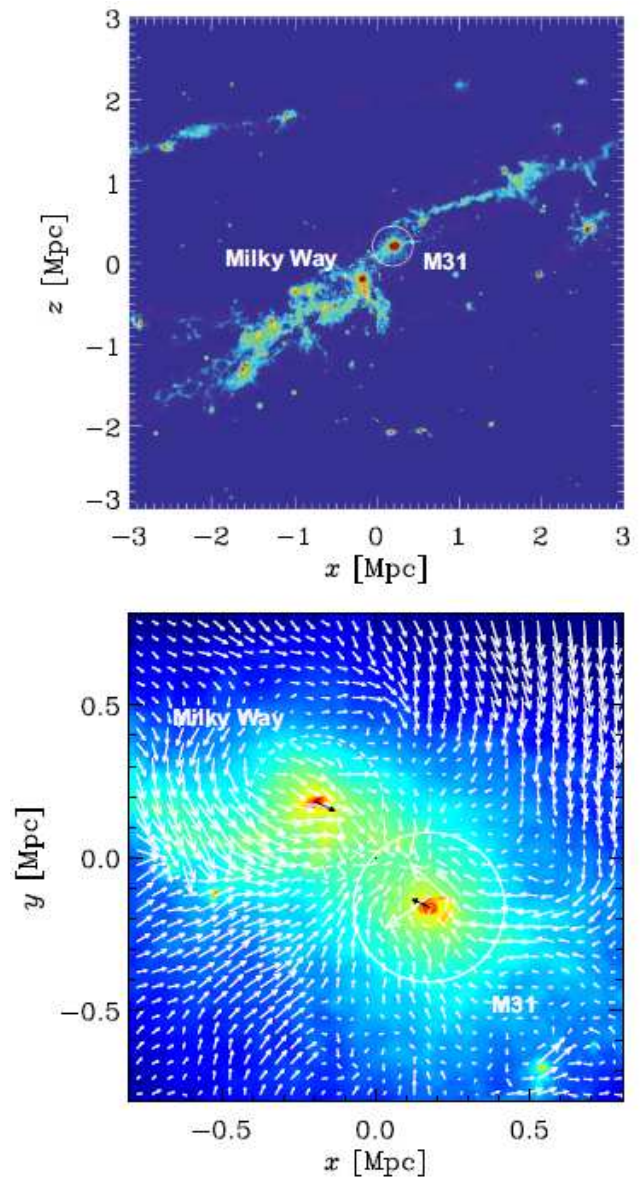}}
\caption[]{
{\it Upper panel:}
spatial distribution of neutral hydrogen gas at $T<10^5$ K gas in
the large-scale environment of the Local Group, as predicted from
a constrained cosmological simulation from the CLUES project
(N14). The color scale reflects
the total hydrogen gas density, where the green filamentary structures outside
the virial radii of the LG member galaxies lie in the density range
$n_{\rm H}=10^{-5}-10^{-3}$ cm$^{-3}$.
Positions and virial radii of
Milky Way and M31 are indicated with white dashed (Milky Way)
and solid (M31) circles.
{\it Lower panel:}
zoom-in of the gas distribution in and around the simulated
MW/M31 galaxies and the bulk motions of that gas with respect
to the Local Group barycenter in the $x/y$ plane.
The white arrows show the gas velocity field, with the largest ones
representing a velocity of $130$ km\,s$^{-1}$.
The black arrows indicate the velocities of the MW and M31 galaxies.
Their absolute
space velocities are $67$ and $76$ km\,s$^{-1}$, respectively. 
The bright colors again indicate regions
of enhanced hydrogen density.
}
\end{center}
\end{figure}


\subsection{Comparison with Local Group simulations}

Cosmological hydrodynamical simulations predict that group 
environments within the overall cosmic web are filled with large amounts of
diffuse gas whose mass exceeds that of gas inside the galaxies (e.g., Raouf, Khosroshabi \& Dariush 2016). 
In their recent study, N14 have explored the large-scale 
distribution and overall physical properties of gas in the Local Group based on 
simulation data from the Constrained Local UniversE Simulations (CLUES) project 
(www.clues-project.org), which aims at simulating a realistic LG environment including local
features and the most prominent surrounding cosmological structures.
To evaluate the large-scale properties of LG gas and its possible signatures
in FUV absorption-line data we here reanalyze the simulations of N14.

We focus particularly on a side-aspect of the N14 study, namely 
the spatial distribution of LG gas {\it outside}
the virial radius of the Milky Way. We also want to explore the kinematics of the LG gas 
with respect to the space motion of the MW disk and its rotation
and compare it with the
velocity distribution of the HVCs seen in our COS data.

The initial matter distribution in the CLUES simulations 
consists of a cubic box of 64 $h^{-1}$ Mpc side length with a high-resolution spherical region 
of 2 $h^{-1}$ Mpc radius located at its center. The masses of the gas and dark matter particles
inside the high-resolution region are $M_{\rm gas}=3.89\times 10^5\,h^{-1}\,M_{\sun}$ and
$M_{\rm DM}=1.97\times 10^6\,h^{-1}\,M_{\sun}$, respectively.
Further details on the setup and realization of the CLUES simulations as well as a detailed
description of the overall results from that simulation can be found in
N14 and references therein. Note that the following results are specific
to the CLUES simulations and their cosmological representation of the LG. 
It would need to be tested whether they could be reproduced using different simulations 
with different initial conditions and physical recipes.

In the upper panel of Fig.\,20 we show the spatial distribution of neutral
hydrogen (H\,{\sc i} at $T<10^5$ K) in
the large-scale environment of the LG at $z=0$. 
The gas distribution is relatively patchy and forms a bar-like structure that 
follows closely the elongated shape of the cosmological filament that hosts the LG. 
Milky Way and M31 (white circles denoting their virial radii) reside in the center of this structure.
Obviously, there are substantial amounts of diffuse warm gas outside the virial radii 
of MW and M31 that potentially could give rise to UV absorption if seen from
the MW disk.

To display the large-scale kinematics of the gas around MW and M31 we
show in the lower panel of Fig.\,20 the velocity vectors (with respect to the LG barycenter)
of the simulated gas particles (white arrows) and the two galaxies (black arrows) 
in the $x/y$ plane.
The galaxies move towards the LG barycenter while the ambient gas is circulating 
around MW and M31 in a complex pattern of infall- and outflow channels.
Because of the complex gas motions, the velocity components in the direction of the
LG barycenter are smaller for the gas particles than for the two galaxies.
As a result, there is a systematic net-motion of both galaxies towards the LG barycenter 
with respect to the ambient LG gas: MW and M31 are ramming into
LG gas that lies in-between them near the LG barycenter, while they are moving 
away from gas that lies behind them in the opposite direction in the outer regions
of the LG filament.


\begin{figure*}[ht!]
\begin{center}
\resizebox{0.75\hsize}{!}{\includegraphics{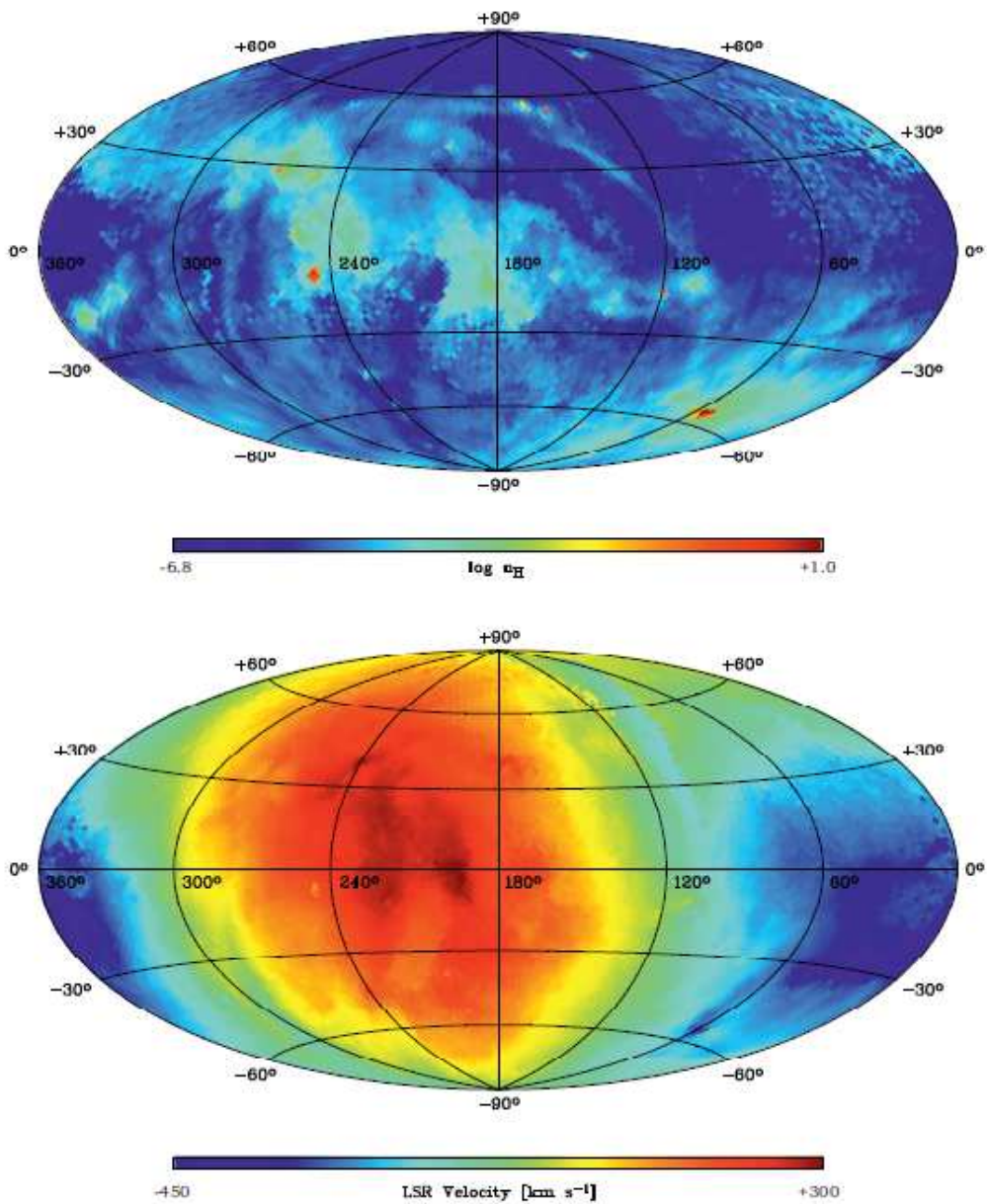}}
\caption[]{
{\it Upper panel:}
Hammer-Aitoff projection of the gas density in the Local Group
(for gas beyond the virial radius of the Milky Way), as seen
in the CLUES simulation from the position of the Sun.
M31 appears as a red/yellow spot near $(l,b)=(60\degree,-55\degree)$.
The gaseous LG filament is indicated with the greenish colour, which displays gas at
densities in the range log $n_{\rm H}=-4$ to $-3$. The filament reaches from the
general M31 direction roughly to its antipole on the sky.
Two smaller dwarf galaxies are visible as red spots.
{\it Lower panel:}
velocity distribution of the gas in the same projection, given
in LSR. The observed dipole pattern separates gas with high positive velocities
at $l=120\degree -320\degree$ from gas at high negative velocities
($l<120\degree, l>320\degree$).
}
\end{center}
\end{figure*}


To predict the velocities of UV absorption based on this simulation we 
take into account the rotational motion of the MW disk. 
We assume that the simulated MW disk has a circular orbit speed at the position of the Sun of
$v_{\rm rot}=255$ km\,s$^{-1}$, as recently derived by Reid et al.\,(2014). 
Taking all this spatial and kinematic information into account we prepared 
maps for the sky distribution of the LG gas from the simulation in a similar way
as we prepared the maps for the observational data. 
In the upper panel of Fig.\,21 we show the density map of LG gas {\it outside} the MW
virial radius. The bar-like gaseous LG filament (see Fig.\,20, upper panel) reaches 
from the general M31 direction (in the simulation: $(l,b)=(60\degree,-55\degree)$ 
roughly to its antipole on the sky (see also N14).
In the lower panel, we plot the expected radial LSR velocity of the LG
gas as a function of $(l,b)$ including the effect of Galactic rotation. A 
dipole pattern is clearly visible. Only at low latitudes ($|b|<30\degree$)
is Galactic rotation relevant for $|v_{\rm LSR}|\geq 200$ km\,s$^{-1}$.
The observed velocity dipole for gas at higher (absolute) latitudes
and velocities instead is a result of the relative
motion of the Milky Way with respect to the surrounding gas in the
direction of the LG barycenter.
From the combination of the velocity and density 
information in the two panels in Fig.\,21 
follows that LG gas at $|b|>30\degree$, if it were cold and dense enough
to be observed in UV absorption, would imprint a dipole absorption pattern on the
sky, with redshifted high-velocity absorption ($v_{\rm LSR}\geq 200$ km\,s$^{-1}$) 
at $l>200\degree, b>0\degree$ and blue-shifted 
high-velocity absorption ($v_{\rm LSR}\leq -200$ km\,s$^{-1}$)
at $l<120\degree, b<0\degree$. 

As discussed earlier, such a dipole
is indeed seen in our COS survey (NIR, SIR; Sect.\,5.4; Figs.\,2,8,11).
It also has been recognized in our previous 
O\,{\sc vi} survey (Sembach et al.\,2003; Wakker et al\,2003) and
other studies of intermediate and high ions (e.g., Collins, Shull \& Giroux 2005).
A definite answer, whether the observed dipole-absorption features 
are indeed related to LG gas yet cannot be given here, 
because the MS similarly imprints an emission/absorption 
dipole pattern on the sky, as evident from Fig.\,8. 
Since a reliable distance determination of this gas appears to be out of reach,
a {\it precise} measurement of its chemical composition (in comparison with abundances
found in the MS) might shed light on its origin.

In summary, the CLUES simulations support (but do not require) the idea that 
some of the observed absorbers in the NIR and SIR represent ionized LG gas 
beyond the virial radius of the Milky Way that follows the large-scale flow
of matter in the LG.


\section{Conclusions}

In this paper, we present the largest available survey of UV absorption 
in Galactic HVCs in the velocity range $|v_{\rm LSR}|= 100-500$ km\,s$^{-1}$, 
tracing the extended gaseous halo and CGM of the Milky Way and gas in the Local Group.
Our survey is based on the analysis of various lines from the metal ions 
Si\,{\sc ii}, Si\,{\sc iii},  C\,{\sc ii}, and C\,{\sc iv}
in 270 spectra of extragalactic UV background sources observed 
with HST/COS. 
We supplement our COS spectra with H\,{\sc i}
21 cm data from GASS and EBHIS, as well as with data from
constrained cosmological simulations from the CLUES project. We update previous results
from smaller HVC surveys (e.g., Lehner et al.\,2012; Herenz et al.\,2013) concerning 
the HVC's sky distribution, their internal structure, their ionization properties, 
and estimate the total HVC mass and accretion rate. Based on our excellent database,
we further present several new results on HVC aspects that have not yet been studied
before. We analyze the angular two-point correlation function of HVC absorption and
analyze small-scale variations in the UV absorption patterns.
We further investigate the relation 
between high-velocity UV absorption and gas in other LG galaxies and the LG's
intragroup medium. We systematically compare the UV absorption 
characteristics of Milky Way disk and halo gas along antipodal sightlines with 
those of low-redshift DLAs. Finally, we discuss the role of HVCs in a cosmological 
context by comparing the observed gas properties with those seen in 
constrained simulations of the LG.

We summarize the main results of our study as follows:\\
\\
(1) {\it Sky-covering fraction of HVCs.}
The overall all-sky covering fraction of high-velocity absorption
in our survey is $f_{\rm c}=77\pm 6$ percent, as derived from the analysis of the most 
sensitive ion, Si\,{\sc iii} for $N_{\rm lim}($Si\,{\sc iii}$)=12.1$. For the other three ions 
we obtain covering fractions of 
$f_{\rm c}($Si\,{\sc ii}$)=69\pm 6$ percent for $N_{\rm lim}($Si\,{\sc ii}$)=12.3$,
$f_{\rm c}($C\,{\sc ii}$)=70\pm6$ percent for $N_{\rm lim}($C\,{\sc ii}$)=13.2$, and
$f_{\rm c}($C\,{\sc iv}$)=58\pm7$ percent $N_{\rm lim}($C\,{\sc iv}$)=12.9$.
The Si\,{\sc iii} detection rate is $\sim 4-5$
times higher than that for H\,{\sc i} for
log $N$(H\,{\sc i}$)\geq 18.7$ along the same sightlines, 
as estimated from the
complementary 21 cm data from GASS and EBHIS. Most of the detected high-velocity
absorbers can be kinematically attributed to known H\,{\sc i} 21 cm gas complexes,
such as the MS, Complex C, Complex A, and others. This implies that
the large cross section of high-velocity UV absorption is mostly due to the 
the ionized envelopes of the neutral HVC cores.
The overall HVC absorption fraction is systematically lower in the northern sky than
in the southern hemisphere 
(e.g., $f_c($Si\,{\sc iii}$)=73\pm 7$ in the north vs. 
$f_c($Si\,{\sc iii}$)=89\pm 13$ in the south). We attribute this asymmetry to the widespread presence 
of diffuse (predominantly ionized) gas from the MS at negative radial velocities 
that covers a significant 
fraction of the sky at $b<0\degree$. We identify two
antipodal regions at $l>240\degree, b > 60\degree$ in the northern
sky (northern ionized region, NIR) and at $l<120\degree ,b=-60 \degree -0\degree$ 
in the south (southern ionized region, SIR) that show prominent UV absorption at very 
high (absolute) radial velocities ($|v_{\rm LSR}|\geq 200$ km\,s$^{-1}$), but with 
only very little H\,{\sc i} 21 cm emission.
\\
\\
(2) {\it The multi-phase nature of HVCs.}
The simultaneous detection of Si\,{\sc ii}, Si\,{\sc iii}, C\,{\sc ii},
C\,{\sc iv}, and H\,{\sc i} in many high-velocity absorption components implies the
presence of multiple gas phases (from neutral to highly ionized gas) in the CGM 
along each sightline.
The substantially higher covering fraction of high-velocity UV absorption
compared to high-velocity H\,{\sc i} 21 cm emission demonstrates that hydrogen is
predominantly ionized in these structures. Many regions that exhibit high-velocity UV absorption
without corresponding 21 cm emission represent the ionized outer layers of nearby
neutral HVC complexes, as indicated by the same large-scale kinematics. From the
observed spatial and kinematic variations of the column-density ratios
Si\,{\sc ii}/Si\,{\sc iii}, C\,{\sc ii}/Si\,{\sc iii}, and
C\,{\sc iv}/Si\,{\sc iii} follows that there is substantially more ionized
gas at high negative radial velocities than at high positive velocities. Again, this trend is
related to diffuse gas associated with the MS. Particularly high
C\,{\sc iv}/Si\,{\sc iii} are found in the NIR and SIR, indicating that
these regions contain mostly diffuse, ionized gas.
\\
\\
(3) {\it Spatial clustering of HVCs and their internal structure.}
We analyze the angular distribution of high-velocity absorption 
by constructing the two-point correlation function, $\xi$,
for HVC components in our QSO sample. We find that $\xi(\theta)$
exhibits a strong clustering signal for angular separations of 
$\theta \leq 30 \degree$, while $\xi$ drops quickly for larger
angles. From the analysis of 12 sightline pairs with small angular separations
(down to $\theta \sim 0.5 \degree$) we find significant variations
in the equivalent widths of high-velocity features, while
the overall velocity structure in the HVC absorption patterns
remains mostly unchanged at the spectral resolution of COS.
For typical HVC distances in the range $d=5-55$ kpc these findings
indicate that high-velocity absorbers in the Milky Way halo 
represent coherent gas structures on scales of a few up to
a few hundred kpc that follow a large-scale kinematics and that 
exhibit internal variations in gas density/physical conditions 
on pc scales.
\\
\\ 
(4) {\it Relation between HVCs and other LG galaxies.}
We systematically investigate sightlines that pass the outskirts of
several Local Group galaxies outside the virial radius of the 
MW to search for absorption signatures
that might be related to the CGM of these galaxies.
Along seven sightlines, high-velocity absorption from extended
halo gas around M31 most likely is present in the
LSR velocity range between $-300$ and $-150$ km\,s$^{-1}$, 
in line with the findings recently presented by Lehner et al.\,(2015).
No CGM absorption is found at impact parameters $\rho\leq 2R_{\rm vir}$
and velocities close to the systemic galaxy velocities for
NGC\,3109, Antila, Sextans\,A, Aquarius (one sightline per galaxy, respectively),
and Leo\,I (12 sightlines). Absorption is found in a sightline passing
IC\,1613 at $\rho\sim 1.5 R_{\rm vir}$ near the systemic velocity of
IC\,1613, but this absorption most likely is associated with gas
from the MS.
Therefore, only for M31 and the Milky Way the COS data provides
compelling evidence for the presence of an extended CGM,
while for the LG dwarfs/satellites there are no hints for a 
circumgalactic gas component.
\\
\\
(5) {\it Local Group gas.}
We further investigate the distinct gas properties and kinematics of the 
NIR and SIR (see definition above) by relating 
the observed HVC absorption characteristics 
in these regions with the kinematics of LG galaxies and with predictions from
hydrodynamical simulations of LG gas. The NIR and SIR, for which no distance
information are available, form a dipole on the 
sky with opposite signs for the observed absorption velocities. Because of 
the observed high radial velocities and the high absolute Galactic latitudes of 
these antipodal regions, the observed dipole pattern cannot be produced by the 
rotation of the Galactic disk. We show that the galaxies outside the virial radius 
of the Milky Way form a similar dipole-velocity pattern on the sky, which reflects 
the general flow of galaxies in the LG towards the LG barycenter, the Milky
Way being part of that flow. Using constrained cosmological
simulations of the LG and its gaseous environment from the CLUES project
(N14) we further demonstrate that LG gas, that circulates around 
LG member galaxies within the overall matter flow, is indeed expected to form a 
dipole-velocity pattern that matches well the observed one for the NIR/SIR. 
This, together with the well-established presence of high-velocity 
O\,{\sc vi} in these directions at similar high radial velocities 
(Sembach et al.\,2003; Wakker et al.\,2003), leads us to speculate that
the NIR and the SIR traces LG gas that follows the large-scale kinematics
of matter in the cosmological filament that forms the LG.
\\
\\
(6) {\it Total mass and infall rate of HVCs.}
From the observed ion covering fractions, previously measured HVC distances
and metallicities as well as the measured total Si and C column 
densities in HVCs we estimate a total gas mass of the Milky Way's HVCs of
$M_{\rm HVC}\geq 3.0\times 10^9\,M_{\sun}$. This is $\geq 43$ percent
of the ISM gas mass in the Galactic disk. 
We find that the total gas accretion rate in the form of
neutral and ionized HVCs is $\geq 6.1\,M_{\sun}$\,yr$^{-1}$. 
The MS, here assumed
to be located at a distance of $d=55$ kpc, dominates
the total HVC mass and gas-accretion rate by far, while all other 
HVCs contribute with only $\leq 10$ percent.
If much of the MS would be located at distances $d=100-150$ kpc
the gas accretion rate would be even higher by a factor of $\sim 2-3$.
The infall of HVCs from the halo to the disk thus represents 
a significant mass inflow that may substantially
influence the future star-formation rate in the Milky Way.
\\
\\
(7) {\it The Milky Way as a DLA.}
From our data set we identify 24 sightline-pairs with antipodal directions
within $6\degree$. We create composite spectra for these antipodal sightlines
to study the integrated absorption characteristics of the Milky Way disk and halo gas
from different vantage points and compare the results with recent studies
of QSO absorption-line systems.
The total column densities along the 24 antipodal sightlines show a substantial scatter,
ranging from log $N$(H\,{\sc i})$=20.31$ to $21.25$, with 71 percent of the sightlines having
log $N$(H\,{\sc i}$)\geq 20.5$. Thus, the Milky Way disk and halo gas (at the position
of the Sun) would be seen as a DLA from an external vantage point, with a total
gas column that depends (non-systematically) on the viewing angle.
We find that also the total velocity width, $\Delta v$, of Si\,{\sc iii} absorption in 
the various disk and halo absorption
components shows a huge spread along the 24 antipodal directions. 
$\Delta v$ varies from $100$ to $800$ km\,s$^{-1}$ and the $\Delta v$
distribution shows a peak at $300$ km\,s$^{-1}$.
With such large velocity spreads, the Milky Way disk/halo gas
lies at the extreme end of the $\Delta v$ distribution of low-$z$
DLAs, but still fits to the general velocity-width vs.\,metallicity relation for
DLAs and sub-DLAs (Som et al.\,2015).
The dominant contribution to the large velocity spread of Milky Way gas comes from
the MS, again indicating the dominant role of the Stream for the Milky Way's
CGM absorption properties.
Since the cosmological cross section of merger-related gas streams
around galaxies is small, the MW thus would represent an {\it a-typical} DLA if
seen from far away. 
\\
\\


\begin{acknowledgements}
This research used the facilities of the Canadian Astronomy Data Centre
operated by the National Research Council of Canada with the support of
the Canadian Space Agency. 
Based on observations obtained with the NASA/ESA
Hubble Space Telescope, which is operated by the Space
Telescope Science Institute (STScI) for the Association of
Universities for Research in Astronomy, Inc., under NASA
contract NAS5D26555.
S.\,E.\,N. acknowledges support from the
Deutsche Forschungsgemeinschaft (DFG) under the grant NU\,332/2$-$1.
Support was provided by NASA through grants HST-GO-12604.01-A (Wakker),
HST-GO-13448.01-A (Wakker), HST-GO-13840.05-A (Wakker),
HST-GO-12982 (Lehner), HST-AR-12854 (Lehner), and HST-AR-12846.001 
(Charlton) from the Space Telescope Science Institute, which is operated by 
the Association of Universities for Research in Astronomy, Incorporated, 
under NASA contract NAS5-26555. We thank an anonymous referee for
helpful comments and suggestions.
\end{acknowledgements}



\begin{appendix}

\section{Supplementrary discussions, tables and figures}

\subsection{Absorption fraction without MS contribution}

To investigate the role of the MS in the velocity-dependent absorption 
fraction (Fig.\,5) we show in Fig.\,A.1 the absorption fraction of 
Si\,{\sc iii} $\lambda 1206.5$ (red),
Si\,{\sc ii} $\lambda 1260.4$ (blue),
Si\,{\sc ii} $\lambda 1193.3$ (green), and
C\,{\sc iv} $\lambda 1548.2$ (orange)
as a function of LSR velocity excluding the regions
covered by the MS (see Table 3). By doing this, the asymmetry between 
negative and positive velocities (Fig.\,5) is reduced, with 
now almost similar absorption fractions in Si\,{\sc iii} on
both sides of the distribution. This trend demonstrates that the MS and
its environment contribute substantially to the absorption fraction of
the CGM at high negative velocites. 


\begin{figure}[h!]
\begin{center}
\resizebox{1.0\hsize}{!}{\includegraphics{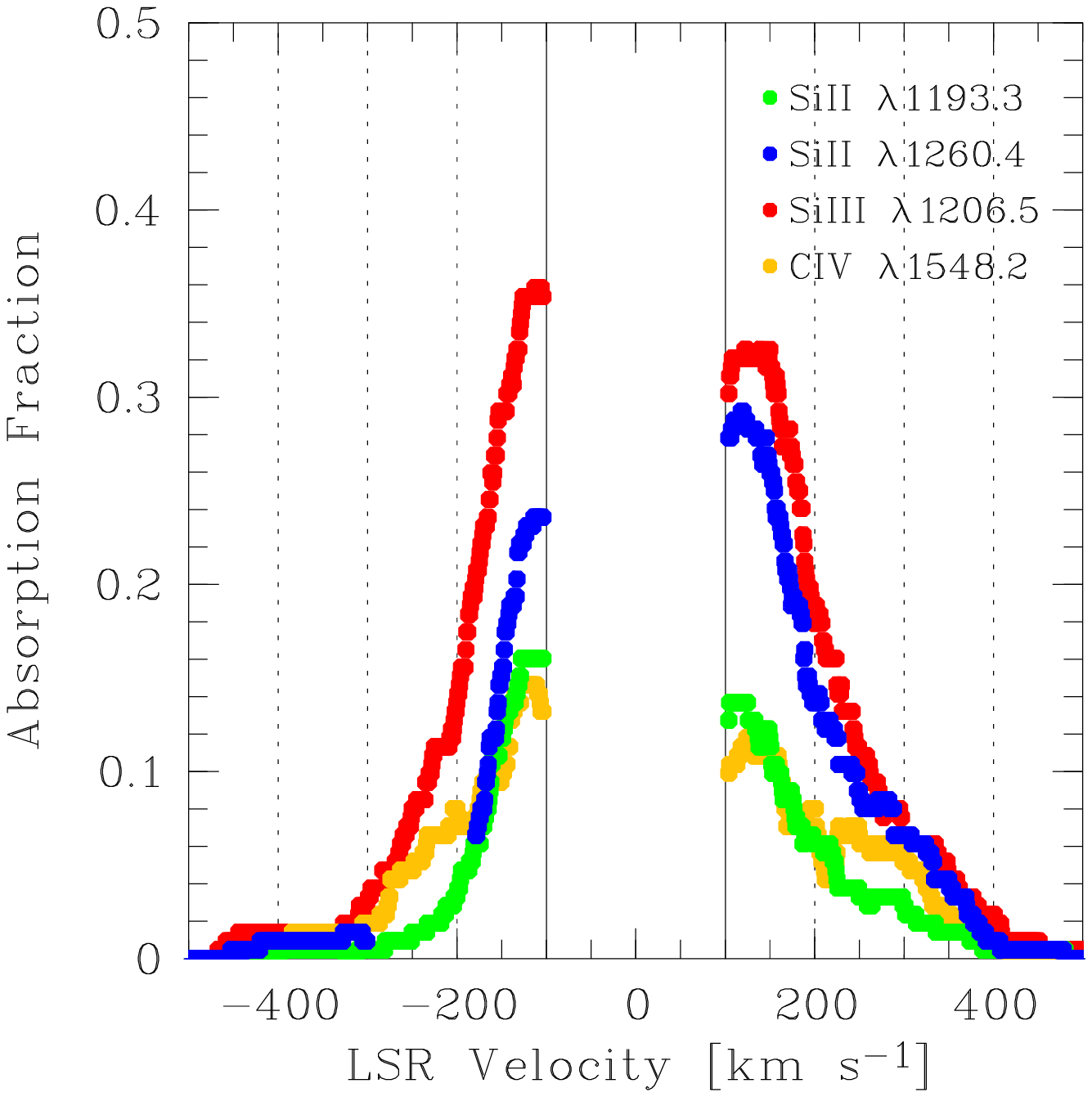}}
\caption[]{
Same as Fig.\,5, but without considering the area covered by the MS (see Table 3
for adopted (l,b) ranges).}
\end{center}
\end{figure}


\subsection{Equivalent-width/column-density ratios vs. LSR velocity}


\begin{figure*}[h!]
\begin{center}
\resizebox{0.70\hsize}{!}{\includegraphics{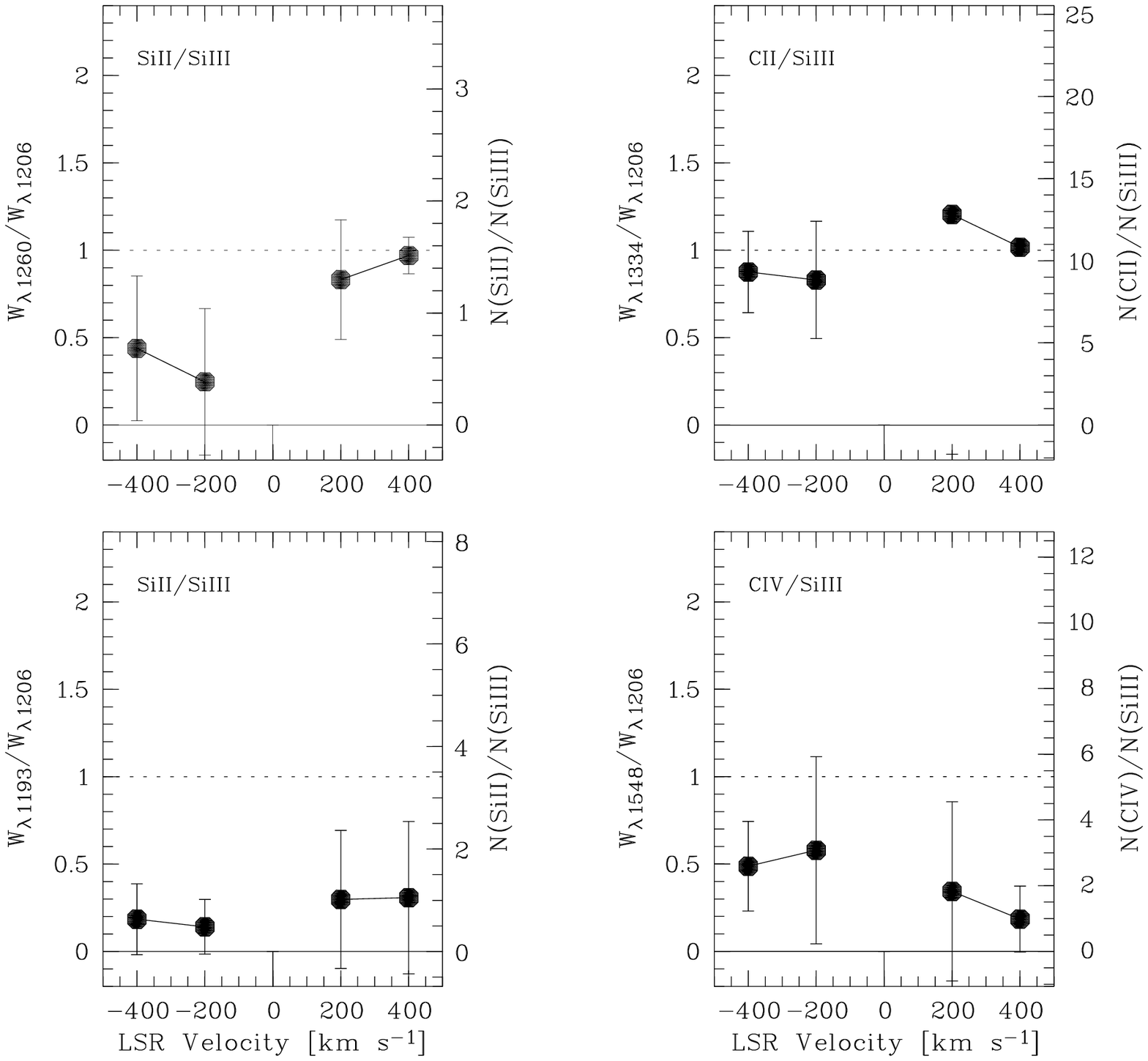}}
\caption[]{
Equivalent-width ratios and column density ratios
for different transitions of low, intermediate
and high ions as a function of LSR velocity. The data are
averaged in 200 km\,s$^{-1}$ wide bins. The error bars
indicate the standard deviation from the mean value.
}
\end{center}
\end{figure*}


In Fig.\,A.2 we show the distribution of the same
equivalent-width/column-density ratios as
presented in Sect.\,4.2, but now as a function of LSR velocity.
For this plot we have averaged the equivalent-width ratios over the entire
sky in velocity bins of $200$ km\,s$^{-1}$. What can be seen here is
that the ratio Si\,{\sc ii} $\lambda 1260.4$/Si\,{\sc iii} $\lambda 1206.5$
is small at negative velocities, in particular for $v_{\rm LSR}<200$ km\,s$^{-1}$.
In the same velocity bin the C\,{\sc iv} $\lambda 1548.2$/Si\,{\sc iii} $\lambda 1206.5$
ratio has its maximum, indicating that gas at high negative radial velocities
has, on average, a substantially higher degree of ionization than gas at any other
(high) radial velocity. Since most of extreme negative velocity absorbers are
found in the region, $b<0\degree$, $l<180\degree$ (see Sect.\,4.2), the trends
seen in Fig.\,A.2 are not surprising: they further demonstrate the distinct properties
of negative-velocity gas in this direction.
In addition, it can be concluded from Fig.\,A.2 that the average degree of ionization,
as indicated by the above listed equivalent-width/column-density ratios, is higher for CGM gas that
moves towards the Sun than for gas that moves away from it.

\subsection{Small-scale structure in HVCs}

As can be seen in Fig.\,2, there are two regions in the sky where the projected 
density of background sources is particularly high: $l=80\degree -110\degree$,
$b=30\degree -60\degree$ (tracing
Complex C at $d\approx10$ kpc) and $l=225\degree -255\degree$,
$b=30\degree -60\degree$ (tracing Complex WA at $d=8-20$ kpc;
Table 3).
For these regions/complexes, we constructed equivalent-width maps for the
Si\,{\sc ii} $\lambda 1260.42$ line, as shown in Fig.\,A.3.
On small angular scales ($\theta=4\degree$) the equivalent widths vary by
factors $2-10$, indicating spatial variations in the column densities
of Si\,{\sc ii} on linear scales of $L\sim 0.7$ kpc (Complex C) and
$L<1.4$ kpc (Complex WA).


\begin{figure*}[h!]
\begin{center}
\resizebox{0.80\hsize}{!}{\includegraphics{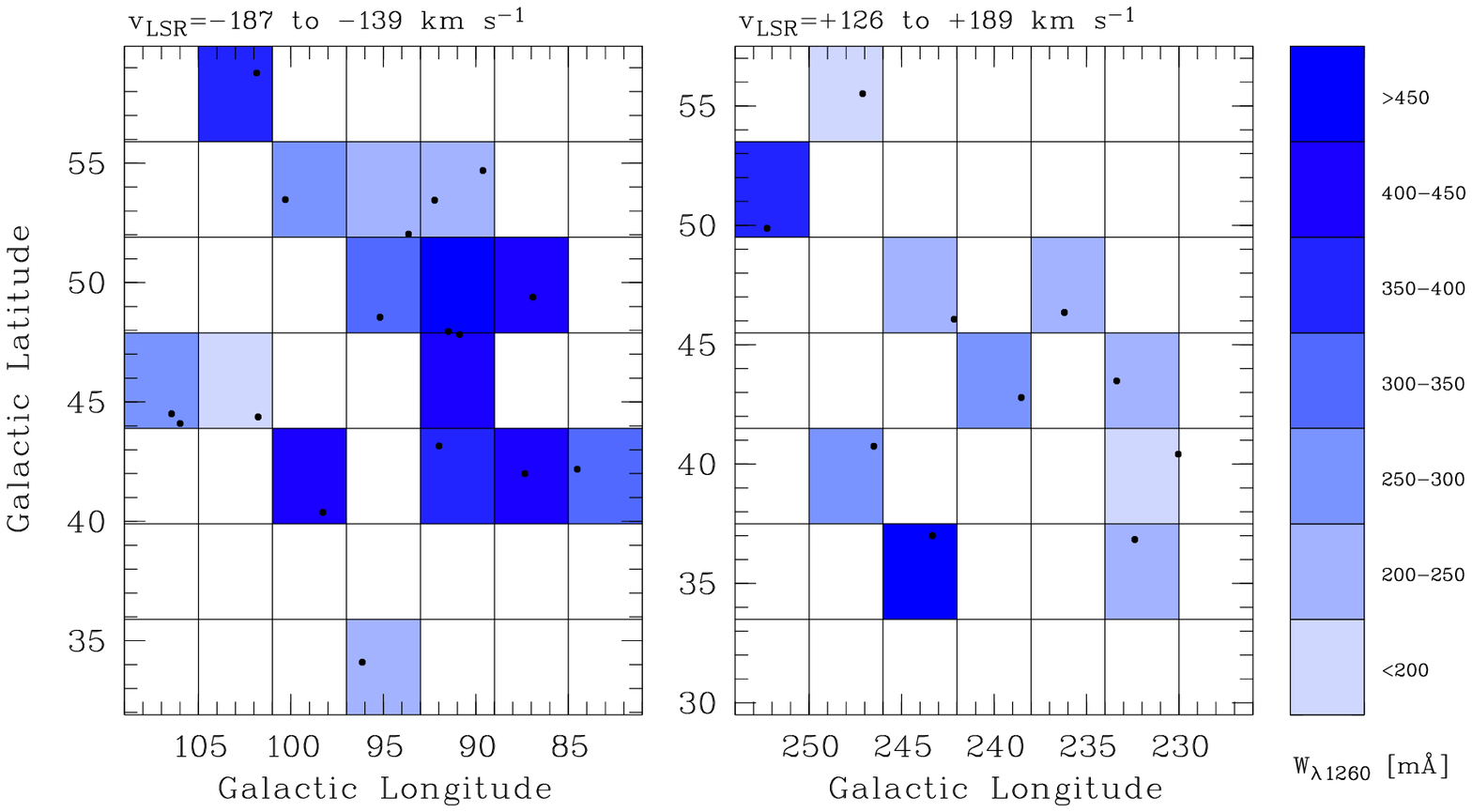}}
\caption[]{
Equivalent-width maps for high-velocity Si\,{\sc ii} $\lambda 1260.4$
absorption for two selected HVC complexes. {\it Left panel:} Complex C,
{\it right panel:} Complex WA. The black dots indicate the positions
of the background QSOs.
}
\end{center}
\end{figure*}


Figs.\,13, and A.3 indicate that the equivalent widths of metal ions can
vary substantially within HVCs on sub-kpc scales.
The largest differences are seen in the
singly-ionized species Si\,{\sc ii} and C\,{\sc ii}, while they are typically
less pronounced in Si\,{\sc iii} and C\,{\sc iv}.
As can be concluded from these figures, the overall velocity
structure in the HVC absorption patterns
remains mostly unchanged along the different sightlines at the (limited)
velocity resolution of the COS instrument ($\sim 20$ km\,s$^{-1}$).
So far, the information on small-scale structure in HVC gas from
absorption-line studies is limited to only a few cloud complexes
(e.g., the HVC towards the LMC; Danforth et al.\,2003; Bluhm et al.\,2001; 
Lehner et al.\,2009; Smoker et al.\,2014).
In contrast, the presence of pc-scale substructure in HVCs in H\,{\sc i} is well
established from high-resolution 21 cm observations (e.g., Wakker et al.\,2002;
Sembach et al.\,2004). The observed variation in the absorption/emission line
strengths imply spatial fluctuations in the local gas density and/or changes in
physical conditions (temperature, chemical composition, radiation
field, etc.). From the few detections of molecular gas in HVCs
(Richter et al.\,1999; Sembach et al.\,2001; Richter et al.\,2001, 2013;
Wakker et al.\,2006; Murray et al.\,2015) follows that substantial 
small-scale structure in the gas densities exists down to sub-pc scales. 
The formation and maintenance of small-scale structure in the star-less HVCs
are based on physical mechanisms that must be different from those in
in the disk, where supernova explosions and other stellar feedback
processes are known to stir up the ISM and form (mostly short-lived)
interstellar structures on sub-pc/AU scales (de\,Avillez \& Breitschwert 2007).


\begin{figure*}[h!]
\begin{center}
\resizebox{0.8\hsize}{!}{\includegraphics{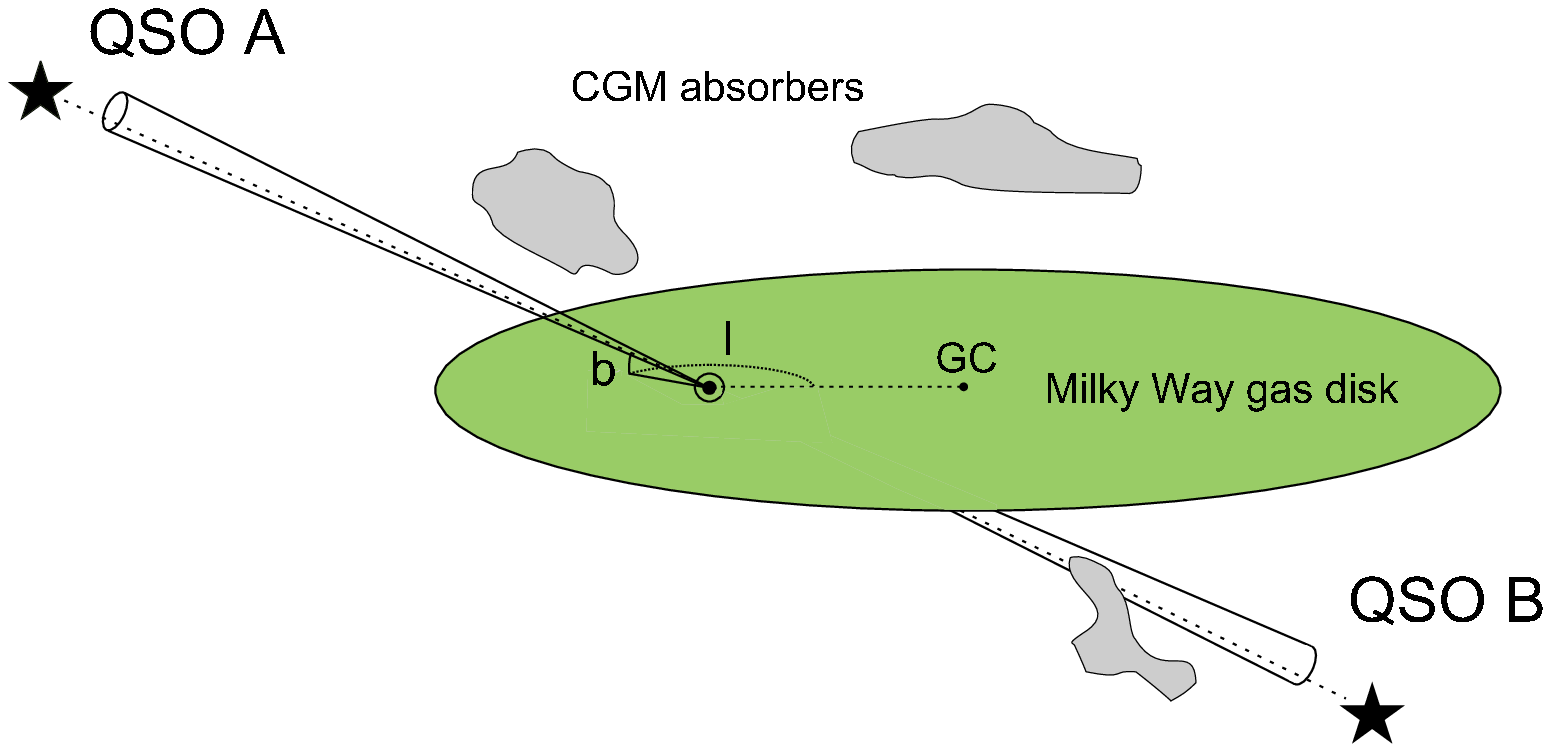}}
\caption[]{
Illustration of concept of identifying antipodal QSO sightlines
at $(l_1,b_1)$,$(l_2,b_2)$ passing the Galactic disk and CGM, as seen
from an exterior vantage point of the Milky Way.
We consider antipodal sightline pairs in cones with an opening angles of
$\theta=6\degree$ in our sample.
}
\end{center}
\end{figure*}


\begin{table*}[h!]
\caption[]{COS QSO sample$^{\rm a}$}
\begin{tabular}{llrr}
\hline
No. & QSO (COS name) & $l$         & $b$ \\
    &                & [$\degree$] & [$\degree$] \\
\hline
        1 & J142947.03+064334.9       &        356.3  &   59.0   \\
        2 & RX\,J2154.1-4414          &        355.2  &  $-$50.9 \\
        3 & 2MASS-J14294076+0321257   &        351.8  &   56.6   \\
        4 & PKS2005-489               &        350.4  &  $-$32.6 \\
        5 & LBQS-1435-0134            &        348.7  &   51.4   \\
        6 & RBS1892                   &        345.9  &  $-$58.4 \\
        7 & QSO-B1435-0645            &        344.0  &   47.2   \\
        8 & VV2006-J140655.7+015713   &        341.8  &   59.0   \\
        9 & SDSS-J135726.27+043541.4  &        340.8  &   62.5   \\
       10 & IRAS-F22456-5125          &        338.5  &  $-$56.6 \\
       11 & ESO-141-55                &        338.2  &  $-$26.7 \\
       12 & HE2347-4342               &        336.0  &  $-$69.6 \\
       13 & SDSSJ134206.56+050523.8   &        333.9  &   64.9   \\
       14 & 1H-2129-624               &        331.1  &  $-$42.5 \\
       15 & VV2006-J131545.2+152556   &        329.9  &   77.0   \\
\hline
\end{tabular}
\\
{\small $^{\rm a}$\,The full version of Table A.1 is only available in electronic form at the CDS via anonymous ftp to\\
{\tt cdsarc.u-strasbg.fr} (130.79.128.5) or via {\tt http://cdsweb.u-strasbg.fr/cgi-bin/qcat?J/A+A/}}.\\
\end{table*}


\begin{table*}[h!]
\caption[]{Summary of HVC absorption-line measurements$^{\rm a}$}
\begin{small}
\begin{tabular}{lllrrrrrrrr}
\hline
No. & QSO (COS name) & Ion & $\lambda_0$ & S/N$^{\rm b}$ & $v_{\rm min}$  & $v_{\rm max}$  & $W_{\rm \lambda}^{\rm c}$ & 
$\Delta W_{\lambda}$ & log $N^{\rm d}$ & $\Delta$(log $N$) \\
    &                &     & [\AA]       &               & [km\,s$^{-1}$] & [km\,s$^{-1}$] & [m\AA]           & [m\AA]               &         & \\ 
\hline
        1 & J142947.03+064334.9       &     CII    &   1334.5 &  6  &   -155  &   -100  &    145  &     27 &  $>$14.03  &        \\
        2 & J142947.03+064334.9       &     CIV    &   1548.2 &  6  &   -165  &    -99  &    123  &     35 &     13.77  &   0.13 \\
        3 & J142947.03+064334.9       &     CIV    &   1550.8 &  6  &   -131  &   -100  &     42  &     22 &     13.55  &   0.16 \\
        4 & J142947.03+064334.9       &     SiII   &   1190.4 &  5  &   -133  &   -103  &     50  &     19 &     13.17  &   0.20 \\
        5 & J142947.03+064334.9       &     SiII   &   1193.3 &  5  &   -149  &   -100  &     94  &     26 &  $>$13.32  &        \\
        6 & J142947.03+064334.9       &     SiII   &   1260.4 &  6  &   -155  &   -100  &    124  &     23 &  $>$13.15  &        \\
        7 & J142947.03+064334.9       &     SiII   &   1526.7 &  5  &   -142  &    -99  &     59  &     32 &     13.64  &   0.26 \\
        8 & J142947.03+064334.9       &     SiIII  &   1206.5 &  4  &   -167  &    -99  &    122  &     39 &  $>$12.81  &        \\
\hline
        9 & PKS2005-489               &     CII    &   1334.5 & 24  &    101  &    202  &    125  &     11 &     13.89  &   0.06 \\
       10 & PKS2005-489               &     CIV    &   1548.2 & 18  &    101  &    167  &     91  &     13 &     13.50  &   0.10 \\
       11 & PKS2005-489               &     SiII   &   1193.3 & 25  &    101  &    187  &     68  &      9 &     13.11  &   0.08 \\
       12 & PKS2005-489               &     SiII   &   1260.4 & 34  &     99  &    206  &    123  &      8 &     13.07  &   0.04 \\
       13 & PKS2005-489               &     SiIII  &   1206.5 & 25  &    101  &    203  &    143  &      9 &     13.00  &   0.06 \\
\hline
       14 & LBQS-1435-0134            &     CII    &   1334.5 & 31  &   -136  &   -100  &     57  &      4 &     13.59  &   0.07 \\
       15 & LBQS-1435-0134            &     CIV    &   1548.2 & 35  &   -183  &   -100  &    118  &      8 &     13.58  &   0.05 \\
       16 & LBQS-1435-0134            &     SiII   &   1190.4 & 32  &   -121  &   -100  &         &        &            &        \\
       17 & LBQS-1435-0134            &     SiII   &   1260.4 & 39  &   -156  &   -100  &     59  &      4 &     12.75  &   0.05 \\
       18 & LBQS-1435-0134            &     SiIII  &   1206.5 & 35  &   -180  &   -100  &     49  &      6 &     12.44  &   0.07 \\
\hline
\end{tabular}
\end{small}
\\
{\small $^{\rm a}$\,The full version of Table A.2 is only available in electronic form at the CDS via anonymous ftp to\\
{\tt cdsarc.u-strasbg.fr} (130.79.128.5) or via {\tt http://cdsweb.u-strasbg.fr/cgi-bin/qcat?J/A+A/}}.\\
{\small $^{\rm b}$\,S/N per resolution element.}\\
{\small $^{\rm c}$\,A ``t'' in front of the equivalent width indicates a tentative HVC detection.}\\
{\small $^{\rm d}$\,For each ion, the maximum value for $N$ is adopted for the further column-density analysis (see Sect.\,2.2).}
\end{table*}


\begin{table*}[t!]
\caption[]{Compilation of Local Group galaxies}
\begin{normalsize}
\begin{tabular}{lllllllrrrr}
\hline
No. & Galaxy name & Alternative name & $\alpha(2000)$ & $\delta(2000)$ & Type &
$d$ & $M_V$ & $v_{\rm rad}$ & $l$ & $b$ \\
&  & & [h\,m] & [$\degree$\,m] & & [kpc] & [mag] & [km\,s$^{-1}$] & [$\degree$] & [$\degree$]\\
\hline
 1 & M31      &   NGC 224 & 00 40.0 & +40 59 & Sb  &   770  & $-21.1$ & $-299$ &   121.2 &  $-21.6$ \\
 2 & M33      &   NGC 598 & 01 31.1 & +30 24 & Sc  &   850  & $-18.9$ & $-180$ &   133.6 &  $-31.3$ \\
 3 & LMC      &           & 05 24.0 & -69 48 & Irr &   49   & $-18.1$ &  $270$ &   280.5 &  $-32.9$ \\
 4 & IC 10    &           & 00 17.7 & +59 01 & Irr &   660  & $-17.6$ & $-343$ &   119.0 &   $-3.3$ \\
 5 & NGC 6822 &   DDO 209 & 19 42.1 & -14 56 & Irr &   540  & $-16.4$ &  $-49$ &    25.3 &  $-18.4$ \\ 
 6 & M32      &   NGC 221 & 00 40.0 & +40 36 & E2  &   770  & $-16.4$ & $-205$ &   121.2 &  $-22.0$ \\
 7 & NGC 205  &           & 00 37.6 & +41 25 & E5  &   830  & $-16.3$ & $-239$ &   120.7 &  $-21.7$ \\ 
 8 & SMC      &           & 00 51.0 & -73 06 & Irr &   58   & $-16.2$ &  $163$ &   302.8 &  $-44.3$ \\
 9 & NGC 3109 &   DDO 236 & 10 00.8 & -25 55 & Irr &   1260 & $-15.8$ &  $403$ &   262.1 &  $+23.1$ \\
10 & NGC 185  &           & 00 36.2 & +48 04 & E3  &   620  & $-15.3$ & $-208$ &   120.8 &  $-14.5$ \\
11 & IC 1613  &   DDO 8   & 01 02.2 & +01 51 & Irr &   715  & $-14.9$ & $-236$ &   129.8 &  $-60.6$ \\
12 & NGC 147   &  DDO 3   & 00 30.5 & +48 14 & E4  &   755  & $-14.8$ & $-193$ &   119.8 &  $-14.3$ \\
13 & Sextans A &  DDO 75  & 10 08.6 & -04 28 & Irr &   1450 & $-14.4$ &  $325$ &   246.2 &  $+39.9$ \\
14 & Sextans B &  DDO 70  & 09 57.4 & +05 34 & Irr &   1300 & $-14.3$ &  $301$ &   233.2 &  $+43.8$ \\
15 & WLM       &   DDO 221& 23 59.4 & -15 45 & Irr &   940  & $-14.0$ & $-116$ &    75.9 &  $-73.6$ \\
16 & Sagittarius &        & 18 51.9 & -30 30 & dE7 &   24   & $-14.0$ &  $140$ &     5.6 &  $-14.1$ \\
17 & Fornax     &         & 02 37.8 & -34 44 & dE3 &   131  & $-13.0$ &   $53$ &   237.1 &  $-65.7$ \\
18 & Pegasus    & DDO 216 & 23 26.1 & +14 28 & Irr &   760  & $-12.7$ & $-181$ &    94.8 &  $-43.5$ \\
19 & And VII    & Cas  Dw & 23 24.1 & +50 25 & dE3 &   760  & $-12.0$ & $-307$ &   109.5 &  $-10.0$ \\
20 & Leo I      & DDO 74  & 10 05.8 & +12 33 & dE3 &   251  & $-12.0$ &  $285$ &   226.0 &  $+49.1$ \\
21 & Leo A      & DDO 69  & 09 56.5 & +30 59 & Irr &   692  & $-11.7$ &   $26$ &   196.9 &  $+52.4$ \\
22 & And II     &         & 01 13.5 & +33 09 & dE3 &   680  & $-11.7$ & $-188$ &   128.9 &  $-29.2$ \\
23 & And I      &         & 00 43.0 & +37 44 & dE0 &   790  & $-11.7$ & $-380$ &   121.7 &  $-24.9$ \\
24 & And VI     & Peg  Dw & 23 49.2 & +24 18 & dE3 &   815  & $-11.3$ & $-354$ &   106.1 &  $-36.3$ \\
25 & SagDIG     &         & 19 27.9 & -17 47 & Irr &   1150 & $-11.0$ &  $-79$ &    21.1 &  $-16.3$ \\
26 & Aquarius   & DDO 210 & 20 44.2 & -13 01 & Irr &   950  & $-10.9$ & $-131$ &    34.0 &  $-31.3$ \\
27 & Antlia     &         & 10 01.8 & -27 05 & dE3 &   1150 & $-10.7$ &  $361$ &   263.1 &  $+22.3$ \\
28 & Sculptor   &         & 00 57.6 & -33 58 & dE  &   78   & $-10.7$ &  $107$ &   287.5 &  $-83.2$ \\
29 & And III    &         & 00 32.6 & +36 12 & dE6 &   760  & $-10.2$ & $-355$ &   119.3 &  $-26.2$ \\
30 & Leo II     & DDO 93  & 11 10.8 & +22 26 & dE0 &   230  & $-10.2$ &   $76$ &   220.2 &  $+67.2$ \\
31 & Cetus      &         & 00 23.6 & -11 19 & dE4 &   800  & $-10.1$ &  $-87$ &   101.4 &  $-72.8$ \\
32 & Sextans    &         & 10 10.6 & -01 24 & dE4 &  90    & $-10.0$ &  $224$ &   243.4 &  $+42.2$ \\
33 & Phoenix    &         & 01 49.0 & -44 42 & Irr &   390  & $-9.9$ &    $56$ &   272.2 &  $-68.9$ \\
34 & LGS 3      &         & 01 01.2 & +21 37 & Irr/dE & 810 & $-9.7$ &  $-277$ &   126.8 &  $-40.9$ \\
35 & Tucana     &         & 22 38.5 & -64 41 & dE5 &   900  &  $-9.6$ &  $182$ &   322.9 &  $-47.4$ \\
36 & Carina     &         & 06 40.4 & -50 55 & dE4 &   87   &  $-9.2$ &  $223$ &   260.1 &  $-22.2$ \\
37 & And V      &         & 01 07.3 & +47 22 & dE  &   810  &  $-9.1$ & $-403$ &   126.2 &  $-15.1$ \\
38 & Ursa Minor & DDO 199 & 15 08.2 & +67 23 & dE5 &   69   &  $-8.9$ & $-250$ &   105.0 &  $+44.8$ \\
39 & Draco      & DDO 228 & 17 19.2 & +57 58 & dE3 &   76   &  $-8.6$ & $-289$ &    86.4 &  $+34.7$ \\
40 & And IX     &         & 00 50.0 & +42 57 & dE  &   780  &  $-8.3$ & $-208$ &   123.2 &  $-19.7$ \\
\hline
\end{tabular}
\noindent
\\
\end{normalsize}
\end{table*}


\begin{table*}[h!]
\caption[]{Neutral/total hydrogen column densities and neutral gas fractions in HVC absorbers$^{\rm a}$} 
\begin{tabular}{llrrrr}
\hline
No. & QSO & $\langle v_{\rm HVC}\rangle$ & log $N$(H\,{\sc i}) & log $N$(H) & log $f_{\rm HI}$ \\
    &     & [km\,s$^{-1}$]               &                     &            & \\
\hline
        1 & J142947.03+064334.9         &         -133 &  $<18.70$   &      $>19.19$    &     $<-0.49$ \\
        2 & PKS2005-489                 &          152 &  $<18.70$   &      $ 18.85$    &     $<-0.15$ \\
        3 & LBQS-1435-0134              &         -140 &  $<18.70$   &      $ 18.46$    &     $< 0.24$ \\
        4 & QSO-B1435-0645              &         -119 &  $<18.70$   &      $ 18.55$    &     $< 0.15$ \\
        5 & VV2006-J140655.7+015713     &         -131 &  $<18.70$   &      $ 18.55$    &     $< 0.15$ \\
        6 & SDSS-J135726.27+043541.4    &         -116 &  $<18.70$   &      $>18.62$    &     $< 0.08$ \\
        7 & IRAS-F22456-5125            &          146 &  $<18.70$   &      $ 18.23$    &     $< 0.47$ \\
        8 & SDSSJ134206.56+050523.8     &         -106 &  $<18.70$   &      $ 17.86$    &     $< 0.84$ \\
        9 & 1H-2129-624                 &          186 &  $<18.70$   &      $ 18.62$    &     $< 0.08$ \\
       10 & 1H-2129-624                 &          129 &  $<18.70$   &      $ 18.31$    &     $< 0.39$ \\
       11 & SDSSJ134251.60-005345.3     &         -114 &  $<18.70$   &      $ 18.94$    &     $<-0.24$ \\
       12 & HARO11                      &          142 &  $ 18.94$   &      $ 18.94$    &     $< 0.00$ \\
       13 & HARO11                      &         -164 &  $ 18.92$   &      $>19.16$    &     $<-0.22$ \\
       14 & RXS-J00057-5007             &          130 &  $<18.70$   &      $>19.19$    &     $<-0.49$ \\
       15 & RBS144                      &          156 &  $ 19.77$   &      $ 19.77$    &     $0.00$   \\
\hline
\end{tabular}
\noindent
\\
{\small $^{\rm a}$\,The full version of Table A.4 is only available in electronic form at the CDS via anonymous ftp to\\
{\tt cdsarc.u-strasbg.fr} (130.79.128.5) or via {\tt http://cdsweb.u-strasbg.fr/cgi-bin/qcat?J/A+A/}}.\\
\end{table*}


\begin{table*}[h!]
\caption[]{Antipodal sightline pairs}
\begin{tabular}{llrrl}
\hline
Pair no. & QSO & $l$         & $b$            & HVC Complex \\
         &     & [$\degree$] & [$\degree$]    & \\
\hline
         1 & 2MASS-J14294076+0321257        &     351.8   &    56.6  &  Complex L \\
         1 & HB89-0232-042                  &     174.5   &   -56.2  &            \\
         2 & PKS2005-489                    &     350.4   &   -32.6  &  MagellanicStream \\
         2 & SDSSJ080908.13+461925.6        &     173.3   &    32.3  &  Complex A \\
         3 & QSO-B1435-0645                 &     344.0   &    47.2  &  Complex L \\
         3 & MRK595                         &     164.8   &   -46.5  &            \\
         4 & IRAS-F22456-5125               &     338.5   &   -56.6  &  MagellanicStream \\
         4 & 1ES1028+511                    &     161.4   &    54.4  &  Complex M \\
         5 & IRAS-F22456-5125               &     338.5   &   -56.6  &  MagellanicStream \\
         5 & 1SAXJ1032.3+5051               &     161.4   &    54.6  &  Complex M \\
         6 & 1H-2129-624                    &     331.1   &   -42.5  &  MagellanicStream \\
         6 & SDSSJ092837.98+602521.0        &     154.1   &    42.4  &  Complex A \\
         7 & VV2006-J130524.3+035731        &     311.7   &    66.6  &            \\
         7 & LBQS-0107-0235                 &     134.0   &   -64.8  &  MagellanicStream \\
         8 & UKS-0242-724                   &     291.8   &   -42.4  &  MagellanicStream \\
         8 & SDSSJ145907.58+714019.9        &     110.0   &    42.1  &  Complex C \\
         9 & ESO-031--G-008                 &     290.3   &   -40.8  &  MagellanicStream \\
         9 & SDSSJ145907.58+714019.9        &     110.0   &    42.1  &  Complex C \\
        10 & HE1159-1338                    &     285.1   &    47.2  &            \\
        10 & PG0003+158                     &     107.3   &   -45.3  &  MagellanicStream \\
        11 & CAL-F-COPY                     &     277.2   &   -35.4  &  MagellanicStream \\
        11 & SDSSJ171737.95+655939.3        &      96.1   &    34.1  &  Complex C \\
        12 & CAL-F-COPY                     &     277.2   &   -35.4  &  MagellanicStream \\
        12 & HS1700+6416                    &      94.4   &    36.2  &  Complex C \\
        13 & IRAS-L06229-6434               &     274.3   &   -27.3  &  MagellanicStream \\
        13 & H1821+643                      &      94.0   &    27.4  &  Complex C \\
        14 & QSO-B1126-041                  &     267.6   &    52.7  &            \\
        14 & NGC7714                        &      88.2   &   -55.6  &  MagellanicStream \\
        15 & IRAS-F04250-5718               &     267.0   &   -42.0  &  MagellanicStream \\
        15 & SBS1624+575                    &      87.3   &    42.0  &  Complex C \\
        16 & IRAS-F04250-5718               &     267.0   &   -42.0  &  MagellanicStream \\
        16 & PG1626+554                     &      84.5   &    42.2  &  Complex C \\
        17 & 2DFGRSS393Z082                 &     226.6   &   -65.0  &  MagellanicStream \\
        17 & SDSS-J145424.33+304658.3       &      48.1   &    62.9  &                   \\
        18 & SDSSJ094733.21+100508.7        &     225.4   &    43.5  &  Complex WAWB \\
        18 & PHL1811                        &      47.5   &   -44.8  &  Complex GCN \\
        19 & TONS210                        &     225.0   &   -83.2  &  MagellanicStream \\
        19 & SDSSJ133045.15+281321.4        &      42.4   &    81.2  & \\
        20 & PKS0405-123                    &     204.9   &   -41.8  & \\
        20 & SDSS-J160519.70+144852.2       &      27.8   &    43.4  & \\
        21 & PKS0405-123                    &     204.9   &   -41.8  & \\
        21 & 1ES1553+113                    &      21.9   &    44.0  & \\
        22 & BZBJ1001+2911                  &     199.5   &    52.6  &  Complex M \\
        22 & PKS2155-304                    &      17.7   &   -52.2  &  Complex GCN \\
        23 & FBQSJ1010+3003                 &     198.4   &    54.6  &  Complex M \\
        23 & PKS2155-304                    &      17.7   &   -52.2  &  Complex GCN \\
        24 & 2MASS-J09591486+3203573        &     194.7   &    52.5  &  Complex M \\
        24 & PKS2155-304                    &      17.7   &   -52.2  &  Complex GCN \\
\hline
\end{tabular}
\noindent
\\
\end{table*}


\section{Sightline pairs and HVC finding charts}


\begin{figure}[h!]
\begin{center}
\caption[]{
Full set of velocity plots for all 12 sightline pairs with angular
separations $\leq 1$ deg (see Fig.\,13 for a detailed description;
FULL SET OF VELOCITY PLOTS WILL BE MADE AVAILABLE ON REQUEST).
}
\end{center}
\end{figure}


\begin{figure}[h!]
\begin{center}
\caption[]{
Full set for HVC finding charts for all 270 sightlines in our survey
(see Fig.\,1 for a detailed description). LG galaxies close to the
sightlines have been indicated with the red ID numbers (Table A.3, first
row; Sect.\,5.3) and red tic marks in the upper panels. Note that
the continuum shown here (and in Fig.\,1) represents the global continuum
derived from the automated continuum fitting procedure (Sect.\,2.2),
but not the exact continuum was used for the measurements
(FULL SET OF HVC FINDING CHARTS WILL BE MADE AVAILABLE ON REQUEST).
}
\end{center}
\end{figure}


\end{appendix}


\end{document}